\documentclass[longauth]{aa}

\usepackage{txfonts}
\usepackage{graphicx}
\usepackage{subfigure}
\usepackage[colorinlistoftodos]{todonotes}
\usepackage{color, colortbl, xcolor}
\usepackage[colorlinks=true, allcolors=blue]{hyperref}
\definecolor{Red}{rgb}{0.9,0,0}
\definecolor{Blue}{rgb}{0,0,0.9}
  \definecolor{Green}{rgb}{0,0.5,0}
\definecolor{Black}{rgb}{0,0,0}

\usepackage{natbib}
\bibpunct{(}{)}{;}{a}{}{,} 
\usepackage{ulem}
\usepackage{soul}
\setstcolor{red}
\usepackage{multirow}
\usepackage{booktabs}
\usepackage{dcolumn}

\begin{document}

   \title{Refined physical parameters for Chariklo's body and rings from stellar occultations observed between 2013 and 2020} 
   \titlerunning{Chariklo's system from stellar occultations.}
   \authorrunning{B. E. Morgado, B. Sicardy, F. Braga-Ribas et al.,}

\author{B. E. Morgado\inst{1,2,3} 
\and
B. Sicardy\inst{1} 
\and
F. Braga-Ribas\inst{4,3,2,1} 
\and
J. Desmars\inst{5,6,1} 
\and
A. R. Gomes-Júnior\inst{7,2} 
\and
D. Bérard\inst{1} 
\and
R. Leiva\inst{8,9,1} 
\and
J. L. Ortiz\inst{10} 
\and
R. Vieira-Martins\inst{3,2,11} 
\and
G. Benedetti-Rossi\inst{1,2,7} 
\and
P. Santos-Sanz\inst{10} 
\and
J. I. B. Camargo\inst{3,2} 
\and
R. Duffard\inst{10} 
\and
F. L. Rommel\inst{3,2,4} 
\and
M. Assafin\inst{11,2} 
\and
R. C. Boufleur\inst{2,3} 
\and
F. Colas\inst{6} 
\and
M. Kretlow\inst{12,10,13} 
\and
W. Beisker\inst{12,13} 
\and
R. Sfair\inst{7} 
\and
C. Snodgrass\inst{14} 
\and
N. Morales\inst{10} 
\and
E. Fernández-Valenzuela\inst{15} 
\and
L. S. Amaral\inst{16} 
\and
A. Amarante\inst{17} 
\and
R. A. Artola\inst{18} 
\and
M. Backes\inst{19,20} 
\and
K-L. Bath\inst{12,13} 
\and
S. Bouley\inst{21} 
\and
M. W. Buie\inst{22} 
\and
P. Cacella\inst{23} 
\and
C. A. Colazo\inst{24} 
\and
J. P. Colque\inst{25} 
\and
J-L. Dauvergne\inst{26} 
\and
M. Dominik\inst{27} 
\and
M. Emilio\inst{28,3} 
\and
C. Erickson\inst{29} 
\and
R. Evans\inst{19} 
\and
J. Fabrega-Polleri\inst{30} 
\and
D. Garcia-Lambas\inst{18} 
\and
B. L. Giacchini\inst{31,32} 
\and
W. Hanna\inst{33} 
\and
D. Herald\inst{12,33} 
\and
G. Hesler\inst{34} 
\and
T. C. Hinse\inst{35,36} 
\and
C. Jacques\inst{37} 
\and
E. Jehin\inst{38} 
\and
U. G. J{\o}rgensen\inst{39} 
\and
S. Kerr\inst{33,40} 
\and
V. Kouprianov\inst{41,42} 
\and
S. E. Levine\inst{43,44} 
\and
T. Linder\inst{45} 
\and
P. D. Maley\inst{46,47} 
\and
D. I. Machado\inst{48,49} 
\and
L. Maquet\inst{6} 
\and
A. Maury\inst{50} 
\and
R. Melia\inst{18} 
\and
E. Meza\inst{1,51,52} 
\and
B. Mondon\inst{34} 
\and
T. Moura\inst{7} 
\and
J. Newman\inst{33} 
\and
T. Payet\inst{34} 
\and
C. L. Pereira\inst{4,3,2} 
\and
J. Pollock\inst{53} 
\and
R. C. Poltronieri\inst{54,16} 
\and
F. Quispe-Huaynasi\inst{3} 
\and
D. Reichart\inst{41} 
\and
T. de Santana\inst{1,7} 
\and
E. M. Schneiter\inst{18} 
\and
M. V. Sieyra\inst{55} 
\and
J. Skottfelt\inst{56} 
\and
J. F. Soulier\inst{57} 
\and
M. Starck\inst{58} 
\and
P. Thierry\inst{59} 
\and
P. J. Torres\inst{60,61} 
\and
L. L. Trabuco\inst{49} 
\and
E. Unda-Sanzana\inst{25} 
\and
T. A. R. Yamashita\inst{7} 
\and
O. C. Winter\inst{7} 
\and
A. Zapata\inst{62,63} 
\and
C. A. Zuluaga\inst{44} 
}

\institute{LESIA, Observatoire de Paris, Université PSL, CNRS, Sorbonne Université, Univ. Paris Diderot, Sorbonne Paris Cité, 5 place Jules Janssen, 92195 Meudon, France\\
\email{morgado.fis@gmail.com}
\newpage
\and
Laboratório Interinstitucional de e-Astronomia - LIneA, Rua Gal. José Cristino 77, Rio de Janeiro, RJ 20921-400, Brazil
\and
Observatório Nacional/MCTI, R. General José Cristino 77, CEP 20921-400 Rio de Janeiro - RJ, Brazil
\and
Federal University of Technology - Paraná (UTFPR / DAFIS), Rua Sete de Setembro, 3165, CEP 80230-901, Curitiba, PR, Brazil
\and
Institut Polytechnique des Sciences Avancées IPSA, 63 boulevard de Brandebourg, 94200 Ivry-sur-Seine, France
\and
Institut de Mécanique Céleste et de Calcul des Éphémérides, IMCCE, Observatoire de Paris, PSL Research University, CNRS,Sorbonne Universités, UPMC Univ Paris 06, Univ. Lille, 77, Av. Denfert-Rochereau, 75014 Paris, France
\and
UNESP - São Paulo State University, Grupo de Dinâmica Orbital e Planetologia, CEP 12516-410, Guaratinguetá, SP, Brazil
\and
Université Côte d’Azur, Observatoire de la Côte d’Azur, CNRS, Laboratoire Lagrange, Bd de l’Observatoire, CS 34229, 06304 Nice Cedex 4, France
\and
Departamento de Astronomía, Universidad de Chile, Camino del Observatorio 1515, Las Condes, Santiago, Chile
\and
Instituto de Astrofísica de Andalucía, IAA-CSIC, Glorieta de la Astronomía s/n, 18008 Granada, Spain
\and
Universidade Federal do Rio de Janeiro - Observatório do Valongo, Ladeira Pedro Antônio 43, CEP 20.080-090 Rio de Janeiro - RJ, Brazil
\and
International Occultation Timing Association / European Section, Am Brombeerhag 13, D-30459 Hannover, Germany
\and
Internationale Amateursternwarte e.V. (IAS), Mittelstrasse 6, D-15749 Mittenwalde, Germany
\and
Institute for Astronomy, University of Edinburgh, Royal Observatory, Edinburgh EH9 3HJ, UK
\and
Florida Space Institute, University of Central Florida, 12354 Research Parkway, Partnership 1, Orlando, FL, USA
\and
BRAMON—Brazilian Meteor Observation Network, Nhandeara, Brazil.
\and
Grupo de Matemática Aplicada e Processamento de Sinais, State University of Mato Grosso do Sul - UEMS, Cassilândia, CEP 79540-000, MS, Brazil
\and
IATE-OAC, Universidad Nacional de Córdoba-CONICET, Laprida 854, X5000 BGR, Córdoba, Argentina
\and
Department of Physics, Chemistry and Material Science, University of Namibia, Private Bag 13301, Windhoek, Namibia
\and
Centre for Space Research, North-West University, Potchefstroom 2520, South Africa
\and
Association T60, Observatoire Midi-Pyrénées, 14 avenue Edouard Belin, 31400 Toulouse, France
\and
Southwest Research Institute, 1050 Walnut Street, Suite 300, Boulder, CO 80302, USA
\and
Dogsheaven Observatory, SMPW Q25 CJ1 LT10B Brasilia, Brazil
\and
Association of Argentine Observatories of Minor Bodies (AOACM))
\and
Centro de Astronomía (CITEVA), Universidad de Antofagasta, Av. Angamos 601, Antofagasta, Chile
\and
Ciel \& Espace, Paris, France
\and
University of St Andrews, Centre for Exoplanet Science, SUPA School of Physics \& Astronomy, North Haugh, St Andrews, KY16 9SS, United Kingdom
\and
Universidade Estadual de Ponta Grossa (UEPG), Ponta Grossa, Brazil
\and
Summit Kinetics Inc, United States
\and
Panamanian Observatory in San Pedro de Atacama - OPSPA
\and
Department of Physics, Southern University of Science and Technology, Shenzhen 518055, China
\and
Centro Brasileiro de Pesquisas Físicas, Rua Dr Xavier Sigaud 150, Rio de Janeiro 22290-180, Brazil
\and
Trans-Tasman Occultation Alliance (TTOA), Wellington PO Box 3181, New Zealand
\and
Association Réunionnaise pour l'Etude du Ciel Austral - ARECA, Sainte-Marie, La Réunion, France
\newpage
\and
Institute of Astronomy, Faculty of Physics, Astronomy and Informatics, Nicolaus Copernicus University, Grudziadzka 5, 87-100, Torun, Poland
\and
Department of Astronomy and Space Science, Chungnam National University, 34134, Daejeon, Republic of Korea
\and
Observatório SONEAR, Brazil
\and
STAR Institute, Université de Liège, Allée du 6 août, 19C, 4000 Liège, Belgium
\and
Centre for ExoLife Sciences (CELS), Niels Bohr Institute, Øster Voldgade 5, 1350 Copenhagen, Denmark
\and
Astronomical Association of Queensland, 5 Curtis Street, Pimpama QLD 4209, Australia
\and
Department of Physics and Astronomy, University of North Carolina, Chapel Hill
\and
Central (Pulkovo) Observatory of the Russian Academy of Sciences, 196140, 65/1 Pulkovskoye Ave., Saint Petersburg, Russia
\and
Lowell Observatory, 1400 W Mars Hill Road, Flagstaff, AZ 86001, USA
\and
Massachusetts Institute of Technology, Department of Earth, Atmospheric and Planetary Sciences, 77 Massachusetts Avenue, Cambridge, MA 02139, USA)
\and
The Astronomical Research Institute, Ashmore, IL
\and
International Occultation Timing Association (IOTA), PO Box 7152, WA, 98042, USA
\and
NASA Johnson Space Center Astronomical Society, Houston, TX, USA
\and
Universidade Estadual do Oeste do Paraná (Unioeste), Avenida Tarquínio Joslin dos Santos 1300, Foz do Iguaçu, PR, 85870-650, Brazil
\and
Polo Astronômico Casimiro Montenegro Filho/FPTI-BR, Avenida Tancredo Neves 6731, Foz do Iguaçu, PR, 85867-900, Brazil
\and
San Pedro de Atacama Celestial Explorations - SPACE, Chile
\and
Comisión Nacional de Investigación y Desarrollo Aeroespacial del Perú - CONIDA, Luis Felipe Villarán 1069, San Isidro, Lima, Perú.
\and
Observatorio Astronómico de Moquegua, CP Cambrune, Carumas, Moquegua, Perú
\and
Physics and Astronomy Department, Appalachian State University, Boone, NC 28608, USA
\and
Astrocan Clube de Astronomia  Nhandeara
\and
Centre for mathematical Plasma Astrophysics, Department of Mathematics, KU Leuven, Celestijnenlaan 200B, B-3001 Leuven, Belgium
\and
Centre for Electronic Imaging, Department of Physical Sciences, The Open University, Milton Keynes, MK7 6AA, UK
\and
Association Des Étoiles pour Tous, 19 Rue Saint Laurent, Maisoncelles, F-77320 Saint Martin du Boschet, France
\and
Observatorio Astronómico Córdoba UNC, Laprida 854, Córdoba, Argentina.
\and
(AGORA observatoire des Makes, AGORA, 18 Rue Georges Bizet, Observatoire des Makes, 97421 La Rivière, France
\and
Institute of Astrophysics, Pontificia Universidad Catolica de Chile, Av. Vicuña Mackenna 4860, Santiago, Chile
\and
 Millennium Institute of Astrophysics, Chile
\and
Centre of Astro-Engineering, Pontificia Universidad Catolica de Chile, Av. Vicuña Mackenna 4860, Santiago, Chile.
\and
Department of Electrical Engineering, Pontificia Universidad Catolica de Chile, Av. Vicuña Mackenna 4860, Santiago, Chile}

   \date{Received 14/06/2021; accepted 12/07/2021}

 
  \abstract
   {The Centaur (10199) Chariklo has the first rings system discovered around a small object. It was first observed using stellar occultation in 2013. Stellar occultations allow the determination of sizes and shapes with kilometre accuracy and obtain characteristics of the occulting object and its vicinity. 
   } 
   {Using stellar occultations observed between 2017 and 2020, we aim at constraining Chariklo’s and its rings physical parameters. We also determine the rings' structure, and obtain precise astrometrical positions of Chariklo.
   }
   {We predicted and organised several observational campaigns of stellar occultations by Chariklo. Occultation light curves were measured from the data sets, from which ingress and egress times, and rings' width and opacity were obtained. These measurements, combined with results from previous works, allow us to obtain significant constraints on Chariklo's shape and rings' structure.
   }
   {We characterise Chariklo's ring system (C1R and C2R), and  obtain radii and pole orientations that are consistent with, but more accurate than, results from previous occultations. We confirmed the detection of W-shaped structures within C1R and an evident variation of radial width. The observed width ranges between 4.8 and 9.1 km with a mean value of 6.5 km. One dual observation (visible and red) does not reveal any differences in the C1R opacity profiles, indicating ring particle's size larger than a few microns. The C1R ring eccentricity is found to be smaller than 0.022 (3$\sigma$), and its width variations may indicate an eccentricity higher than $\sim$0.005. We fit a tri-axial shape to Chariklo's detections over eleven occultations and determine that Chariklo is consistent with an ellipsoid with semi-axes of $143.8^{+1.4}_{-1.5}$, $135.2^{+1.4}_{-2.8}$ and $99.1^{+5.4}_{-2.7}$ km. Ultimately, we provided seven astrometric positions at a milliarcseconds accuracy level, based on Gaia EDR3, and use it to improve Chariklo's ephemeris.}
   {}
   \maketitle 
%

\section{Introduction} \label{intro}

The Centaur (10199) Chariklo is a small object in our Solar System moving on an elliptical orbit between Saturn and Uranus, at heliocentric distances varying from 13.1 to 18.9 au. It was discovered in 1997 \citep{Ticha_1997MPEC}, it is the largest Centaur known to date. From thermal infrared observations, its surface equivalent radius ranges between 109 and 151 km \citep{Jewitt_1998, Altenhoff_2001, Campins_2002, Sekiguchi_2012, Bauer_2013, Fornasier_2013, Fornasier_2014, Lellouch_2017}, with a most recent solution of 121~$\pm$~4 km \citep{Lellouch_2017} obtained with the Atacama Large Millimeter/submillimeter Array (ALMA\footnote{\url{https://www.almaobservatory.org/en/home/}}). 

From a stellar occultation observed in 2013, \cite{Braga-Ribas_2014} reported the discovery of two dense and narrow rings (named 2013C1R and 2013C2R, C1R and C2R for short) surrounding Chariklo at 390 and 405~km from the body centre, respectively. This was the first time that rings were observed elsewhere than around giant planets. Being narrow and dense, they bear resemblance with some of Uranus' rings \citep{Elliot_1984, French_1991}. More on Chariklo's rings properties (orbital radii and pole, width and opacity) is given by \cite{Berard_2017} and are reviewed in \cite{Sicardy_2018}.

Another ring was later-on discovered around the dwarf planet (136108) Haumea \citep{Ortiz_2017}, and there are evidences (yet to be confirmed) of a ring (or a dust shell) around the Centaur (2060) Chiron \citep{Ortiz_2015,Sickafoose_2020}. This raises new questions on these ring systems origins, dynamical evolution and stability, and this has been studied by several authors, see for example \cite{Pan_2016, Hyodo_2016, Melita_2017, Michikoshi_2017, Araujo_2018, Sicardy_2018, Sicardy_2019} and references therein. Dynamical models are essential to address these questions and better understand the orbital evolution of ring particles around these small objects. On the other hand, new observations can be used to constrain these dynamical models and provide new insights into the origin and evolution of rings around small bodies.

Chariklo's ring system has a diameter of $\sim$800 km, subtending about 80 milliarcseconds (mas) when projected onto the sky plane as seen from Earth. As a consequence, resolved ground-based imaging of this system is extremely challenging with currently available equipment. For instance, \cite{Sicardy_2015DPS} observed this system with the SPHERE, an Adaptive Optics system attached at the 8.2-m Very Large Telescope (VLT/ESO) and obtained Point Spread Function (PSFs) of 30-40 mas (corresponding to $\sim$300-400 km at Chariklo's distance). These authors also observed the system with the Hubble Space Telescope (HST), with typical PSFs of 30 mas. None of those observations were able to detect Chariklo's rings, or find an extended system of cometary jets or tenuous dust around the body. In contrast, stellar occultations can achieve resolutions at km-level and provide stringent constraints on the size and shape of the main body and its astrometric position at mas level. Moreover, optical depths and inner structures of the rings can be revealed.

Between 2013 and 2016, thirteen stellar occultations by Chariklo's system were observed, from those only the occultation of 2013-06-03 \citep{Braga-Ribas_2014, Berard_2017} has more than two chords detections over C1R. As there was no occultation with three or more detections of the main body, \cite{Leiva_2017} used five of single-chord and double-chords occultations to derive possibles 3D shapes of Chariklo's main body. \cite{Berard_2017} mentions that a two-chord occultation by CR1 was observed from Australia on 2016-10-01, see Fig.~3 of that paper. This event was also a two-chord on the main body, and used by \cite{Leiva_2017}.

These authors derived four possible solutions for Chariklo's shape: $(i)$ a sphere with radius $R=129 \pm 3$~km; $(ii)$ a Maclaurin spheroid with semi-axes $a=b=143^{+3}_{-6}$~km and $c=96^{+14}_{-4}$~km; $(iii)$ a Jacobi ellipsoid with  $a=157 \pm 4$~km, $b=139 \pm 4$~km and $c=86 \pm 1$~km; and $(iv)$ a triaxial ellipsoid with $a=148^{+6}_{-4}$~km, $b=132^{+6}_{-5}$~km and $c=102^{+10}_{-8}$~km.

\cite{Berard_2017} confirmed the presence of Chariklo's rings and their geometry as published by \cite{Braga-Ribas_2014}. Moreover, they provided for the first time resolved opacity profiles of CR1, revealing a W-shaped structure, and found significant azimuthal width variations for that ring, with values ranging between 5.0 and 7.5~km. This is reminiscent of the ring width variations found among Uranus' dense rings, in particular its widest one, the $\epsilon$ ring \citep{Nicholson_2018}.

In this paper, we present the results of eight unpublished stellar occultations by Chariklo and its ring system observed between 2017 and 2020. It contains the analysis of three multi-chord detections of Chariklo's main body, two multi-chord detections of C1R and one multi-chord detection of C2R. The parameters obtained here can be used to further constrain the dynamical models of the rings, and the size and shape of Chariklo.

In Section \ref{sc:pred} we describe our prediction process and the observational campaigns. Section \ref{sc:redu} details the reduction process. In Section \ref{sc:rings} we present our results concerning Chariklo's rings, with the estimation of pole orientation (\ref{sc:C1R_pole}), C1R structures and global parameters (\ref{sc:C1R_structure}) and C2R global parameters (\ref{sc:C2R}). Section \ref{sc:main_body} contains the analysis of Chariklo's 3D shape considering 11 stellar occultation chords obtained between 2013 and 2020. In Section \ref{sc:astrometry} we present Chariklo's astrometric solution derived from our work. Conclusions are provided in Section \ref{sc:conclusion}.

\section{Predictions and Observational Campaigns} \label{sc:pred}

All the events presented here were predicted in the framework of the European Research Council (ERC) \textit{Lucky Star} project\footnote{\url{https://lesia.obspm.fr/lucky-star/index.php}}, and are publicly available for the community. After 2016, the Gaia DR1 (GDR1) catalogue \citep{Gaia_Mission, Gaiadr1_2016} was the source of the stellar positions for our prediction pipeline \citep{Assafin_2012, Camargo_2014, Desmars_2015}. After December 2018, it was replaced by the Gaia DR2 (GDR2) catalogue \citep{Gaiadr2_2018}, and by Gaia EDR3 (GEDR3) after the beginning of 2021 \citep{Gaia_EDR3}. Regular astrometric programs were carried out at the European Southern Observatory (ESO), the Observatório do Pico dos Dias (OPD/LNA) and the Observatorio de Sierra Nevada (OSN), among others, to improve the ephemerides of various objects. 

Selected events triggered observational campaigns, involving both fixed and portable telescopes that were deployed along the predicted shadow paths. Besides professional astronomers, they also involved citizen astronomer communities, such as the International Occultation Timing Association (IOTA/Europe\footnote{\url{https://www.iota-es.de/}}), Planoccult\footnote{\url{http://vps.vvs.be/mailman/listinfo/planoccult}}, and Occult Watcher\footnote{\url{https://www.occultwatcher.net/}}. 

Chariklo's ephemeris was pinned down to the five-milliarcsec (mas) level ($\sim$50 km at Chariklo's distance, \citealt{Desmars_2017DPS}) using previous stellar occultations published by \cite{Braga-Ribas_2014,Leiva_2017,Berard_2017}, GDR2, and the \textsc{NIMA} \citep[Numerical Integration of the Motion of an Asteroid][]{Desmars_2015} integrator. This uncertainty, being smaller than the size of Chariklo, allowed the detection of multi-chord stellar occultations by Chariklo and its rings between 2017 and 2020.

Table \ref{tb:predictions} provides information related to the observed events. It lists the geocentric closest approach time (UTC), the Gaia EDR3 source identifier, the geocentric ICRS coordinates of the stars at occultation epoch, applying proper motion, parallax, and radial velocity (when available). The last column of the table lists the (Gaia) G magnitude of the stars, which can be compared to Chariklo's V magnitude, $\sim$18.9.

\begin{table*}[h]
\begin{center}
\caption{Occulted stars parameters for each observed event as obtained from Gaia EDR3.}
\begin{tabular}{ccccc}
\hline
\hline
Occ. date and time & Gaia EDR3          & Right Ascension$^1$ & Declination$^1$ & G mag \\
UTC                & source identifier  &                     &                 &  \\
\hline
\hline

2020-06-19 15:51 & 6852214948874014720 & 20$^h$ 02$^m$ 31$^s$.99612 $\pm$ 0.21 mas & -22$^\circ$ 20' 41''.4784 $\pm$ 0.14 mas & 15.569  \\
2019-09-04 08:03 & 6766511277365829760 & 19$^h$ 27$^m$ 42$^s$.15144 $\pm$ 0.09 mas & -25$^\circ$ 29' 21''.2488 $\pm$ 0.06 mas & 13.117  \\
2019-08-08 21:41 & 6766365630734160896 & 19$^h$ 31$^m$ 49$^s$.78464 $\pm$ 0.11 mas & -25$^\circ$ 34' 26''.7724 $\pm$ 0.09 mas & 15.158 \\
2019-08-02 10:02 & 6766356933423996544 & 19$^h$ 33$^m$ 07$^s$.47470 $\pm$ 0.42 mas & -25$^\circ$ 34' 42''.3668 $\pm$ 0.35 mas & 17.766  \\
2017-08-24 02:59 & 6737213175138988672 & 18$^h$ 42$^m$ 35$^s$.22729 $\pm$ 0.18 mas & -31$^\circ$ 09' 50''.6837 $\pm$ 0.14 mas & 17.468  \\
2017-07-23 05:58 & 6737020112089260672 & 18$^h$ 48$^m$ 09$^s$.22138 $\pm$ 0.04 mas & -31$^\circ$ 26' 32''.4591 $\pm$ 0.03 mas & 14.005  \\
2017-06-22 21:18 & 6760223758801661440 & 18$^h$ 55$^m$ 15$^s$.65251 $\pm$ 0.04 mas & -31$^\circ$ 31' 21''.6706 $\pm$ 0.03 mas & 14.223 \\
2017-04-09 02:24 & 6757456459825426560 & 19$^h$ 04$^m$ 03$^s$.62801 $\pm$ 0.03 mas & -31$^\circ$ 17' 15''.2304 $\pm$ 0.03 mas & 14.094  \\

\hline
\hline
\multicolumn{5}{l}{\rule{0pt}{3.0ex} $^1$ The stars coordinates (RA and Dec.) and their uncertainties were propagated to the occultation epoch with the formalism}\\

\multicolumn{5}{l}{proposed by \cite{Butkevich_2014} using the parameters (proper motion, parallax, radial velocity, etc) from Gaia}\\

\multicolumn{5}{l}{EDR3 \citep{Gaia_EDR3}.}\\

\label{tb:predictions}
\end{tabular}
\end{center}

\end{table*} 

Our campaigns involved sites in South and North America, Africa, and Oceania. The observations used a wide range of telescope sizes, from small portable telescopes (0.25-0.30 m) to large observatories facilities such as the Very Large Telescope (VLT, 8.20 m), Gemini North (8.10 m), Southern Astrophysical Research (SOAR, 4.10 m),  Observat\'orio Pico dos Dias (OPD, 1.60 m), and Danish/ESO (1.54 m). The observational circumstances (telescopes, instruments, stations, observers, etc.) are given in Appendix \ref{App:Observations}.

\section{Data Analysis} \label{sc:redu}

We used classical photometric pipelines to extract the occultation light curves, as detailed in Section \ref{sc:photometry}. Abrupt opaque edges models, including the effects of diffraction, finite band width, exposure time and stellar diameter, were fitted to the star dis- and re-appearances behind Chariklo's main body. For ring occultations, the apparent opacity is also an adjustable parameter. The timings obtained at the various stations define occultation chords, that are in turn used to constrain Chariklo's shape (Section \ref{sc:time_projection}) and the ring size and orientation is space (Section \ref{sc:ring_redu}).

\subsection{Calibration and aperture photometry} \label{sc:photometry}

All the images were converted to the standard \texttt{fits} format. The video files (\texttt{ser} or \texttt{avi}) were converted using \textsc{tangra v3.7.3} \citep{Pavlov_2020} or Python codes based on \textsc{astropy v4.0.1} \citep{Astropy_2013}. When available, calibration images (bias, dark and flat-field) were used to correct the original images using standard \textsc{iraf} \citep[Image Reduction and Analysis Facility][]{Butcher_1981} procedures.

The flux from the target stars was obtained by aperture photometry using the \textsc{praia} \citep[Platform for Reduction of Astronomical Images Automatically][]{Assafin_2011} package. Note that during the occultation, the star and Chariklo images are blended together in the same aperture. The total flux from the  star and Chariklo was normalised to unity outside the  occultation, using a polynomial fit just before and after the event. Finally, nearby stars were used as photometric calibrators to correct for low-frequency sky transparency fluctuations. The resulting light curves are displayed in Appendix~\ref{App:Lightcurves}.

\subsection{Times and projection in the sky plane} \label{sc:time_projection}

The immersion ($t_{\rm imm}$) and emersion ($t_{\rm eme}$) times were determined using a Monte Carlo approach with uniformly distributed parameters to test a vast number of simulated models ($\sim$100,000) for which chi-squared statistics are computed. This was done with the \textsc{sora v0.1.2} (Stellar Occultation Reduction Analysis)\footnote{Website: \url{https://sora.readthedocs.io/}} package. The models consider a sharp-edge occultation model convolved with Fresnel diffraction, apparent stellar diameter (at Chariklo's distance), and finite integration time. For more details, see \cite{Braga-Ribas_2013} and references therein.

\begin{figure}[h]
\centering
\includegraphics[width=0.50\textwidth]{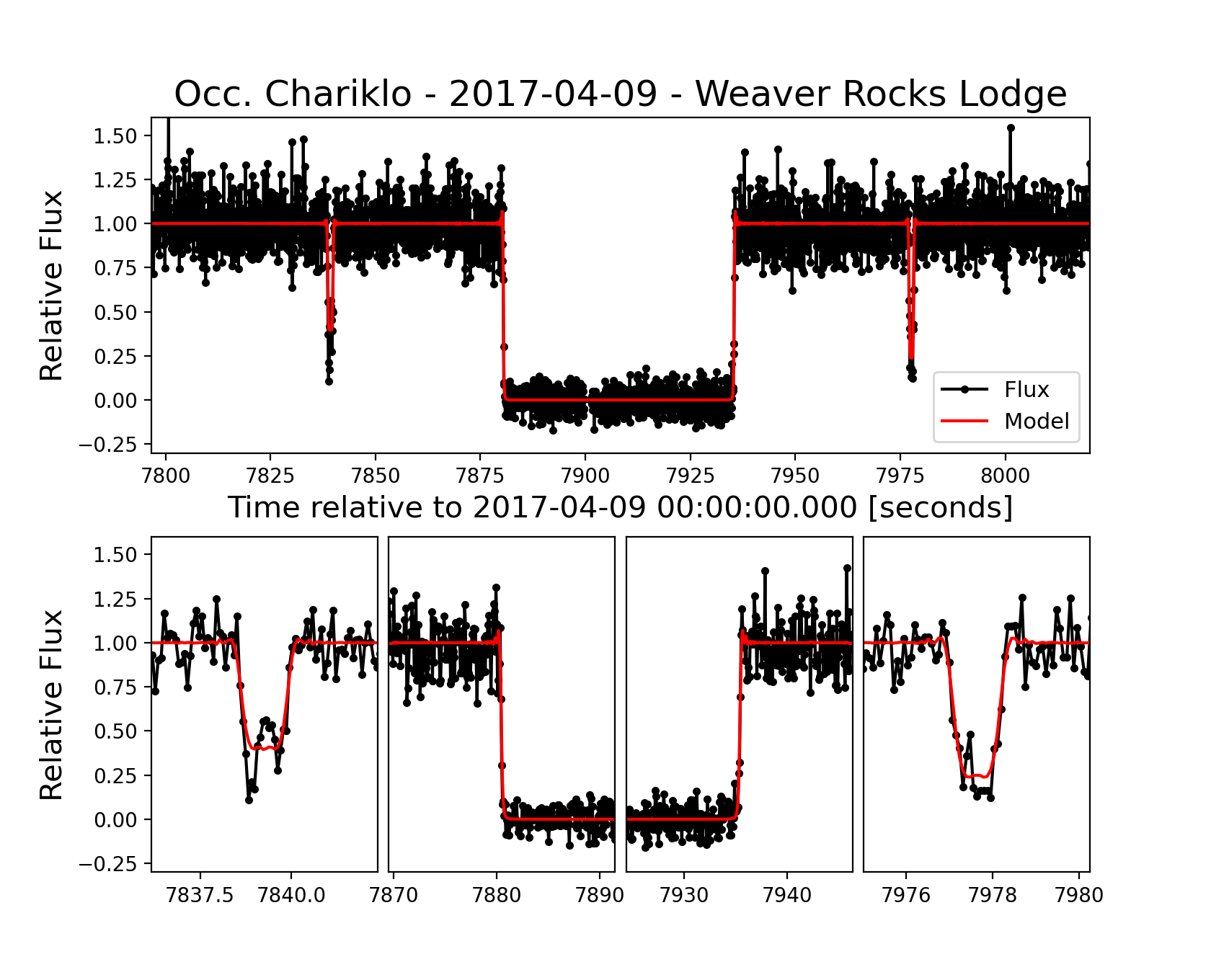}
\caption{Normalised light curve obtained at Weaver Rocks Lodge on 2017-04-09. The black dots show the observational data, while the red lines represent the fitted model. The bottom panels contain a zoom of 20 seconds centred on the immersion and emersion times and 2.5 seconds centred in each C1R detection. In this data-set, there is no clear detection of C2R. All the light curves of this work can be found in Appendix~\ref{App:Lightcurves}.}
\label{Fig:lc_example}
\end{figure}

To determine the rings' parameters, we also need to fit the apparent opacity ($p'$) of the flux drop \citep{Berard_2017}. If insufficiently sampled, there is a large correlation between the apparent opacity of the rings and the duration of the detection ($\Delta t = t_{\rm eme} - t_{\rm imm}$). Thus, the positive ring detections have been divided into two groups: resolved detections, with more than three points showing a significant stellar flux drop during the occultation (higher than $3\sigma$); and unresolved detections, with less than three significant points. The three parameters ($t_{\rm imm}$, $t_{\rm eme}$ and $p'$) can be fitted for the resolved detections, while the unresolved detections allow only the unambiguous determination of the central time ($t_{\rm mean} = \frac{1}{2}(t_{\rm eme} - t_{\rm imm})$). The obtained parameters for the main body and rings are listed in Tables \ref{tb:times_param_1} and \ref{tb:times_param_2}, and an example of analysed light curve is provided in Fig.~\ref{Fig:lc_example}.

Using Chariklo ephemeris (\textsc{nima v.17}), the GEDR3 stars' position propagated to the event epoch and the observer's position on Earth (latitude, longitude, and height), each occultation time is associated with a stellar position relative to Chariklo $(f,~g)$, as projected in the sky plane. The position $(f,~g)$ is expressed in kilometers, $f$ (resp. $g$) being counted positively towards the local east (resp. north) celestial direction. Note that an offset in right ascension and declination, $(f_0,~g_0)$, must be applied to Chariklo's ephemeris to account for errors on both the ephemeris and the star position. This point is discussed in the sub-section \ref{sc:ring_redu}. Thus, the final Chariklo-centric stellar position in the sky plane is given by $(f-f_0,~g-g_0)$. Those positions are used in turn to determine Chariklo's rings orbital parameters and Chariklo's size and shape, see Sections~\ref{sc:rings} and \ref{sc:main_body}.

\subsection{Projection in the ring plane} \label{sc:ring_redu}

The ring parameters obtained in the sky-plane (as derived from $t_{\rm imm}$, $t_{\rm eme}$ and $p'$) are projected in the ring plane to derive the relevant parameters, more precisely: $(i)$ the ring radial width $W_r$;  $(ii)$ its distance $r$ to Chariklo's centre; and $(iii)$ its normal opacity, $p_N$. The quantities $W_r$ and $r$ are calculated by projecting the corresponding $(f-f_0,~g-g_0)$ in the ring plane, once the ring pole orientation $(\alpha_{p}$, $\delta_{p})$ is known. 

Finally, the ring opacity normal to its plane, $p_N$,  is given by
\begin{equation}
p_N = p'~.~|\sin{(B)}| \label{eq:p_n}, 
\end{equation}
assuming a monolayer disk, where $B$ is the ring opening angle ($B= 0^\circ$ and $B= 90^\circ$ corresponding to edge-on and pole-on viewings, respectively).

A relevant integral quantity can also be derived, the equivalent width ($E_p = W_r~.~p_N$), which is related to the amount of material present in a radial cut of the ring. General discussions concerning those parameters are provided in \cite{Elliot_1984,French_1991}, and a more specific application to Chariklo's ring is presented in \cite{Berard_2017}. The derived ring parameters are listed in Tables~\ref{tb:ring_parameters_1}-\ref{tb:ring_parameters_2}.

\section{Rings' structures and parameters} \label{sc:rings}

\subsection{Ring Pole} \label{sc:C1R_pole}

From the multi-chord ring detection of 2013-06-03, \cite{Braga-Ribas_2014} derived  two possible ring pole orientations, assuming the ring to be circular. In that case, the apparent ring oblateness $\epsilon’$ projected in the sky plane is given by
\begin{equation}
    \label{eq:Bxepsilon}
    \epsilon' = 1 - \sin(B).
\end{equation}

Long-term photometric trends \citep{Duffard_2014} allowed to remove this ambiguity, providing $\alpha_{p}=$ 10$^{\rm h}$ 05$^{\rm m} \pm 2^{\rm m}$ and $\delta_{p}=$ +41$^\circ$ $29' \pm 13'$. This solution was later confirmed by stellar occultations observed between 2014 and 2016 \citep{Berard_2017}. Note that the ring pole orientation is in principle defined as being in the direction of the ring angular momentum, which in turn depends on the (unknown) direction of motion for the ring particles. This allows the possible solution $\alpha_{p}+180^\circ$ and $-\delta_{p}$. Here, we choose arbitrarily the solution with $\delta_{p}>0$. This choice can be modified when the direction of motion is known, and do not influence our results.

The ring orbital parameters are obtained by fitting an ellipse to the occultation points, as seen in the sky plane. Five parameters are adjusted: the centre of the ellipse ($f_{0}$, $g_{0}$); the apparent semi-major axis ($a’$); the apparent oblateness ($\epsilon’ = 1 - b'/a'$); and the position angle ($P$) of the apparent semi-minor axis ($b'$), counted positively from the celestial north direction towards the celestial east. 

In this work, and to avoid underdetermined fits, we only consider the occultations which have at least six C1R detections, plus the discovery observation on 2013. These events were observed on 2017-06-22 and 2017-07-23, and are used to pin down the pole orientation and assess possible ring eccentricities.

The fit uses a Monte Carlo approach with uniformly distributed parameters to test $\sim$10,000,000 ellipses, thus providing a $\chi^2$ statistics, as well as the best fit corresponding to the minimum value, $\chi_{\rm min}^2$. The range $\chi_{\rm min}^2$ to $\chi_{\rm min}^2+1$ then provides the marginal 1$\sigma$ error bar for each parameter, i.e. the 68.3\% confidence for that parameter, ignoring the other four parameters, see \cite{Souami_2020} and references therein for more details. In this project, this step was done with the \textsc{sora v0.1.2} package.

Assuming that C1R is circular with radius $r_{C1R}$ equals the fitted $a'$, we derive the pole orientation $(\alpha_{p},\delta_{p})$ from $B$ and $P$, using the equations
\begin{eqnarray}
\sin(\delta_{p}) &=& \cos(B) \cos(P) \cos(\delta) - \sin(B) \sin(\delta),
\label{eq:pole_1} \\
\cos(\alpha - \alpha_{p}) &=& - [\sin(B) + \sin(\delta) \sin(\delta_{p})]/[\cos(\delta) \cos(\delta_{p})], \label{eq:pole_2} \\
\sin(\alpha - \alpha_{p}) &=& -\cos(B)\sin(P)/\cos(\delta_{p}), 
\label{eq:pole_3}
\end{eqnarray}
where $(\alpha,\delta)$ is Chariklo's ICRS position at epoch.

\begin{figure}[h]
\centering
\includegraphics[width=0.50\textwidth]{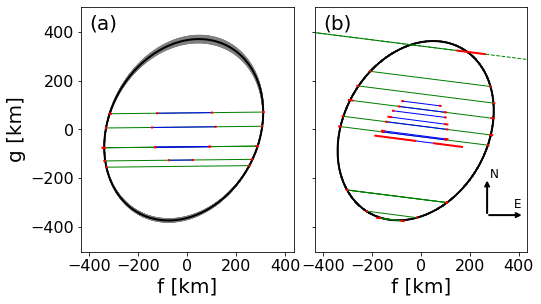}
\caption{Chords in the sky-plane relative to Chariklo's main body (in blue), C1R (in green), and their uncertainties (in red). Ellipses were fitted to C1R points, the black line stands for the best-fitted ellipse, and in gray is all the ellipses in the 1-sigma region. The left panel (a) is the event on 2017-06-22 and in the right panel (b) the event on 2017-07-23. For the sake of clarity, some redundant chords on 2017-07-23 were not plotted.}
\label{Fig:pole}
\end{figure}

Figure \ref{Fig:pole} displays the best-fitted ellipse on C1R for the events on 2017-06-22 and 2017-07-23. It also shows the sky-plane projection of the obtained chords of the main body and rings. The obtained values are given in Table \ref{tb:pole}.

\begin{table}[h]
\begin{center}
\caption{%
Ring pole orientation (ICRS) obtained from events with multi-chords detections of CR1 and their 1$\sigma$ uncertainties. Where $r_{C1R}$, $P$ and $B$ stands for the C1R radius, the position angle of the pole and the ring opening angle, respectively.
}%
\begin{tabular}{cccc}
\hline
\hline
Parameter & 2013-06-03$^a$ & 2017-06-22 & 2017-07-23  \\
\hline
\hline

$r_{C1R}$ (km) & 390.6 (03.3) &  386.6 (12.3)            & 385.9 (00.4)\\ 
$P$ ($^{\circ}$) & -61.54 (0.10) & -62.13 (3.46)         & -60.85 (0.11) \\
$B$ ($^{\circ}$) & +33.77 (0.41) & +52.33 (1.97)         & +49.91 (0.11) \\
$\alpha_{p}$ ($^\circ$) & 151.25 (0.50) &  149.46 (2.60) & 151.03 (0.14) \\
$\delta_{p}$ ($^\circ$) & +41.48 (0.22) &  +40.98 (2.12) & +41.81 (0.07) \\
\hline
n.$^{\circ}$ det. & 14 & 12 & 23$^b$ \\
$\chi^2_{pdf}$ & 1.48 & 1.88 & 2.55 \\
\hline
\hline
\multicolumn{4}{l}{\rule{0pt}{3.0ex} $^a$ Values from \cite{Braga-Ribas_2014}}\\
\multicolumn{4}{l}{$^b$ The grazing chords were not used in the fit.}\\
\label{tb:pole}
\end{tabular}
\end{center}
\end{table} 

The results of Table \ref{tb:pole} show a $\sim$1$\sigma$ consistency with the discovery results of 2013-06-03. Moreover, the values from 2017-07-23 are more accurate than previously published by \cite{Braga-Ribas_2014}. The invariance of the derived pole orientation of C1R over time indicates that no significant eccentricity is detected for that ring. This is discussed more quantitatively later.

\subsection{C1 Ring structure} \label{sc:C1R}

\subsubsection{Radial profiles and eccentricity constraints} \label{sc:C1R_structure}

\begin{figure*}
\centering
\includegraphics[width=1.00\textwidth]{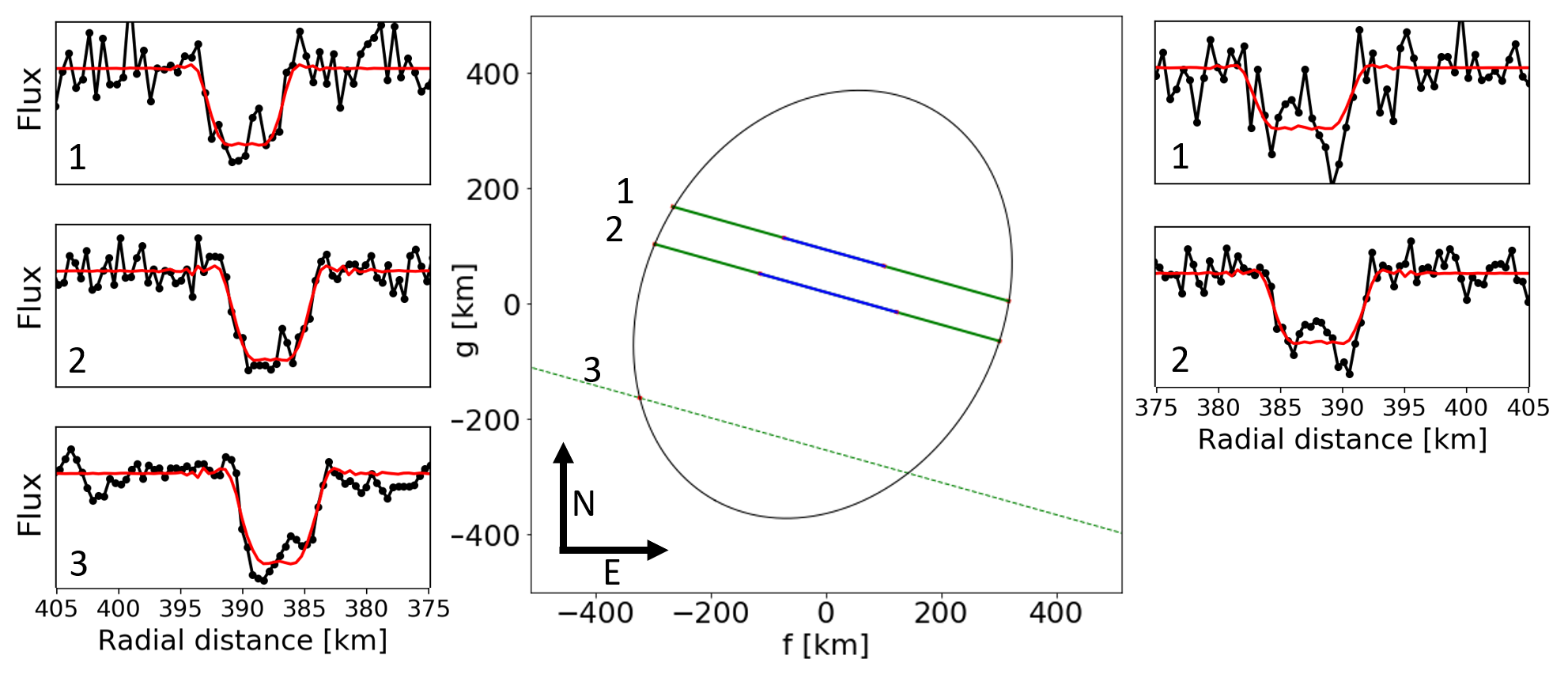}
\caption{%
Results of the 2017-04-09 event. The central plot displays the occultation chords projected in the sky-plane for the main body (in blue), C1R (in green), and their uncertainties (red segments). The black line is the best-fitting ellipse to the C1R point, considering a fixed pole orientation (as determined in Section \ref{sc:C1R_pole}). The side panels display the normalised radial ring profiles, projected in the ring-plane, numbered as follows: Wabi Lodge (1), Weaver Rocks Lodge (2), and Hakos (3), the right panels contain the first detection (immersion) at each station and the left panels contain the seconds detections (emersion). As mentioned in Appendix \ref{App:Observations} and shown in Appendix \ref{App:Lightcurves}, the Hakos light curve contains only the emersion detection of the ring due to bad weather conditions at immersion. We call attention to the unambiguous detection of W-shaped structures within the C1R.
}%
\label{Fig:structure_april}
\end{figure*} 

The high resolution data sets presented here allow the study of the fine structure of C1R. The event of 2017-04-09, observed from Namibia, was special for this purpose. In this event, we had five detections of C1R, as this is unsuitable for fitting all the ellipse's parameters, we fixed the pole orientation as obtained at 2017-07-23 to determine the centre position ($f_0$, $g_0$). With the fitted centre, it was possible to project the chords in the sky-plane to the rings' plane. 

The untypical low velocity of the 2017-04-09 occultation ($\sim$5 km/s) allowed to probe Chariklo's ring with a resolution ($\Delta r$) of $\sim$0.5 km per data point for the observations made at Wabi Lodge, Weaver Rocks Lodge, and Hakos. In this event, clear W-shaped structures appear in the different observations, see Figure~\ref{Fig:structure_april} for more details.

\begin{figure*}
\centering
\includegraphics[width=1.00\textwidth]{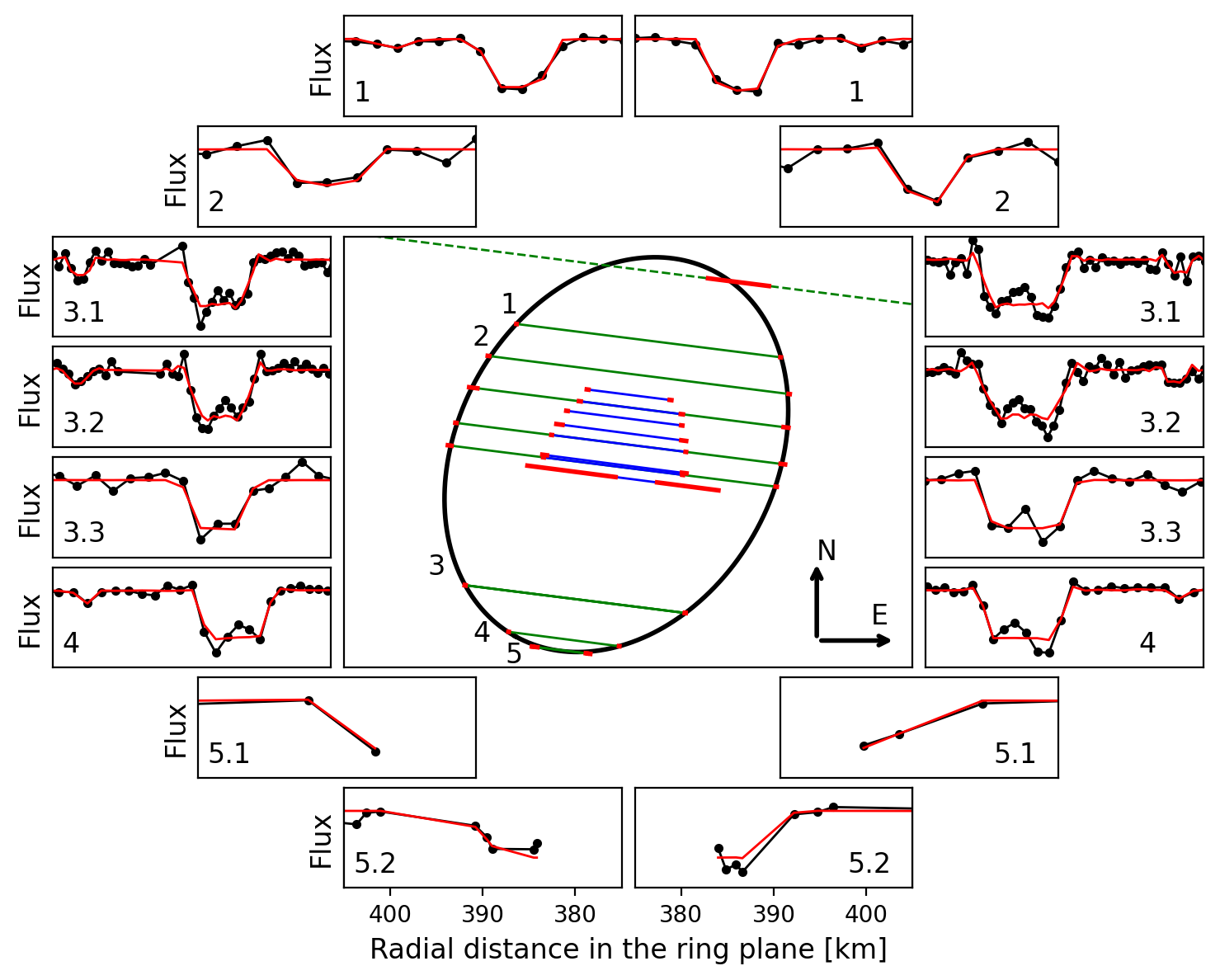}
\caption{%
Same as Figure \ref{Fig:structure_april} for the 2017-07-23 event. The side panels are numbered as follows: Cerro Paranal (1), Tolar Grande (2), La Silla (3), Observat\'orio Pico dos Dias (4) and Cerro Tololo (5). The observations on La Silla were made using the Danish telescope dual experiment in the Visual (3.1) and Red bands (3.2) and a 1-meter telescope (3.3). The left panels contain the first detection (immersion) at each station and the right panels contain the seconds detections (emersion). The observations on Cerro Tololo were grazing over C1R and they were made using the SARA (5.1) and PROMPT (5.2). For sake of clarity, some redundant chords were visually suppressed.
}%
\label{Fig:structure_july}
\end{figure*}

From the occultation of 2017-07-23, the data sets from Danish telescope (Visual and Red filters) and from OPD allowed clear detection of similar W-shaped structures within C1R, see Figure \ref{Fig:structure_july} for details. The occultation's radial velocity at these stations allowed to probe Chariklo's ring with a resolution ($\Delta r$) of $\sim$1.0 km in the ring-plane per data point. The VLT telescope had $\Delta r$ on C1R of 2.2 km (exposure time of 0.1 seconds), so no clear structures were detected.

\begin{figure}[h]
\centering
\includegraphics[width=0.50\textwidth]{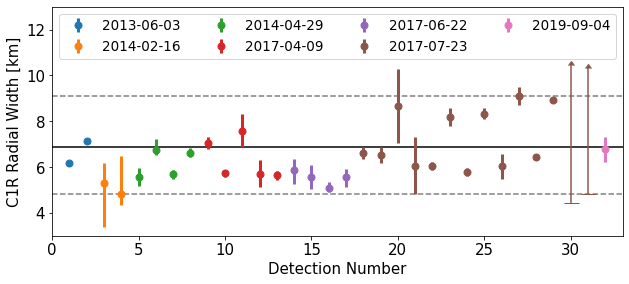}
\caption{%
C1R radial width for all the obtained resolved detections. Note that the PROMPT and SARA values only provide lower limits to the radial width. The black line represents the mean value of $W_r$ (6.9~km) with the gray dashed line as the minimum and maximum values (4.8 and 9.1~km).
}%
\label{Fig:Wr_L}
\end{figure}

Using the resolved C1R profiles, we derive a mean radial width of $\sim$6.9 km, with extreme values varying between $\sim$4.8 and $\sim$9.1 km for C1R, see Figure \ref{Fig:Wr_L}.

For order of magnitude estimation, we may assume that part of the width variation is associated with a $m=1$ mode, in which case $W_r \sim [1 - q_e \cos(f)] \delta a$, with $q_e = e + a \partial e /\partial a$, where $\delta e$ and $\delta a$ are the variations of eccentricity ($e$) and semi-major axis ($a$) across the ring. By analogy with Uranus and Saturn's eccentric ringlets, we may assume that $a \partial e /\partial a \gg e$ \citep{French_1986}. Adopting $\Delta W_{r}= 3.9$~km and $a= r_{C1R}= 385.6$~km, we obtain $\delta e \sim \delta W_{r}/2a \approx 0.005$. Relying again on Uranus and Saturn analogues, we expect $e$ to be a few times $\delta e$, i.e. of the order of 0.01-0.02. This corresponds to full radial excursions of some $\sim$10~km for the C1R ring. This is an easily detectable quantity, but with the difficulty that Chariklo's centre of mass must be pinned down at better than that distance. Those numbers are model dependent and should be taken with caution.

\subsubsection{Two-band resolved profiles of C1R} \label{sc:C1R_Danish}

\begin{figure}[h]
\centering
\includegraphics[width=0.50\textwidth]{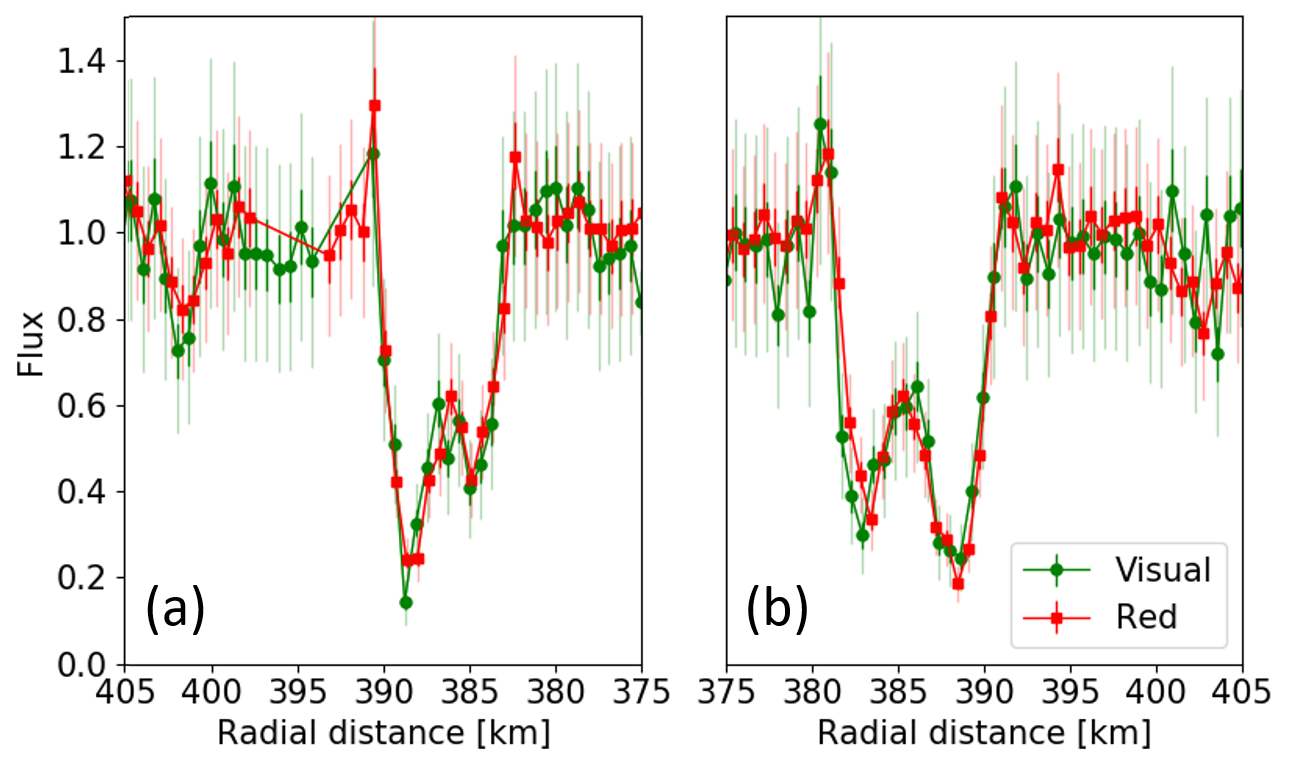}
\caption{%
The normalised radial profiles of C1R and C2R projected in the ring plane, as observed from the 1.54-m Danish telescope during the 2017-07-23 event. Green points: the visual channel (0.45-0.65 $\mu$m); red points: the red channel (0.70-1.0 $\mu$m). The darker (resp. lighter) error bars are for the 1$\sigma$ (resp. 3$\sigma$) level stemming from the SNR of the target source and the calibrators. The (a) and (b) panels stands for the first detection (immersion) and the second detection (emersion), respectively. We call attention to the agreement between the visual and the red channels in both detections.
}%
\label{Fig:Danish_VR}
\end{figure}

The 2017-07-23 event was observed at the 1.54 meters Danish telescope at La Silla, using a dual system with a visible band (0.45-0.65 $\mu$m) and a red band (0.70-1.00 $\mu$m), see \cite{Skottfelt_2015}. Figure \ref{Fig:Danish_VR} displays the C1R and C2R radial profiles at immersion and emersion in both bands. 

A visual inspection of Figure \ref{Fig:Danish_VR} clearly reveals the W-shaped structure of the C1R radial profiles. No significant differences between the two bands appear, as confirmed by the fitted parameters of the rings (see Tables \ref{tb:times_param_1}, \ref{tb:times_param_2} and \ref{tb:ring_parameters_1}), where no significant differences are noted at 1$\sigma$ level.

The similarity of the radial profiles of C1R in the two bands indicates that this ring contains mostly particles larger than a few microns \citep{Elliot_1984}. This is expected by analogy with the dense and narrow Saturnian or Uranian rings. In fact, the local Keplerian shear of the velocity field inside Chariklo's rings is similar to that encountered in Saturn's or Uranus' rings \citep{Sicardy_2018}. Moreover, water ice seems to be a major component of Chariklo's main ring \citep{Duffard_2014}, as it is the case for Saturn's (and probably Uranus') rings. Thus, comparable collisional processes as those found in Saturn's or Uranus' rings are expected in C1R, and consequently comparable particle size distributions. In particular, the opacities of the dense Saturn's and Uranus' rings are dominated by cm- to meter-sized particles, with a very small amount of dust for both Saturn's \citep{Cuzzi_2018} and Uranus' \citep{Esposito_1991} rings, which are in line with our finding for C1R.

\subsection{C2 Ring structure} \label{sc:C2R}

Chariklo's C2R is narrower than C1R, with an orbital radius that was previously estimated to 405~km and an equivalent width that can vary between 0.1 and 1 km \citep{Braga-Ribas_2014, Berard_2017}. This small size makes it very difficult to detect this ring. In this work we present unresolved detections, but most of the previous observations did not detect it. We obtained, here, the first multi-chord detection of C2R during the 2017-07-23 occultation.

As there is no resolved detection of C2R in this work, only the mid-times of each detection could be obtained. The ring radial width and opacity have large uncertainties and are highly correlated. A better defined quantity is then the equivalent width, for which we obtain a mean value $0.117 \pm 0.08$~km from the values listed in Table~\ref{tb:ring_parameters_2}, all obtained during the 2017-07-23 event.

\begin{figure}[h]
\centering
\includegraphics[width=0.50\textwidth]{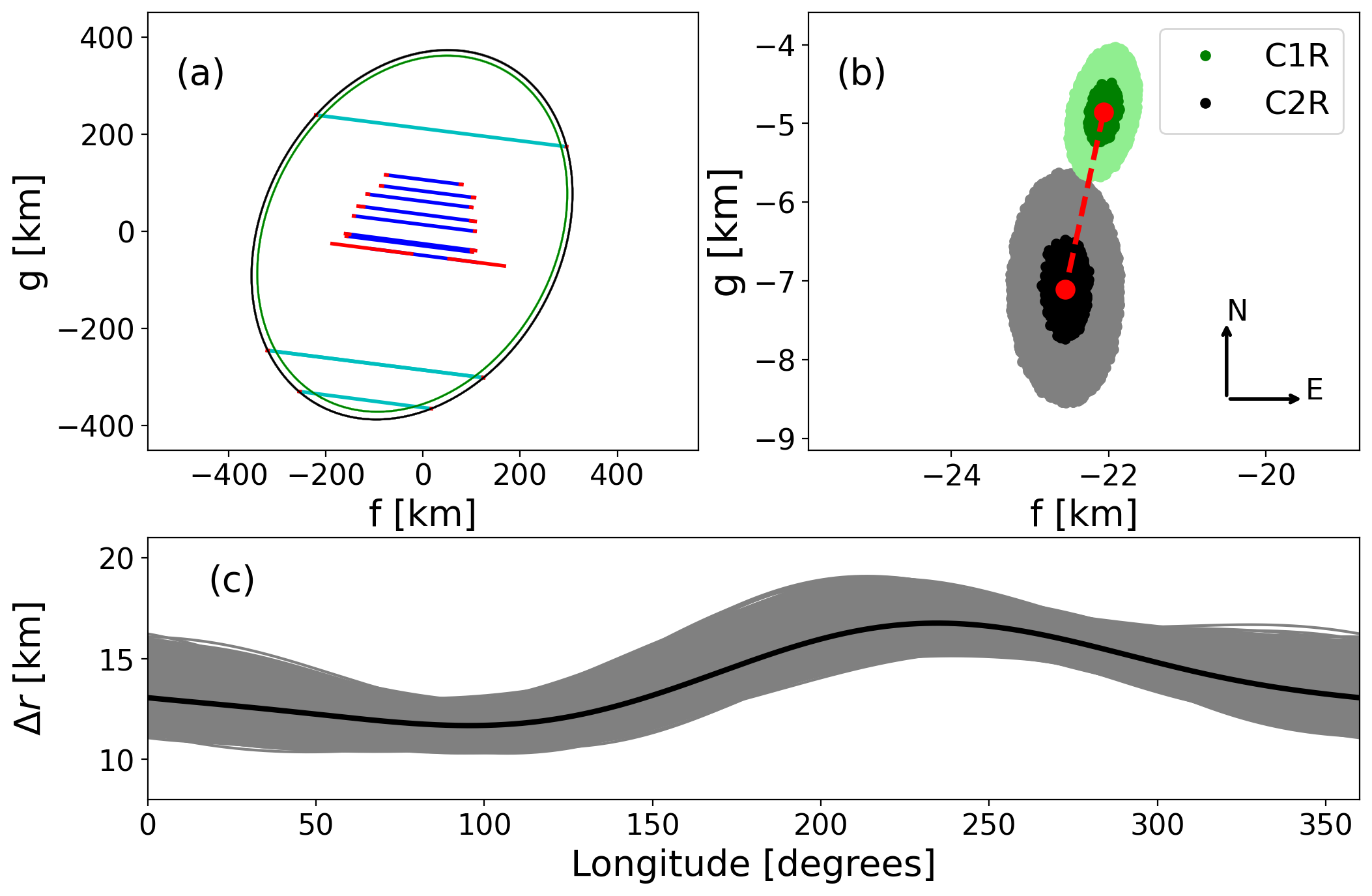}
\caption{%
Panel (a): the occultation chords of the 2017-07-23 event detected, for Chariklo's main body (blue) and for the C2R ring (cyan). The uncertainties at the extremities of the chords are indicated in red. For sake of clarity, the C1R chords are not reproduced here, and can be seen in Figures \ref{Fig:pole} and \ref{Fig:structure_july}. The fitted ellipses to the C2R is drawn in black. The 1$\sigma$ uncertainty region of that fit is indistinguishable from the black line thickness at this scale. The best fitted ellipse to the C1R ring detections is shown in green. Panel (b): the  fitted centres of C1R (green) and C2R (black) in the sky plane. The darker colours stands for the simultaneous 1$\sigma$ uncertainty regions (corresponding to the $\chi^2_{min} + 2.3$ criterion), while the lighter colours stands for the 3$\sigma$ regions ($\chi^2_{min} + 11.4$ criterion). The red dashed line shows a 2.28-km offset between the C1R and C2R centres. Panel (c): the radial difference $\Delta r$ (in the ring-plane) between C2R and C1R vs. the longitude, deduced from the elliptical fits of panel (a). The grey zone delimits the 1$\sigma$ region associated with the uncertainties on C1R and C2R. It shows that at that level, no significant radial distance variations between C1R and C2R are detected.
}%
\label{Fig:C1R_C2R}
\end{figure}

Using the same approach described in sections \ref{sc:C1R_pole} and \ref{sc:C1R_structure}, we fitted the best ellipse to obtain the global orbital parameters of C2R and compared the resulting centre, position angle and pole with those obtained for C1R during the same event. Figure \ref{Fig:C1R_C2R} (panel (a)) displays the C2R chords and the corresponding best-fitted ellipse. A close-in view showing the fitted centres of C1R and C2R in the sky plane (panel (b)) reveals an offset of 2.28~km between those two centres at the $\sim$3$\sigma$ level. The radial offset $\Delta r$ between C2R and C1R in the ring plane vs. the longitude, deduced from the respective elliptical fits, is displayed in panel (c). The grey area contains the 1$\sigma$ uncertainty region for $\Delta r$, considering the respective uncertainties on the C1R and C2R fits.

Table \ref{tb:C1RxC2R} provides the fitted parameters obtained independently for C1R and C2R. We find that the two ring pole orientations are mutually consistent at a 1$\sigma$ level. The 2.28-km offset between the rings centres might be indicative of a differential eccentricity of the order of $2.28/386 \sim 0.006$ between the two rings. However, we estimate that this 3$\sigma$ level detection remains marginally significant. Finally, we find an average radial distance $\Delta R$ between C1R and C2R of 14 km, with a 1$\sigma$ uncertainty interval of $\sim$ 10-19~km which is in agreement with the mean value of 14.8~km found by \cite{Berard_2017}.

\begin{table}[h]
\begin{center}
\caption{Parameters obtained on 2017-07-23 from fitting ellipses over C1R and C2R projections in the sky-plane.}
\begin{tabular}{ccc}
\hline
\hline
Parameter  & C1R & C2R   \\
\hline
\hline
$f_0$ (km)              & -22.0 (0.1)   &  -22.5 (0.2) \\ 
$g_0$ (km)              & -04.8 (0.1)   &  -07.1 (0.4) \\ 
$r$ (km)                & 385.9 (0.4)   &  399.8 (0.6) \\ 
$P$ ($^{\circ}$)        & -60.85 (0.11) &  -61.18 (0.19)   \\
$B$ ($^{\circ}$)        & +49.91 (0.11) &  +49.99 (0.16)   \\
$\alpha_{p}$ ($^\circ$) & 151.03 (0.14) &  150.91 (0.22)  \\
$\delta_{p}$ ($^\circ$) & +41.81 (0.07) &  +41.60 (0.12)  \\
\hline
n.$^{\circ}$ det.       & 23            & 08              \\
$\chi^2_{pdf}$          & 2.55          & 0.68      \\
\hline
\hline
\label{tb:C1RxC2R}
\end{tabular}
\end{center}
\end{table} 

Considering the multi-band detection of C2R with the Danish telescope, similar to the findings for C1R, no significant differences between the two filters appear, but due to its poorer detection level, differences can be hidden within the noise.

\subsection{Search for faint ring material between C1R and C2R} \label{sc:ring_material}

Our best light curves in terms of time resolution and SNR were obtained during the 2017-07-23 event at VLT, OPD, and Danish telescope (La Silla). They can be used to detect or place an upper limit of material orbiting in the gap between C1R and C2R. We assume that this putative semi-transparent material is concentric and co-planar to C1R and C2R.

Following similar procedures as described in \cite{Berard_2017}, we convert each data-point ($i$) from flux to equivalent width ($E_p$) using Equation \eqref{eq:e_p}, where $\phi(i)$ is the normalised light-flux, and $\Delta r (i)$ is the radial resolution in the ring plane travelled by the star during one exposure ($\Delta t (i)$),

\begin{equation} \label{eq:e_p}
    E_{p}(i)~=~\frac{|\sin (B)|}{2}~.~[1-\phi(i)]~.~\Delta r(i)~.
\end{equation}

\begin{figure}[h]
\centering
\includegraphics[width=0.50\textwidth]{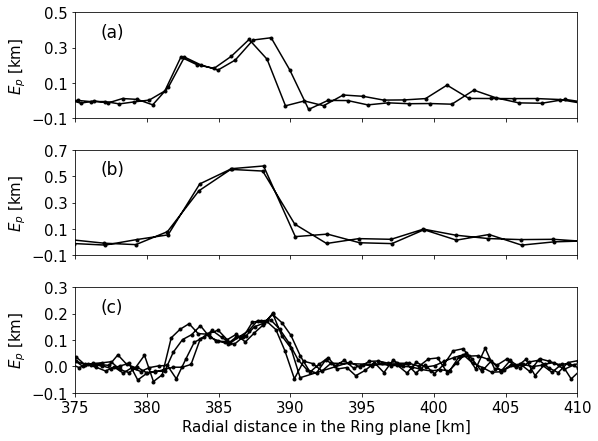}
\caption{Equivalent width of ring material over the radial distance in the ring plane for the observations in OPD/LNA (a), VLT/ESO (b) and Danish/ESO (c). We determine a 3-sigma upper limit of semi-transparent uninterrupted material between C1R and C2R with normal opacity of 0.006.}
\label{Fig:material}
\end{figure}

The top panel (a) of Figure \ref{Fig:material} contains a plot of $E_p(i)$ over the radial distance in the ring plane $R(i)$ for the observation at OPD. The 1-sigma limit for the equivalent width ($E_p(1\sigma)$) for the region between C1R and C2R in the ring plane (between 391 and 399 km) was of 24 meters, this means that if there were an uninterrupted ring between C1R and C2R with normal opacity greater than ($p_n$) of 0.003 (0.009) it would have been detected with a 1-sigma (3-sigma) confidence level. As for the VLT observation with a $\Delta r$ of 2.2 km, it only contains three data points between C1R and C2R for each detection. For this data-set a ring with normal opacity of 0.011 would have been detected within 3-sigma confidence level, as can be seen in the mid panel (b) in Figure \ref{Fig:material}. 

Another approach is the combination of some data to increase the quality of the measurement. As shown in Section \ref{sc:C1R_Danish}, the Danish observations in two different filters are virtually the same (within the errors), meaning that these two observations can be combined to improve the quality of the data. Using the combination of both light curves illustrated in the bottom panel~(c) in Figure \ref{Fig:material} we obtain an $E_p(1\sigma)$ of 16 meters. A ring material between C1R and C2R with normal opacity greater than 0.002 (0.006) would have been detected with at 1-sigma (3-sigma) confidence level. 

\section{Chariklo's size, shape and rotational parameters}\label{sc:main_body}

Between 2013 and 2020, Chariklo was the target of several successful stellar occultations. That allows us to go further than only the determination of its apparent size and shape. It enables us to determine Chariklo's 3D shape. We can also analyse the centre position of Chariklo and how it compares with its rings' centre, thus giving us constraints on Chariklo rings' excentricity. Finally, we can evaluate Chariklo's rotational parameters (rotational light curve amplitude) to see how our model compares with previous observations.
    
\subsection{Chariklo's 3D shape} \label{sc:3d_shape}

\begin{figure*}
\centering
\includegraphics[width=1.00\textwidth]{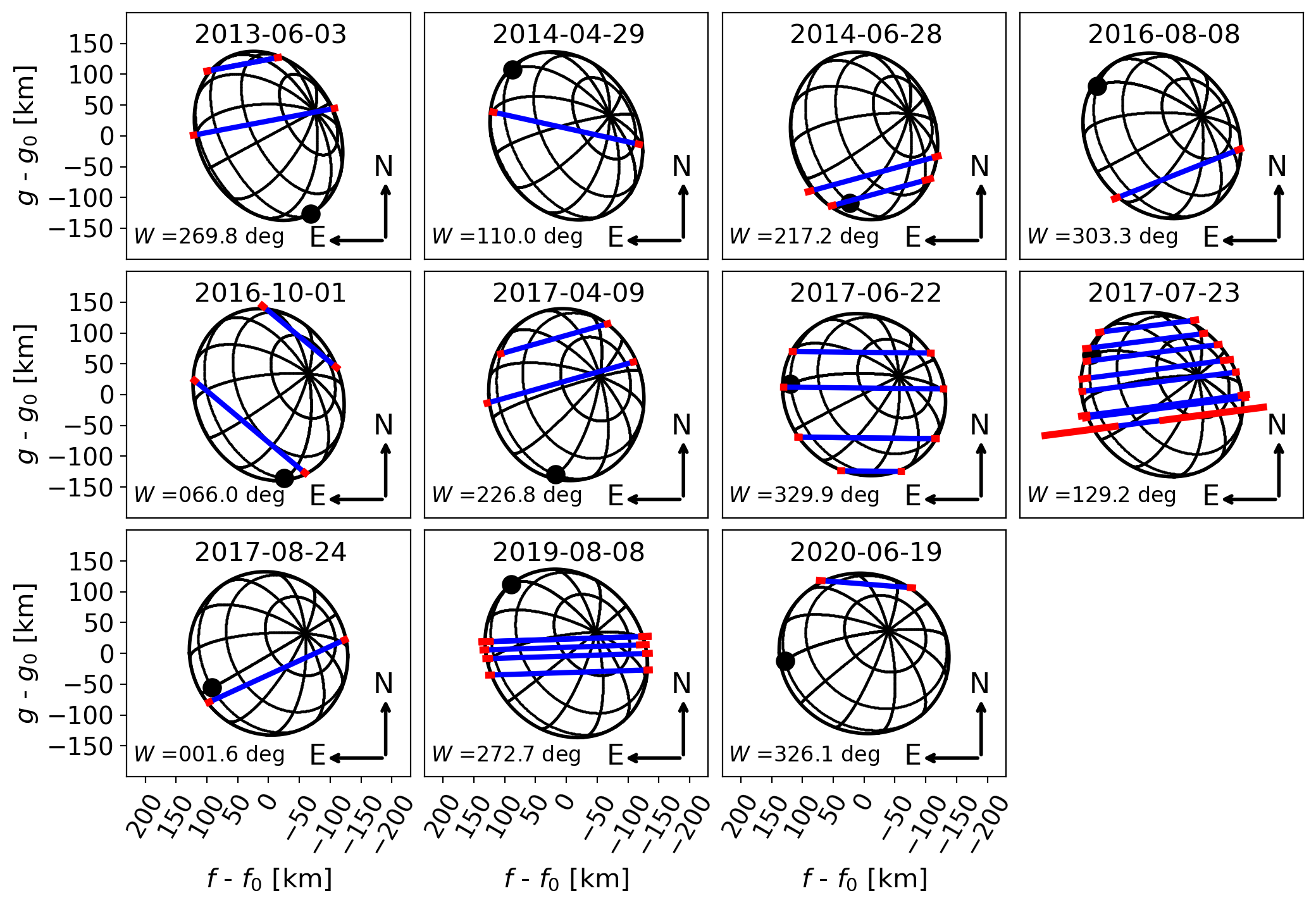}
\caption{%
Ellipsoidal model that best fits the 11 stellar occultations observed between 2013-06-23 and 2020-06-19. Each panel corresponds to an occultation event identified by the time stamp in the upper part of each one. The blue lines stands for the observed chords with their uncertainties in red. The black dot indicates the intersection between the equator and the prime meridian, which is used as the reference to define the rotation angle $W$, written in the lower left of each panel.
}%
\label{Fig:3dfit}
\end{figure*}  

First, let us consider that Chariklo's pole orientation is the same as its rings and, as we show in Section \ref{sc:C1R_pole}, C1R pole orientation can be considered fixed between 2013 and 2017.

As an initial guess, we will consider that the centre of Chariklo and its rings are the same. This assumption will allow us to consider 11 stellar occultations that had at least one chord over Chariklo's main body and at least two ring detections. The only exception was the occultation on 2019-08-08 that had five detections of the main body, but no ring detection. In Section \ref{sc:mb_centre} we will further analyse Chariklo's centre and use it to derive an upper limit to Chariklo's rings excentricities.

We also consider that Chariklo's rotational period is known and equals 7.004 $\pm$ 0.036 \citep{Fornasier_2014}. Even though we know Chariklo’s period, the period uncertainty is large enough that its rotational phase cannot be determined at the moment of each occultation. In other words, it can be any value between 0 and 360 degrees. 

Our analysis considers the rotational elements as described in \cite{Archinal_2018}. The angle $W(t)$ specifies the position of the prime meridian (in this work chosen to be in the direction of the longest axis) on a given date (considering light time between Chariklo and the observer), $W_{0}$ is the value of $W$ at the reference epoch ($t_{\rm ref}$) and $\dot{W}$ stands for the rotation rate, here expressed in degrees per days,
\begin{equation}
W(t) = W_0 + \dot{W}(t - t_{\rm ref}) ~.    \label{eq:wt}
\end{equation}

The reference epoch was chosen as the time of the first stellar occultation considered here, 2013-06-03 06:25:00 UTC (2456446.767361111~JD), to minimise error propagation. Let us consider a given 3D ellipsoidal model with semi-axis $a > b > c$, rotating around its smallest axis ($c$). We project this ellipsoid in the sky-plane and determine its projected limb as the points where gradient of the normal to the direction to the observer is equal to zero.

The model limb is then compared with the occultations chords, and chi-squared statistics are computed using
\begin{equation}
\chi^2 = \sum^{N}_{i} \frac{(r_{i} - r'_{i})^2}{\sigma_i^2 + \sigma_{model}^2},
\end{equation}
where $r_i$ is the radial distance of the $i^{th}$ observed chord extremity, $r'_i$ is the radial distance of the ellipsoid limb, $\sigma_i$ is the observed radial uncertainty of each extremity point and $\sigma_{model}$ consist of the uncertainty of our ellipsoidal model in contrast with Chariklo's real 3D shape, for instance the existence of topographic features within Chariklo's surface. $\sigma_{model}$ was estimated iteratively as 3 km, so the resulting $\chi^2$ per degree of freedom would be close to 1.

We compute this for 10,000,000 models, varying $a$, $b$, $c$, $W_0$ and $\dot{W}$. As we do not intend to fit the rotation rate, we let it vary between 1227.20 and 1239.95 deg/days only. This value was obtained considering a period between 6.968 and 7.040 hours, as provided by \cite{Fornasier_2014}. For the other parameters, we let $a$, $b$, $c$ vary in the range of  125 $\pm$ 50 km and $W_0$ between 0 and 360 degrees.

Our best fitted value was the ellipsoid with semi-axis equals to $143.8^{+1.4}_{-1.5}$ , $135.2^{+1.4}_{-2.8}$ and $99.1^{+5.4}_{-2.7}$ km with $\chi^2_{pdf}$ equal to 0.688. Figure \ref{Fig:3dfit} shows this ellipsoid fitted to the observational data. The standard deviation of the radial residuals was 4.11 km (2.8\% considering a volume equivalent radius of 124.8 km), with a maximum value of 7.78 km and a minimum of -5.52 km, this can indicate topographic features of this range on Chariklo's surface.

\subsection{Chariklo's density} \label{sc:mb_density}

After fitting an ellipsoid to Chariklo, the first thing that becomes clear is that Chariklo does not have a shape corresponding with the equilibrium figure of a Maclaurin spheroid (where $a$ would be equal to $b$). Instead, if we consider the Jacobi figure, the expected uniform density ($\rho$) of Chariklo would be $1.55^{+0.05}_{-0.08}~g/cm^3$, as can be calculated with Equation \eqref{eq:density_1},

\begin{eqnarray}
    \Omega = \dfrac{2\pi}{G \rho T^2} &=& \beta \gamma \int^{\infty}_{0} \dfrac{u}{(1 + u)(\beta^2 + u)\Delta(u, \beta, \gamma)}du ~, \label{eq:density_1} \\
    \Delta(u, \beta, \gamma) &=& \sqrt{(1 + u)(\beta^2 + u)(\gamma^2 +u)} ~, \label{eq:density_2}
\end{eqnarray}
where $G$ is the gravitational constant, $T$ is the rotational period, $\beta$ and $\gamma$ stand for the semi-axis ratios $b/a$ and $c/a$ respectively, more details in \cite{Sicardy_2011}, \cite{Ortiz_2017} and references therein. This density is not consistent with the expected value, between 0.79 and 1.04 $g/cm^3$, of a body with a rotational period ($T$) of 7.004 hours. For a stable Jacobi ellipsoid, the shapes should lie with the dimensionless parameter $\Omega$ between 0.284 and 0.374 \citep{Tancredi_2008}, instead of 0.191 as obtained in our analysis.

On the other hand, one possibility is that Chariklo rings' particles would be close to the 1/3 resonance with the central body \cite{Sicardy_2019}. That means that the particles complete one revolution while the body completes three rotations. In this context, an ellipsoid with a semi-axis of 143.8, 135.2 and 99.1 km should have a density between 0.73 and 0.85 $g/cm^3$, so the 1/3 resonance would be between 385 and 405 km, the region where the Chariklo rings' particles are. These are indications that the hydrostatic equilibrium of a homogeneous body does not dominate Chariklo's shape.

\subsection{Comparing Chariklo's centre with its rings' centre} \label{sc:mb_centre}

As mentioned in Section \ref{sc:3d_shape}, we considered that Chariklo's centre is the same as its rings' centre. Any deviation from this would appear as a systematic effect unique to each stellar occultation. In extreme cases (tens of kilometres), this would also preclude a unique solution with the methodology applied. That is not the case, as can be seen from Figure \ref{Fig:3dfit}.

We used our derived 3D model and fitted it for the centre on the chords of 2017-07-23 multi-chord detection, to check for small systematic effects. Here we compare the fitted value for Chariklo main body and the centre obtained for its rings (values in Table \ref{tb:C1RxC2R}). We obtain a radial difference between the main body and C1R of $2.46^{+2.86}_{-2.07}$ km in the sky plane considering an 1-sigma confidence level. At a 3-sigma level, this value is between zero and 8.41 km. If we consider that this offset is in the ring plane, this could be translated to a $2.51^{+2.84}_{-2.50}$ km difference, in a 3-sigma confidence level ranging between 0 and 8.55 km.

The 1-sigma (3-sigma) region's limit can give us hints about the maximum difference between Chariklo's centre and the centre of its rings. This value cannot be larger than 5.35 (8.55) km at the moment of the detected occultations. These values give us an upper limit of the eccentricity of C1R rings of about 0.014 (0.022).

Applying the same analysis to C2R as observed on 2017-07-23, we obtain a maximum radial distance of 3.27 (6.84) km in the ring plane. This is representing an upper limit of about 0.0082 (0.0171) for C2R eccentricity considering a 1-sigma (3-sigma) confidence level.

\subsection{Rotational light curve amplitude} \label{sc:period}

Considering Chariklo's rotational light curve, we can check if our ellipsoidal model can explain the light curve amplitude provided by \cite{Leiva_2017} as observed by \cite{Davies_1998, Peixinho_2001,Galiazzo_2016, Fornasier_2014, Leiva_2017}, and listed in Table \ref{tb:delta_mag}.

\begin{table}[h]
\begin{center}
\caption{Rotational light curve amplitudes from the literature. The alias column is a simplified reference name, the first letter of the main author's name, plus the two last digits of publication year.}
\begin{tabular}{lclc}
\hline
\hline
\textbf{Date}  & \textbf{$\Delta mag$} & \textbf{Reference} & \textbf{Alias}   \\
\hline
\hline
1997-05 & < 0.02 &           \cite{Davies_1998}    & D98 \\
1999-03 & < 0.05 &           \cite{Peixinho_2001}  & P01 \\
2006-06 & 0.13 $\pm$ 0.03 &  \cite{Galiazzo_2016}  & G16 \\
2013-06 & 0.11 $\pm$ 0.02 &  \cite{Fornasier_2014} & F14 \\
2015-07 & 0.06 $\pm$ 0.02 &  \cite{Leiva_2017}     & L17 \\

\hline
\hline
\label{tb:delta_mag}
\end{tabular}
\end{center}
\end{table} 

\begin{figure}[h]
\centering
\includegraphics[width=0.50\textwidth]{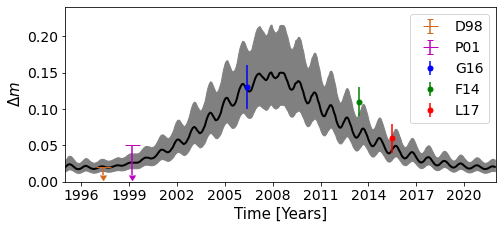}
\caption{%
Rotational light curve's amplitude variations over the years. In black is our best fitted ellipsoidal model, and in grey all the ellipsoids within the 1-sigma region. We compare our model with the observation detailed in Table \ref{tb:delta_mag}.
}%
\label{Fig:delta_mag}
\end{figure}

We calculate the magnitude amplitude using Equation \eqref{eq:delta_mag} as derived by \cite{Fernandez-Valenzuela_2017} and references therein,

\begin{equation}
    \Delta m = -2.5~log\dfrac{A_{min}p_{V} + A_{r}(I/F)}{A_{max}p_{V} + A_{r}(I/F)} ~, \label{eq:delta_mag}
\end{equation}
where $A_{min}$ and $A_{max}$ stands for the smallest/largest covered area by the 3D model, $A_{r}$ is the area covered by Chariklo's rings. Here we consider Chariklo's geometric albedo ($p_{V}$) as 3.7\% and the ring reflectivity ($I/F$) as 4.9\% \citep{Leiva_2017}. As shown in Figure \ref{Fig:delta_mag} our 3D model can explain the observed light curve amplitude. Any slight difference can be associated with albedo variegation.

\section{Astrometric positions} \label{sc:astrometry}

A valuable by-product of stellar occultations is the mas-level precision of the occulting body astrometry, especially since  the Gaia DR2 and EDR3 catalogues provide the stars' positions at sub-mas-level precision \citep{Desmars_2019, Herald_2020, Rommel_2020}. This means a significant improvement of the occulting body ephemerides, which in turn allows better predictions for future events in a virtuous loop. 

We obtained Chariklo's astrometric positions from seven stellar occultations observed between 2017 and 2020, assuming that the centre of Chariklo coincides with the centre of C1R. Note that the 2019-09-04 occultation only gives the egress detection of C1R, so that no astrometric constraints can be derived from that event, more details in Appendix \ref{App:Paul_Maley}. Table \ref{tb:positions} lists the positions obtained and their associated uncertainties. Those uncertainties stem from the uncertainties of the fitted ring centre ($\sigma f_{0},~\sigma g_{0}$) and the uncertainties in the star positions propagated to the occultation epoch (see Table \ref{tb:predictions}). Note that the resulting uncertainties are usually at sub-mas level, corresponding to a few kilometres at Chariklo's distance, where 1 mas$~\sim$10~km.

\begin{table*}
\begin{center}
\caption{Astrometric Chariklo's positions for each event.}
\begin{tabular}{ccc}
\hline
\hline
Date and time UTC & Right Ascension$^a$ & Declination$^a$ \\
\hline
\hline

2020-06-19 15:51:00.000 & 20$^h$ 02$^m$ 32$^s$.0002322 $\pm$ 0.339 mas & -22$^\circ$ 20' 41''.488487 $\pm$ 0.227 mas \\
2019-08-08 21:41:00.000 & 19$^h$ 31$^m$ 49$^s$.7910264 $\pm$ 0.227 mas & -25$^\circ$ 34' 26''.785014 $\pm$ 0.544 mas \\
2019-08-02 10:02:00.000 & 19$^h$ 33$^m$ 07$^s$.4812764 $\pm$ 0.872 mas & -25$^\circ$ 34' 42''.519146 $\pm$ 1.187 mas \\
2017-08-24 02:59:00.000 & 18$^h$ 42$^m$ 35$^s$.2371826 $\pm$ 0.426 mas & -31$^\circ$ 09' 50''.561462 $\pm$ 0.432 mas \\
2017-07-23 05:58:00.000 & 18$^h$ 48$^m$ 09$^s$.2288885 $\pm$ 0.103 mas & -31$^\circ$ 26' 32''.437598 $\pm$ 0.096 mas \\
2017-06-22 21:18:00.000 & 18$^h$ 55$^m$ 15$^s$.6602082 $\pm$ 0.116 mas & -31$^\circ$ 31' 21''.621802 $\pm$ 0.110 mas \\
2017-04-09 02:24:00.000 & 19$^h$ 04$^m$ 03$^s$.6255610 $\pm$ 0.129 mas & -31$^\circ$ 17' 15''.257638 $\pm$ 0.127 mas \\

\hline
\hline
\multicolumn{3}{l}{\rule{0pt}{3.0ex} $^a$ The positions assume the EDR3 star positions given in Table~\ref{tb:predictions}}\\

\label{tb:positions}
\end{tabular}
\end{center}

\end{table*}

Those positions were used to improve Chariklo's ephemeris using the \textsc{nima} software \citep{Desmars_2015}. The corresponding bsp file (\textsc{nima} v.19) can be downloaded from the Lucky Star website\footnote{\url{https://lesia.obspm.fr/lucky-star/obj.php?p=624}}. The updated ephemeris provides a propagated uncertainty on Chariklo's position of 2 mas in right ascension and declination, as of January 2022. This corresponds to roughly 20~km at Chariklo's distance.

\section{Conclusions} \label{sc:conclusion}

This paper presents results from eight unpublished stellar occultations by Chariklo and its rings, observed between 2017 and 2020. Our main results are:

\begin{itemize}
    \item[\textbf{(i)}] For the first time, we detected stellar occultations with three (or more) chords over Chariklo's main body;

    \item[\textbf{(ii)}] We obtained the second and third multi-chord detections of C1R since Chariklo's rings discovery (2013-06-03);

    \item[\textbf{(iii)}] For the first time, we obtained multi-chord detections of C2R;

    \item[\textbf{(iv)}] We obtained the first simultaneous multi-band observation of Chariklo's rings;

    \item[\textbf{(v)}] New astrometric data for the Chariklo system was obtained from all the detected stellar occultations, including two single-chord and one double-chord events.

\end{itemize}

Table \ref{tb:results} contains a summary of our main results. We determined a Chariklo's rings pole orientation consistent with, and more precise than those previously determined by \cite{Braga-Ribas_2014}. We could not refute Chariklo's ring to be circular within our measurements' errors. We estimate an upper 3$\sigma$ limit of 0.022 for the eccentricity of C1R. By analogy with Uranus' and Saturn's ringlets, and from its width variations, we estimate an eccentricity greater than $\sim$0.005 for that ring which remains, however, model dependent.

\begin{table*}[h]
\begin{center}
\caption{Summary of the obtained parameters for Chariklo and its rings.}
\begin{tabular}{clcl}
\hline
\hline
& \textbf{Parameter}  & \textbf{Value} & \textbf{Comments}   \\
\hline
\hline
\rule{0pt}{2.5ex}
            & Radius              & 385.9 $\pm$ 0.4 km            &  See Section \ref{sc:C1R_pole} \\
\rule{0pt}{2.5ex}
            & Pole orientation (RA)  & 151.03 $\pm$ 0.14 deg         &  See Section \ref{sc:C1R_pole} \\
\rule{0pt}{2.5ex}
            & Pole orientation (Dec) & +41.81 $\pm$ 0.07 deg         &  See Section \ref{sc:C1R_pole} \\
\rule{0pt}{2.5ex}
C1 Ring     & Mean width          & 6.86 km                        &  See Section \ref{sc:C1R_structure} \\
\rule{0pt}{2.5ex}
            & Width variation     & 4.8 $\leq$ $W_r$ $\leq$ 9.1 km      &  See Section \ref{sc:C1R_structure} \\
\rule{0pt}{2.5ex}
            & Eccentricity        & 0.005 $\leq$ $e_{C1R}$ $\leq$ 0.022 &  See Section \ref{sc:C1R_structure} and \ref{sc:mb_centre} \\
\rule{0pt}{2.5ex}
            & Particle's size     & > 1.0 micron                  &  See Section \ref{sc:C1R_Danish} \\
\hline
\rule{0pt}{2.5ex}
Gap between     & Radial distance between rings       &  $13.9^{+5.2}_{-3.4}$ km  &  See Section \ref{sc:C2R} \\
\rule{0pt}{2.5ex}
C1R and C2R     & Equivalent width of material in gap & < 0.048 km                &  See Section \ref{sc:ring_material} \\

\hline
\rule{0pt}{2.5ex}
            & Radius                    & 399.8 $\pm$ 0.6 km     &  See Section \ref{sc:C2R} \\
\rule{0pt}{2.5ex}
            & Pole orientation (RA)        & 150.91 $\pm$ 0.22 deg  &  See Section \ref{sc:C2R} \\
\rule{0pt}{2.5ex}
C2 Ring     & Pole orientation (Dec)       & +41.60 $\pm$ 0.12 deg  &  See Section \ref{sc:C2R} \\
\rule{0pt}{2.5ex}
            & Equivalent width          & 0.117 $\pm$ 0.080 km   &  See Section \ref{sc:C2R} \\
\rule{0pt}{2.5ex}
            & Eccentricity              & < 0.017                &  See Section \ref{sc:mb_centre} \\
\hline
\rule{0pt}{2.5ex}
            & a (Semi-major axis)     & $143.8^{+1.4}_{-1.5}$ km     &  See Section \ref{sc:3d_shape} \\
\rule{0pt}{2.5ex}
Chariklo    & b (Semi-median axis)    & $135.2^{+1.4}_{-2.8}$ km     &  See Section \ref{sc:3d_shape} \\
\rule{0pt}{2.5ex}
            & c (Semi-minor axis)     &  $99.1^{+5.4}_{-2.7}$ km     &  See Section \ref{sc:3d_shape} \\
\rule{0pt}{2.5ex}
            & Volume equiv. radius    & $124.8^{+3.0}_{-2.3}$ km     &  See Section \ref{sc:3d_shape} \\
\hline   
\hline
\label{tb:results}
\end{tabular}
\end{center}
\end{table*} 

\begin{figure*}[h]
\centering
\includegraphics[width=1.00\textwidth]{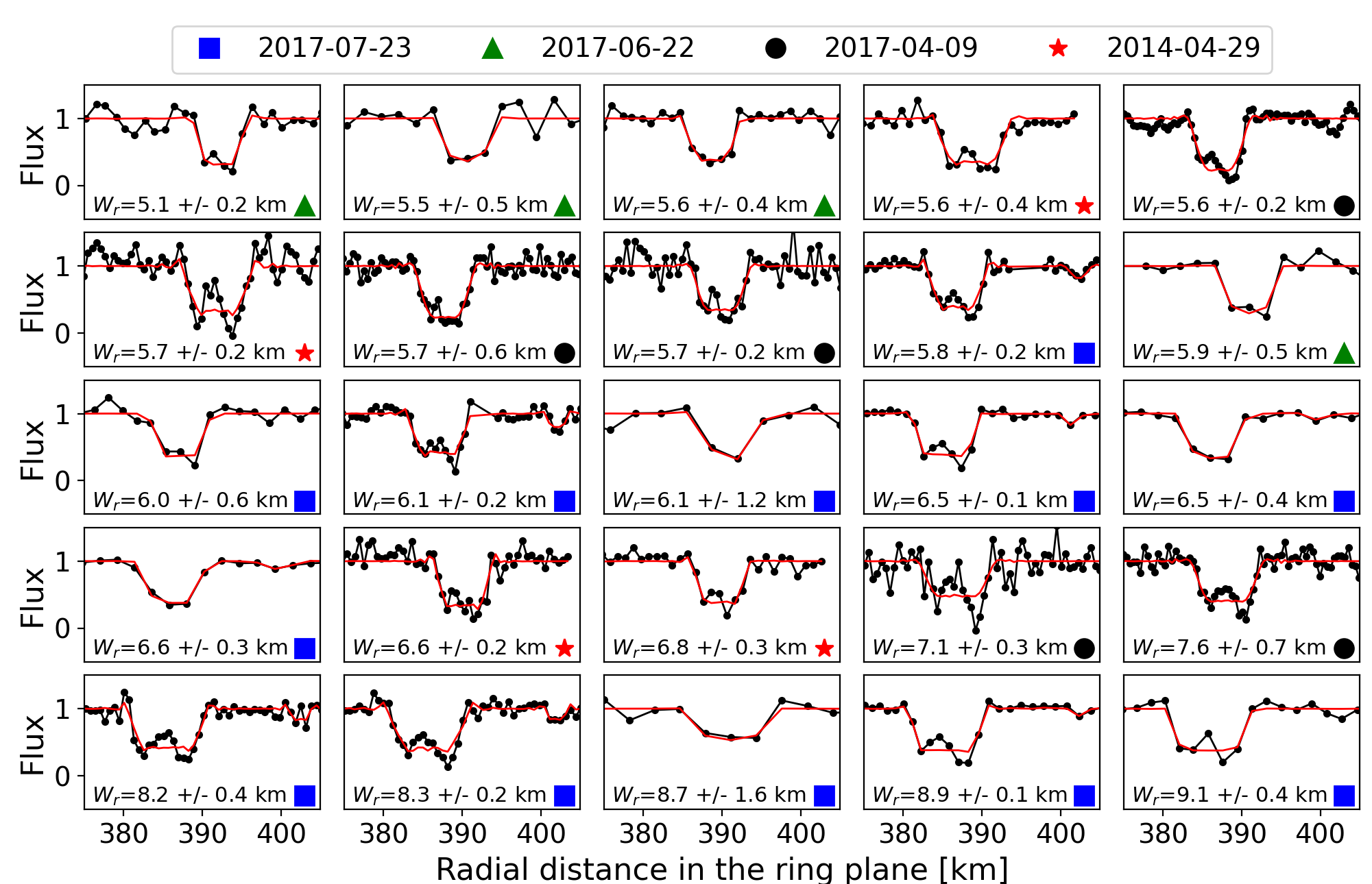}
\caption{%
All resolved detections of Chariklo C1R over radial distance in the ring plane. The black dots represents the observed flux and the red line is the best fitted model. The detections were sorted by radial width, as written in the lower part of each panel. The symbols in the lower right of each panel represents the different occultations. The data and analysis of the 2014-04-29 occultation was obtained from \cite{Berard_2017}.
}%
\label{Fig:structure_all}
\end{figure*}

The observation on 2017-04-09 and some of 2017-07-23 allowed to probe C1R with a radial resolution of $\sim$0.5 and $\sim$1.0 km, respectively. This high resolution allowed precise detection of W-shaped structures within C1R. Also, there is a clear difference between the structures over the detections, as shown in Figure \ref{Fig:structure_all} containing all resolved detections since 2014. From our observations, we detect that the mean radial width of C1R is $\sim$6.9 km and can range between $\sim$4.8 and $\sim$9.1 km.

The Danish 1.54 meters telescope in La Silla observed the event on 2017-07-23 in two different bands, visual (0.45-0.65 microns) and red (0.70-1.00 microns). There is no significant difference between these two observations, what can be interpreted as that Chariklo's ring is mostly made  by particles larger than a few microns.

The observations from the large telescope's facilities on 2017-07-23 allowed for the first time a multi-chord detection of C2R. From this, we can do an independent analysis of C2R and C1R and compare the obtained values. We obtained a 1-sigma agreement between the rings' pole orientation, a strong argument that they are coplanar, as expected. There is a slight difference between the centre position of both rings. That can be explained as a differential eccentricity between both rings, but this is not significant at the 3-sigma level. The determined radial difference in the ring plane between C1R and C2R is $14^{+7}_{-8}$ km. We checked for material between C1R and C2R and determined a 3-sigma upper limit of a ring material with normal opacity of 0.006.

Multiple stellar occultations between 2013 and 2020 combined with statistical analysis allowed the determination of Chariklo's 3D size and shape. We fitted 11 stellar occultations and determined that Chariklo is consistent with a triaxial ellipsoid with semi-axes of $143.8^{+1.4}_{-1.5}$, $135.2^{+1.4}_{-2.8}$ and $99.1^{+5.4}_{-2.7}$, and a volume equivalent radius of $124.8^{+3.0}_{-2.3}$. These values are consistent, within error bars, with previous values derived from occultations \citep{Leiva_2017}. We highlight that the new data, including multi-chord occultations, reduced the uncertainty of the modelled parameters. This 3D-shape is essential for constraining the dynamics of Chariklo's ring system.

Stellar occultations provide Chariklo's astrometric positions at mas-precision level, even for single- or double-chords observations. Those positions have been implemented in the \textsc{nima} software to update Chariklo's ephemeris. It has a propagated uncertainty of $\sim$5 mas ($\sim$50 km at Chariklo's distance) up to January 2022. The 50-km uncertainty is small compared to Chariklo's diameter ($\sim$260~km) and the ring span ($\sim$800~km). This permits a careful planning of future occultations with a high rate of success. In addition to Chariklo's, the ephemerides of other small bodies with comparable accuracies are available on the Lucky Star webpage\footnote{\url{https://lesia.obspm.fr/lucky-star/predictions.php}.}.

The parameters obtained in this work should be useful for constraining dynamical models of Chariklo and its rings, and provide new insights into the formation and evolution of this system.

\begin{acknowledgements}

We would like to thank D. Souami for her useful feedback, which improved the quality of this manuscript.

This work was carried out within the “Lucky Star" umbrella that agglomerates the efforts of the Paris, Granada and Rio teams, which is funded by the European Research Council under the European Community’s H2020 (ERC Grant Agreement No. 669416).

This research made use of \textsc{sora}, a python package for stellar occultations reduction and analysis, developed with the support of ERC Lucky Star and LIneA/Brazil, within the collaboration Rio-Paris-Granada teams.

This work has made use of data from the European Space Agency (ESA) mission Gaia (\url{https://www.cosmos.esa.int/gaia}), processed by the Gaia Data Processing and Analysis Consortium (DPAC, https://www.cosmos.esa.int/web/gaia/dpac/consortium).

Part of this research is suported by INCT do e-Universo, Brazil (CNPQ grants 465376/2014-2). Based in part on observations made at the Laborat\'orio Nacional de Astrof\'isica (LNA), Itajub\'a-MG, Brazil.

The data include observations taken by the MiNDSTEp team at the Danish 1.54m telescope at ESO’s La Silla observatory” and “UGJ  acknowledges funding from the European Union H2020-MSCA-ITN-2019 under grant no. 860470 (CHAMELEON) and from the Novo Nordisk Foundation Interdisciplinary Synergy Programme grant no. NNF19OC0057374.

TRAPPIST-South is funded by the Belgian Fund for Scientific Research (Fond National de la Recherche Scientifique,
FNRS) under the grant PDR T.0120.21.

The authors acknowledge the use of Sonja Itting-Enke’s C14 telescope and the facilities at the Cuno Hoffmeister Memorial Observatory (CHMO).

The following authors acknowledge the respective CNPq grants: BEM 150612/2020-6; FB-R 314772/2020-0; RV-M 304544/2017-f5, 401903/2016-8; JIBC 308150/2016-3 and 305917/2019-6; MA 427700/2018-3, 310683/2017-3, 473002/2013-2. GBR acknowledges CAPES-FAPERJ/PAPDRJ grant E26/203.173/2016 and CAPES-PRINT/UNESP grant 88887.571156/2020-00, MA FAPERJ grant E-26/111.488/2013 and ARGJr FAPESP grant 2018/11239-8.

J.L.O., P.S-S., R.D., and N.M. acknowledge financial support from the State Agency for Research of the Spanish MCIU through the ``Center of Excellence Severo Ochoa'' award to the Instituto de Astrofísica de Andalucía (SEV-2017-0709), from Spanish project AYA2017-89637-R, and from FEDER. P.S-S. acknowledges financial support by the Spanish grant AYA-RTI2018-098657-J-I00 ``LEO-SBNAF'' (MCIU/AEI/FEDER, UE).

E. Jehin is a Belgian FNRS Senior Research Associate.

RS and OCW acknowledge FAPESP grant 2016/24561-0 and CNPq grant 305210/2018-1.

TCH acknowledges financial support from the National Research Foundation (NRF; No. 2019R1I1A1A01059609).

\end{acknowledgements}

\bibliographystyle{aa} 
\bibliography{ref} 

\begin{appendix} 

\section{Observational circumstances}\label{App:Observations}

Table \ref{tb:obs_sites_1} summarise the observational circumstances of each station for the eight stellar occultations presented here (name of station, coordinates, observers, telescope aperture, detector, band, exposure and  cycle times, and the light curve Root Mean Square (RMS) noise. The status of each observation is mentioned (positive or negative detection of the main body, resolved, unresolved or no detection of the rings). Weather overcast and instrumental issues are also indicated.

\begin{table*}
\begin{center}
\caption{Observational stations, technical details and circumstance.}
\begin{tabular}{lccccc}
\hline
\hline
Site   & Longitude  & Observers & Telescope aperture & Exposure Time & Light curve RMS \\
Status Main Body      & Latitude  &           & Detector                & Cycle         & (flux) \\
Status Ring    & Altitude  &           & Filter             & (s)           & \\
\hline
\hline
\vspace{0.05em}\\
\multicolumn{6}{c}{\textbf{2020-06-19 -- Australia}}\\
\hline
\hline
Glenlee & 150$^o$ 30' 01.6" E & S. Kerr & 30 cm         & 0.640 & 0.138 \\
Positive Detection & \phantom{0}23$^o$ 16' 10.1" S &            & Watec 910BD  & 0.640 & \\
Unresolved Detection & \phantom{00}50 m &       & Clear &  \\ 
\hline
\hline
\vspace{0.05em}\\
\multicolumn{6}{c}{\textbf{2019-09-04 -- Hawaii}}\\
\hline
\hline
Gemini North -- Mauna Kea & 155$^o$ 28' 08.0" W & F. Braga-Ribas & 810 cm        & 0.040 & 0.019 \\
Instrumental Issue & \phantom{0}19$^o$ 49' 25.0" N &            & Andor/IXon-EM  & 0.040 & \\
                   & 4184 m &       & Blue-g / Red-i &  \\ 
\hline
Mauna Loa & 155$^o$ 34' 34.9" W & P. Maley & 36 cm         & 0.017 & 0.127 \\
Instrumental Issue & \phantom{0}19$^o$ 32' 10.6" N &  C. Erickson  & Watec 910HX & 0.017 & \\
Resolved Detection$^a$ & 3395 m &       & Clear &  \\ 
\hline
\hline
\vspace{0.05em}\\
\multicolumn{6}{c}{\textbf{2019-08-08 -- La Réunion / Namibia}}\\
\hline
\hline

Ste. Marie / La Réunion  & 55$^o$ 34' 00.2" E & B. Mondon & 28 cm         & 0.500 & 0.269 \\
Positive Detection     & 20$^o$ 53' 48.5" S &     & QHY174M  & 0.500 & \\
No Detection$^b$ & \phantom{00}50 m &       & Clear &  \\ 
\hline

Ste. Marie / La Réunion  & 55$^o$ 31' 08.6" E & G. Hesler & 40 cm         & 0.500 & 0.405 \\
Positive Detection     & 20$^o$ 54' 38.9" S &     & QHY174M  & 0.500 & \\
No Detection & \phantom{00}85 m &       & Clear &  \\ 
\hline

Maido / La Réunion      & 55$^o$ 23' 14.5" E & T. Payet & 30 cm         & 0.500 & 0.286 \\
Positive Detection    & 21$^o$ 04' 15.5" S &     & ASI178MM  & 0.500 & \\
No Detection & 2195 m &       & Clear &  \\ 
\hline

Les Makes / La Réunion & 55$^o$ 24' 36.3" E & P. Thierry & 60 cm         & 0.531 & 0.263 \\
Positive Detection    & 21$^o$ 11' 56.1" S &     & ASI178MM  & 0.531 & \\
No Detection & \phantom{0}990 m &       & Clear &  \\ 
\hline

Langevin / La Réunion             & 55$^o$ 38' 47.7" E & F. Colas & 28 cm     & 0.333 & 0.290 \\
Positive Detection   & 21$^o$ 23' 12.0" S &          & ASI178MM  & 0.333 & \\
No Detection         & \phantom{00}20 m   &          & Clear     &  \\ 
\hline

Vastrap Guestfarm / Namibia  & 27$^o$ 44' 05.2" E & M. Kretlow & 40 cm         & --- & --- \\
Instrumental Issue     & 18$^o$ 25' 49.5" S &     & Merlin/Raptor  & --- & \\
        & 1090 m &       & Clear &  \\ 

\hline
\hline
\vspace{0.05em}\\
\multicolumn{6}{c}{\textbf{2019-08-02 -- Australia}}\\
\hline
\hline
Yass  & 148$^o$ 58' 35.1" E & W. Hanna & 51 cm           & 3.000 & 0.207 \\
Positive Detection & \phantom{0}34$^o$ 51' 50.9" S &     & QHY174GPS  & 3.001 & \\
No Detection & \phantom{0}536 m &       & Clear &  \\ 
\hline

Murrumbateman  & 148$^o$ 59' 54.8" E & D. Herald & 40 cm         & 1.280 & 0.204 \\
Positive Detection & \phantom{0}34$^o$ 57' 31.5" S &     & Watec 910BD  & 1.280 & \\
No Detection & \phantom{0}594 m &       & Clear &  \\ 
\hline

Flynn  & 149$^o$ 02' 57.5" E & J. Newman & 40 cm         & 1.280 & 0.532 \\
No Detection & \phantom{0}35$^o$ 11' 55.3" S&     & Watec 910BD  & 1.280 & \\
No Detection & \phantom{0}657 m &       & Clear &  \\ 
\hline
\hline
\vspace{0.05em}\\
\multicolumn{6}{c}{\textbf{2017-08-24 -- Brazil}}\\
\hline
\hline

OPD -- Brazópolis  & \phantom{0}45$^o$ 34' 57.5" W & J. I. B. Camargo & 160 cm         & 0.500 & 0.063 \\
Positive Detection & \phantom{0}22$^o$ 32' 07.8" S &  F. Quispe-Huaynasi  & Andor/IXon-EM  & 0.513 & \\
Unresolved Detection & 1864 m &       & Clear &  \\ 
\hline
\hline

\multicolumn{6}{l}{\rule{0pt}{3.0ex} $^a$ Due to instrumental issue the original data-set was lost, however a video recorded by the observer's smartphone was used}\\
\multicolumn{6}{l}{to derive the immersion and emersion times of the second detection of the ring.}\\
\multicolumn{6}{l}{\rule{0pt}{3.0ex} $^b$ No detection means that the data-set's features do not allowed the detection of Chariklo's rings.}
\label{tb:obs_sites_1}
\end{tabular}
\end{center}
\end{table*}

\setcounter{table}{0}

\begin{table*}
\begin{center}
\caption{\textbf{[Cont.]} Observational stations, technical details and circumstance.}
\begin{tabular}{lccccc}
\hline
\hline
Site   & Longitude  & Observers & Telescope aperture & Exposure Time & Light curve RMS \\
Status Main Body      & Latitude  &           & Detector                & Cycle         &  (flux) \\
Status Ring    & Altitude  &           & Filter             & (s)           & \\
\hline
\hline

\vspace{0.05em}\\
\multicolumn{6}{c}{\textbf{2017-07-23 -- South America}}\\
\hline
\hline
San Pedro de Atacama / Chile  & 68$^o$ 10' 44.0" W & N. Morales     & 41 cm         & 2.000 & 0.058 \\
Negative Detection            & 22$^o$ 57' 14.0" S &                & SBIG STL11K3  & 4.370 & \\
No Detection                  & 2397 m             &                & Clear &  \\ 
\hline
San Pedro de Atacama / Chile  & 68$^o$ 10' 48.0" W & J. Fabrega & 40 cm         & 1.000 & 0.099 \\
Negative Detection     & 22$^o$ 57' 09.0" S &     & FLI PL16803  & 2.700 & \\
Unresolved Detection     & 2396 m &       & Clear &  \\ 
\hline
San Pedro de Atacama / Chile  & 68$^o$ 10' 48.0" W & J. F. Soulier & 40 cm         & 3.000 & 0.023 \\
Negative Detection     & 22$^o$ 57' 09.0" S &     & QHY9 CCD  & 7.000 & \\
No Detection     & 2396 m &       & Clear &  \\ 
\hline
San Pedro de Atacama / Chile  & 68$^o$ 10' 48.0" W & A. Maury & 50 cm         & 3.000 & 0.045 \\
Negative Detection     & 22$^o$ 57' 09.0" S &     & FLI13803  & 5.650 & \\
No Detection     & 2396 m &       & Clear &  \\ 
\hline
Cerro Burek / Argentina  & 69$^o$ 18' 25.0" W & N. Morales     & 45 cm          & 2.000 & 0.066 \\
Negative Detection       & 31$^o$ 47' 12.0" S &                & SBIG STL11000  & 4.250 & \\
No Detection             & 2665 m             &                & Clear &  \\ 
\hline
Ckoirama / Chile  & 69$^o$ 55' 49.0" W & J. P. Colque & 60 cm         & 1.000 & 0.188 \\
Negative Detection  & 24$^o$ 05' 21.0" S &     & FLI PL16801  & 2.569 & \\
No Detection        & 0964 m &       & Clear &  \\ 
\hline
VLT -- Cerro Paranal / Chile  & 70$^o$ 24' 16.5" W & B. Sicardy & 820 cm         & 0.100 & 0.027 \\
Negative Detection     & 24$^o$ 37' 39.3" S &     & HAWK-I  & 0.100 & \\
Resolved Detection     & 2635 m &       & K$_s$-band &  \\ 
\hline
SSO$^{c}$ -- Cerro Paranal / Chile  & 70$^o$ 24' 16.5" W & E. Jehin & 100 cm         & 1.200 / 1.200 & 0.020 / 0.013 \\
Negative Detection     & 24$^o$ 37' 39.3" S &     & Andor/IKONL  & 2.000 / 2.040 & \\
Unresolved Detection     & 2635 m &       & g' / I + z &  \\ 
\hline
Tolar Grande / Argentina  & 67$^o$ 23' 43.6" W & M. Kretlow & 41 cm         & 0.130 & 0.092 \\
Negative Detection     & 24$^o$ 35' 23.5" S & R. A. Artola     & Merlin/Raptor  & 0.130 & \\
Resolved Detection     & 3524 m &       & Clear &  \\ 
\hline
Cono de Arita / Argentina  & 67$^o$ 46' 00.2" W & M. D. Starck-Cuffini & 25 cm         & 0.500 & 0.161 \\
Positive Detection     & 25$^o$ 01' 59.8" S & M. V. Sieyra     & ASI178MM  & 0.500 & \\
No Detection     & 3532 m &       & Clear &  \\ 
\hline
El Rodeo / Argentina  & 66$^o$ 08' 35.4" W & E. Meza & 36 cm         & 0.150 & 0.358 \\
Positive Detection     & 24$^o$ 52' 37.0" S & R. Melia    & Merlin/Kite  & 0.170 & \\
Unresolved Detection     & 3022 m &       & Clear &  \\ 
\hline
Cachi Adentro / Argentina  & 66$^o$ 14' 16.7" W & E. M. Schneiter & 36 cm         & 0.100 & 0.452 \\
Positive Detection     & 25$^o$ 05' 33.9" S & C. Colazo    & Merlin/Raptor  & 0.100 & \\
No Detection     & 2684 m & M. Buie      & Clear &  \\ 
\hline
Augusto de Lima / Brazil  & 44$^o$ 04' 08.4" W & B. L. Giacchini & 30 cm         & 1.000 & 0.174 \\
Positive Detection     & 18$^o$ 01' 20.7" S & G. Benedetti-Rossi     & Merlin/Raptor  & 1.000 & \\
No Detection     & 0620 m &       & Clear &  \\ 
\hline
El Salvador / Chile  & 69$^o$ 45' 03.8" W & J-L. Dauvergne & 28 cm         & 0.050 & 0.318 \\
Positive Detection     & 26$^o$ 18' 48.3" S &     & Merlin/Raptor  & 0.050 & \\
Unresolved Detection     & 1607 m &       & Clear &  \\ 
\hline
São José do Rio Preto / Brazil  & 49$^o$ 17' 52.2" W & R. Sfair & 35 cm         & 1.000 & 0.213 \\
Positive Detection     & 20$^o$ 42' 43.4" S & A. Amarante    & Merlin/Raptor  & 1.000 & \\
No Detection     & 0482 m & R. Poltronieri      & Clear &  \\ 
\hline
Inca de Oro / Chile  & 69$^o$ 54' 45.6" W & S. Bouley & 25 cm         & 0.100 & 0.267 \\
Positive Detection     & 26$^o$ 44' 45.3" S &     & Merlin/Raptor  & 0.100 & \\
Unresolved Detection     & 1592 m &       & Clear &  \\ 
\hline
Bilac / Brazil  & 50$^o$ 30' 20.9" W & L. S. Amaral & 25 cm         & 6.000 & 0.150 \\
Positive Detection     & 21$^o$ 23' 38.6" S &     & Canon T1i  & 11.000 & \\
No Detection     & 0426 m &       & Clear &  \\ 
\hline
Danish -- La Silla / Chile  & 70$^o$ 44' 20.2" W & C. Snodgrass    & 154 cm         & 0.033 / 0.033 & 0.082 / 0.063 \\
Negative Detection          & 29$^o$ 15' 21.3" S & T. C. Hinse     & Lucky Imager  & 0.034 / 0.034 & \\
Resolved Detection          & 2337 m             & U. G. Jorgensen & Visual \& Red$^{d}$ &  \\ 
                            &                    & M. Dominik      &                     &  \\ 
                            &                    & J. Skottfelt    &                     &  \\ 

\hline
\hline

\multicolumn{6}{l}{\rule{0pt}{3.0ex} $^{c}$ The SSO observations were done at the telescopes Io and Europa.}\\
\multicolumn{6}{l}{\rule{0pt}{3.0ex} $^{d}$ The Danish dual experiment observed in the Visual (0.45-0.65 $\mu$m) and Red (0.7-1.0 $\mu$m) bands.}\\

\label{tb:obs_sites_2}
\end{tabular}
\end{center}
\end{table*}

\setcounter{table}{0}

\begin{table*}
\begin{center}
\caption{\textbf{[Cont.]} Observational stations, technical details and circumstance.}
\begin{tabular}{lccccc}
\hline
\hline
Site   & Longitude  & Observers & Telescope aperture & Exposure Time & Light curve RMS \\
Status Main Body      & Latitude  &           & Detector                & Cycle         & (flux) \\
Status Ring    & Altitude  &           & Filter             & (s)           & \\
\hline
\hline
\vspace{0.05em}\\
\multicolumn{6}{c}{\textbf{2017-07-23 -- South America [Cont.]}}\\
\hline
\hline
1m -- La Silla / Chile  & 70$^o$ 44' 20.2" W & A. Zapata & 100 cm         & 0.100 & 0.074 \\
Negative Detection     & 29$^o$ 15' 21.3" S & P. Torres & Merlin/Raptor  & 0.100 & \\
Resolved Detection     & 2337 m &       & Clear &  \\ 
\hline
TRAPPIST-South -- La Silla / Chile  & 70$^o$ 44' 20.2" W & E. Jehin & 60 cm         & 2.000 & 0.040 \\
Negative Detection     & 29$^o$ 15' 21.3" S &     & FLI PL3041-BB  & 2.795 & \\
Unresolved Detection     & 2337 m &       & Clear &  \\ 
\hline
Foz do Iguaçu / Brazil  & 54$^o$ 35' 37.0" W & D. I. Machado & 28 cm         & 4.000 & 0.038 \\
Negative Detection     & 25$^o$ 26' 05.0" S & L. L. Trabuco  & Merlin/Raptor  & 4.000 & \\
No Detection     & 0184 m &       & Clear &  \\ 
\hline
OPD -- Brazópolis / Brazil  & \phantom{0}45$^o$ 34' 57.5" W & F. Braga-Ribas & 160 cm         & 0.117 & 0.029 \\
Negative Detection & \phantom{0}22$^o$ 32' 07.8" S & F. L. Rommel      & Andor/IXon-EM  & 0.130 & \\
Resolved Detection & 1864 m &       & Clear &  \\ 
\hline
SARA -- Cerro Tololo / Chile  & 70$^o$ 48' 23.0" W & R. Leiva & 60 cm         & 1.000 & 0.094 \\
Negative Detection     & 30$^o$ 10' 11.0" S & S. Levine    & FLI  & 1.679 & \\
Resolved Detection     & 2207 m & C. Zuluaga      & Clear &  \\ 
\hline
PROMPT$^{e}$ -- Cerro Tololo / Chile  & 70$^o$ 48' 23.0" W & J. Pollock    & 40 cm         & 1.000 & 0.037 \\
Negative Detection                    & 30$^o$ 10' 11.0" S & V. Kouprianov & AltaU-47  & 0.615 & \\
Resolved Detection                    & 2207 m             & D. Reichart   & Clear &  \\
                                      &                    & T. Linder     &       &  \\
\hline
Oliveira / Brazil  & 43$^o$ 59' 03.1" W & C. Jacques & 45 cm         & --- & --- \\
Weather Overcast   & 19$^o$ 52' 55.0" S &     & ML FLI16803  & --- & \\
        & 982 m &       & Clear   \\ 
\hline
Guaratinguetá / Brazil  & 45$^o$ 11' 25.0" W & T. Santana       & 40 cm         & --- & --- \\
Weather Overcast      & 22$^o$ 48' 34.0" S & T. Moura     & Merlin/Raptor  & --- & \\
                      & 0567 m             & O. C. Winter       & Clear &  \\ 
                      &                    & T. Akemi    &     &  \\ 
\hline
Brasília / Brazil  & 47$^o$ 54' 39.9" W & P. Cacella & 50 cm         & --- & --- \\
Weather Overcast   & 15$^o$ 53' 29.9" S &     & ASI174MM  & --- & \\
        & 1064 m &       & Clear &  \\ 
\hline
Ponta Grossa / Brazil  & 50$^o$ 05' 56.6" W & C. L. Pereira & 40 cm         & --- & --- \\
Weather Overcast     & 25$^o$ 05' 22.5" S &  M. Emilio    & Merlin/Raptor  & --- & \\
        & 910 m &       & Clear &  \\ 
\hline
SOAR - Cerro Pachón / Chile  & 70$^o$ 44' 21.1" W & J. I. B. Camargo & 410 cm         & --- & --- \\
Weather Overcast     & 30$^o$ 14' 16.9" S &     & Merlin/Raptor  & --- & \\
     & 2694 m &       & Clear &  \\ 

\hline
\hline
\vspace{0.05em}\\
\multicolumn{6}{c}{\textbf{2017-06-22 -- Namibia}}\\
\hline
\hline
Outeniqua Lodge     & 16$^o$ 49' 17.7" E & F. Colas   & 25 cm    & 0.100 & 0.325 \\
Positive Detection  & 21$^o$ 17' 58.2" S & J. Desmars & Merlin/Raptor   & 0.100 & \\
Unresolved Detection  & 1416 m             &            & Clear    &  \\ 
\hline

Onduruquea Guest Farm & 15$^o$ 59' 33.7" E & M. Kretlow & 50 cm     & 0.075 & 0.116 \\
Positive Detection    & 21$^o$ 36' 26.0" S &            & Merlin/Raptor    & 0.080 & \\
Resolved Detection    & 1220 m             &            & Clear     &  \\ 
\hline

Windhoek             & 17$^o$ 06' 31.9" E & M. Backes &  35 cm   & 0.150 & 0.148 \\
Positive Detection   & 22$^o$ 41' 55.2" S &           & Merlin/Raptor   & 0.150 & \\
Unresolved Detection & 1902 m             &           & Clear     &  \\ 
\hline

Windhoek             & 17$^o$ 06' 31.9" E & E. Meza &  40 cm    & 0.200 & 0.151 \\
Positive Detection   & 22$^o$ 41' 55.2" S &         & ASI178MM & 0.203 & \\
Unresolved Detection & 1902 m             &         & Clear     &  \\ 
\hline

Tivoli Lodge         & 18$^o$ 01' 01.2" E & L. Maquet & 40 cm    & 0.075 & 0.175 \\
Positive Detection   & 23$^o$ 27' 40.2" S &           & Merlin/Raptor   & 0.075 & \\
Unresolved Detection & 1344 m             &           & Clear    &  \\ 
\hline

Hakos               & 16$^o$ 21' 41.3" E & W. Beisker & 50 cm   & 0.050 & 0.122 \\
Negative Detection  & 23$^o$ 14' 11.0" S &            & Merlin/Raptor  & 0.050 & \\
Resolved Detection  & 1843 m             &            & Clear   &  \\ 
\hline
\hline

\multicolumn{6}{l}{\rule{0pt}{3.0ex} $^{e}$ The PROMPT observation were the combined data observed by P1, P3 and P5, allowing for a short cycle.}\\

\label{tb:obs_sites_3}
\end{tabular}
\end{center}
\end{table*}

\setcounter{table}{0}

\begin{table*}
\begin{center}
\caption{\textbf{[Cont.]} Observational stations, technical details and circumstance.}
\begin{tabular}{lccccc}
\hline
\hline
Site   & Longitude  & Observers & Telescope aperture & Exposure Time & Light curve RMS \\
Status Main Body      & Latitude  &           & Detector                & Cycle         & (flux) \\
Status Ring    & Altitude  &           & Filter             & (s)           & \\
\hline
\hline

\vspace{0.05em}\\
\multicolumn{6}{c}{\textbf{2017-04-09 -- Namibia}}\\
\hline
\hline

Wabi Lodge  & 17$^o$ 32' 00.6" E & J. L. Dauvergne & 30 cm         & 0.100 & 0.212 \\
Positive Detection     & 20$^o$ 20' 38.9" S &     & Merlin/Raptor  & 0.100 & \\
Resolved Detection     & 1382 m &       & Clear &  \\ 
\hline

Weaver Rocks Lodge  & 16$^o$ 50' 29.9" E & M. Kretlow & 50 cm         & 0.080 & 0.147 \\
Positive Detection     & 20$^o$ 41' 56.3" S &     & Merlin/Raptor  & 0.080 & \\
Resolved Detection     & 1647 m &       & Clear &  \\ 
\hline

Hakos  & 16$^o$ 21' 42.0" E & K. L. Bath & 50 cm         & 0.100 & 0.089 \\
Negative Detection     & 23$^o$ 14' 10.0" S &     & Merlin/Raptor  & 0.100 & \\
Resolved Detection$^f$     & 1881 m &       & Clear &  \\ 
\hline
\hline

\multicolumn{6}{l}{\rule{0pt}{3.0ex} $^{f}$ Only the second detection of the C1R was observed, due to the weather condition.}\\
\label{tb:obs_sites_4}
\end{tabular}
\end{center}
\end{table*}


\section{Occultation maps}\label{App:Maps}

\begin{figure*}[h]
\centering
\includegraphics[width=1.00\textwidth]{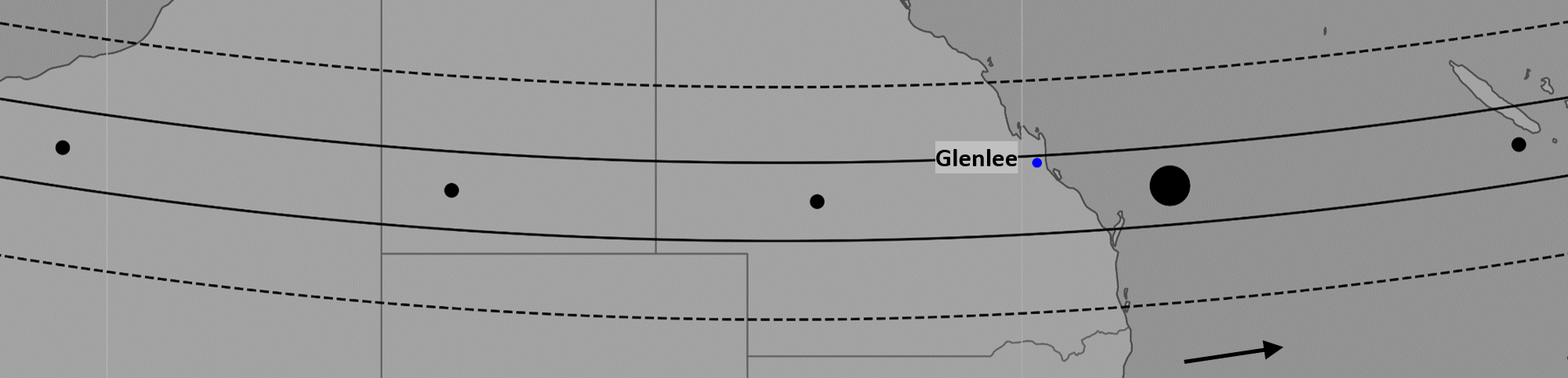}
\caption{Occultation map for the 2020-06-19 event.}
\label{Fig:occ_map_1}
\end{figure*}

\begin{figure*}[h]
\centering
\includegraphics[width=1.00\textwidth]{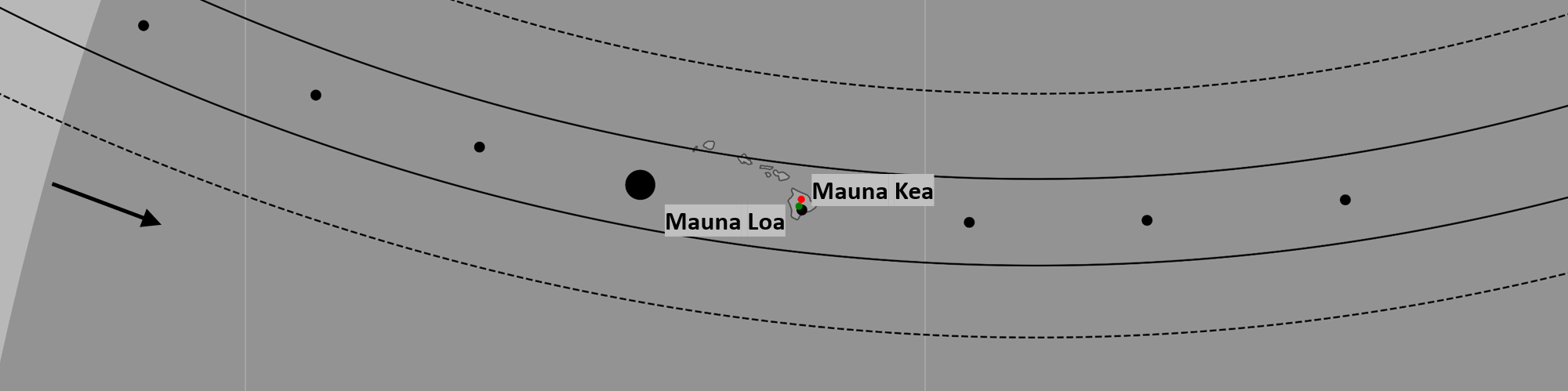}
\caption{Occultation map for the 2019-09-04 event.}
\label{Fig:occ_map_2}
\end{figure*}

\begin{figure*}[h]
\centering
\includegraphics[width=1.00\textwidth]{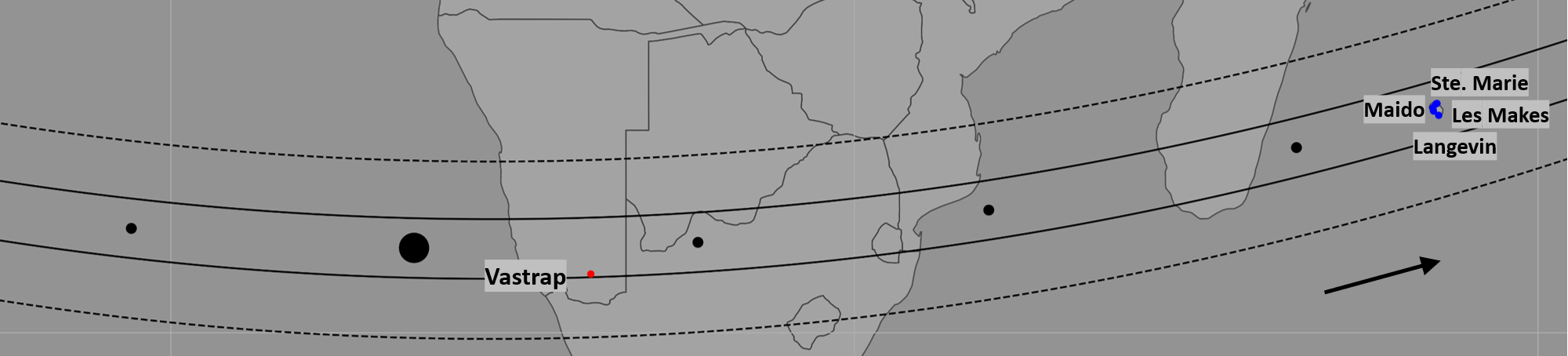}
\caption{Occultation map for the 2019-08-08 event.}
\label{Fig:occ_map_3}
\end{figure*}

\begin{figure*}[h]
\centering
\includegraphics[width=1.00\textwidth]{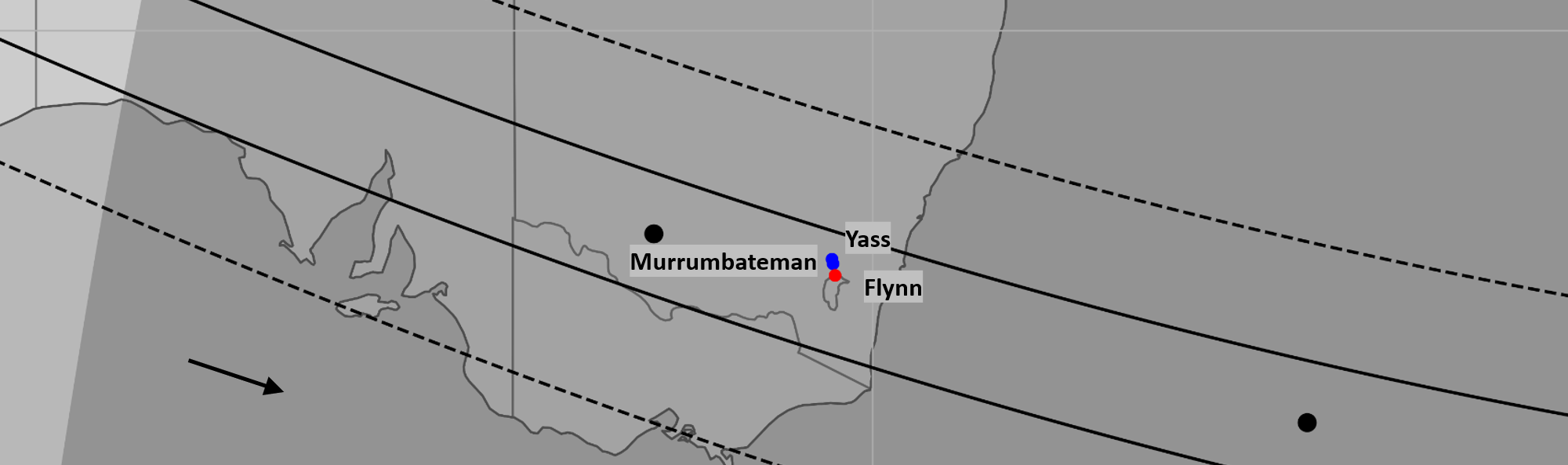}
\caption{Occultation map for the 2019-08-02 event.}
\label{Fig:occ_map_4}
\end{figure*}

\begin{figure*}[h]
\centering
\includegraphics[width=1.00\textwidth]{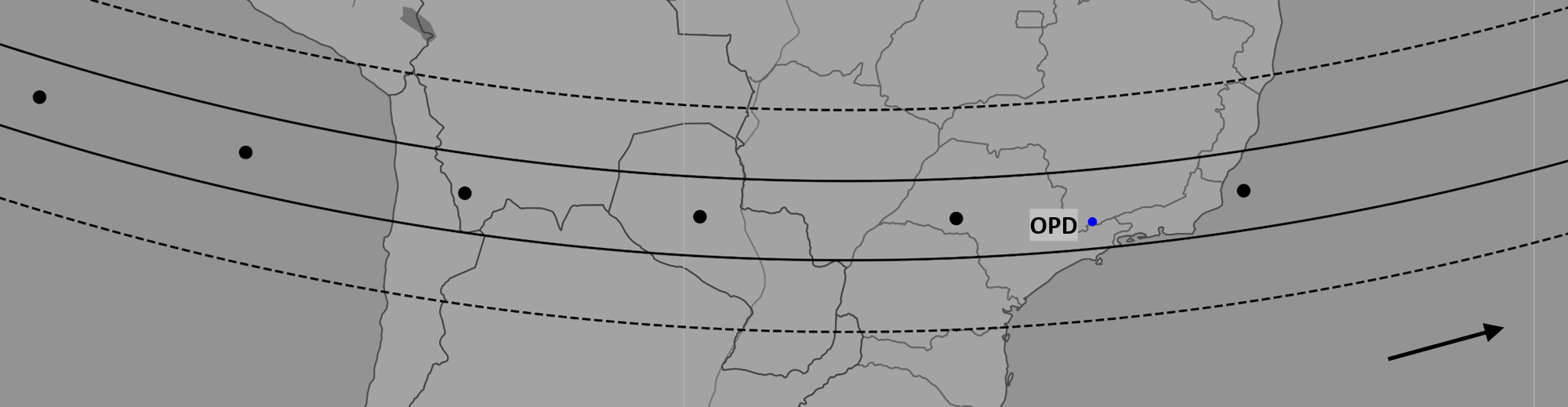}
\caption{Occultation map for the 2017-08-24 event.}
\label{Fig:occ_map_5}
\end{figure*}

\begin{figure*}[h]
\centering
\includegraphics[width=1.00\textwidth]{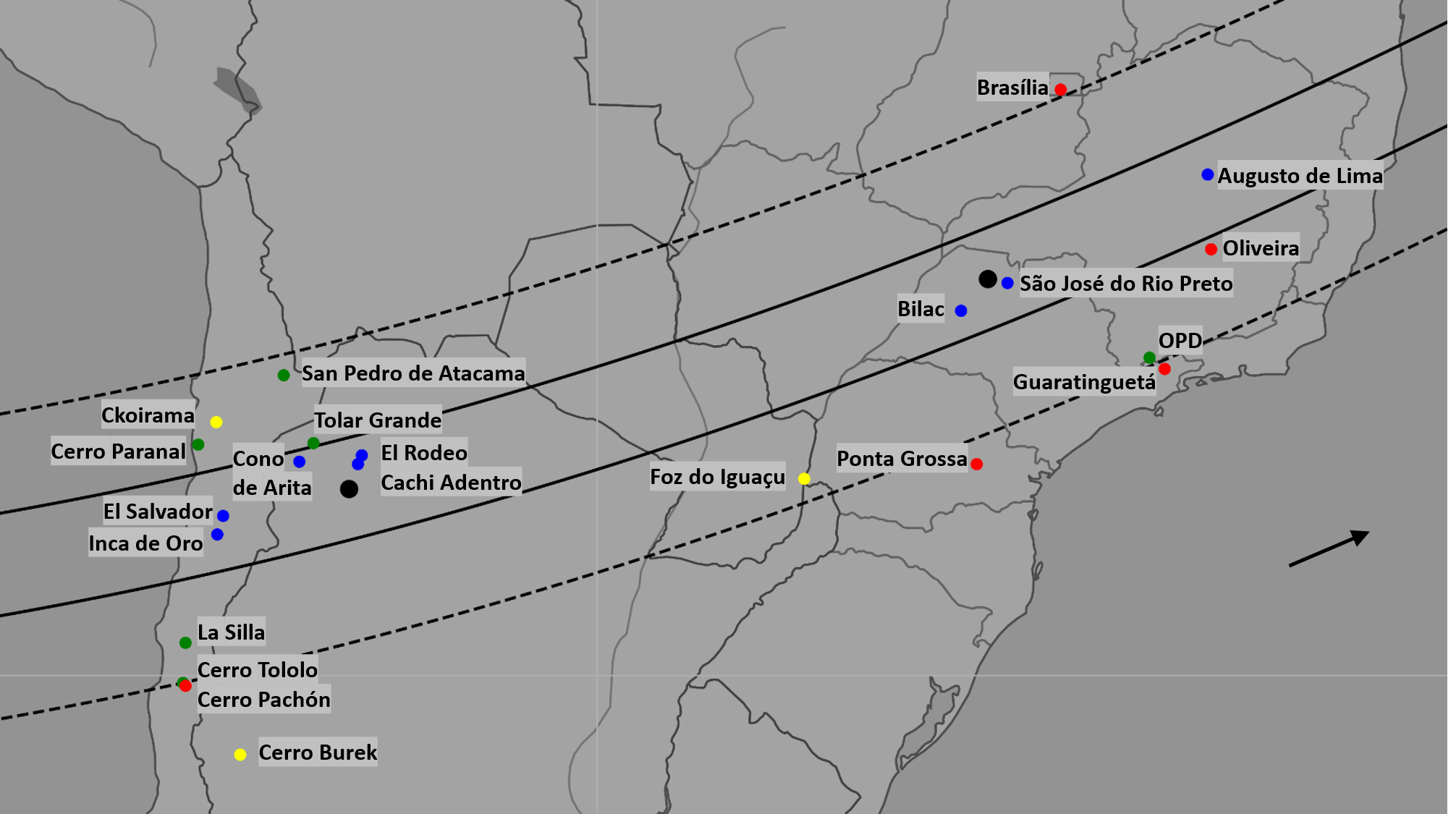}
\caption{Occultation map for the 2017-07-23 event.}
\label{Fig:occ_map_6}
\end{figure*}

\begin{figure*}[h]
\centering
\includegraphics[width=1.00\textwidth]{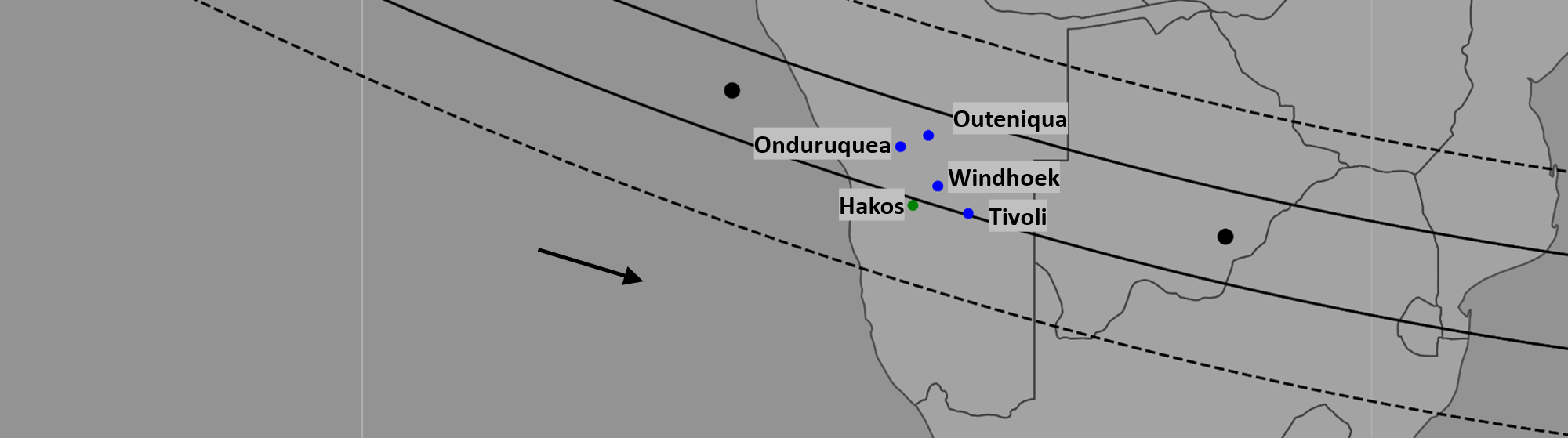}
\caption{Occultation map for the 2017-06-22 event.}
\label{Fig:occ_map_7}
\end{figure*}

\begin{figure*}[h]
\centering
\includegraphics[width=1.00\textwidth]{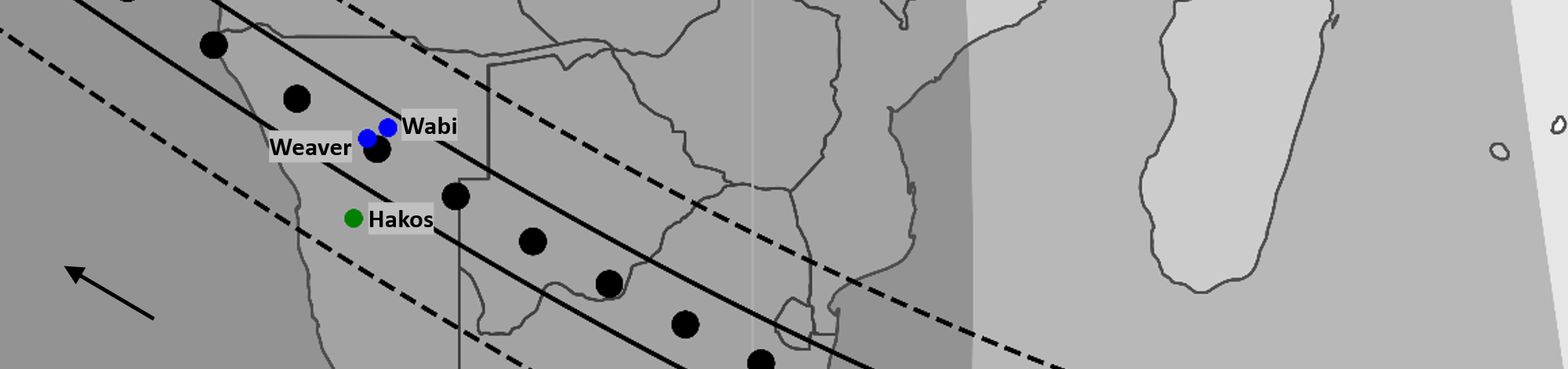}
\caption{Occultation map for the 2019-09-04 event.}
\label{Fig:occ_map_8}
\end{figure*}

Figures \ref{Fig:occ_map_1}, \ref{Fig:occ_map_2}, \ref{Fig:occ_map_3}, \ref{Fig:occ_map_4}, \ref{Fig:occ_map_5}, \ref{Fig:occ_map_6}, \ref{Fig:occ_map_7} and \ref{Fig:occ_map_8} contain the occultation maps for the eight unpublished stellar occultations between 2017 and 2020. The maps were organised in inverse chronological order. The blue dots stand for the stations with a positive detection over the main body, the green dots are the ones with positive detections only for the rings, the yellow dots were stations with data that was unfit to detect the rings and negatives for the main body, and the red dots were stations without data covering the occultation instants (weather overcast or instrumental issues). The black dots stand for the centre of Chariklo every minute, and the big black dot represents the closest approach time. The solid black line stands for the path of Chariklo's shadow and the dashed line for the ring's shadow.

\section{Occultation times and fitted parameters} \label{App:Fitted_params}

Table \ref{tb:times_param_1} lists the fitted times (immersion and emersion), duration of the event ($\Delta t$), apparent opacity ($p'$, in sky plane) and minimum $\chi^2$ per degree of freedom ($\chi^2_{pdf}$) obtained for all the positive detections. The events are presented in reversed chronological order, and the detections are listed from northernmost to southernmost stations. The type of detection is mentioned: MB for Chariklo's main body, C1R and C2R for ring detections, for each we usually have a $1^{\rm st}$ and a $2^{\rm nd}$ (immersion and emersion) detections.

Table \ref{tb:ring_parameters_1} presents the rings parameters projected into the ring plane for each detection: the radial distance ($r$) to the ring centre (assumed to be circular), the radial width ($W_r$), normal opacity ($p_N$) and equivalent width ($E_p$). The number of data-points within each detection (\#) is also given, as well as the SNR of the observed stellar drops, i.e. the ratio of the stellar drop (one minus the bottom flux of the event, $\phi_0$) and the RMS noise of the light curve.  

\begin{table*}[h]
\begin{center}
\caption{Fitted parameters obtained for each light curve with a positive detection.}
\begin{tabular}{lcccccc}
\hline
\hline
Site   & Detection  & Immersion time & Emersion time  & Duration & Apparent & $\chi^2_{pdf}$\\
       &            & (hh:mm:ss.sss) & (hh:mm:ss.sss) & (s)      & opacity  & (imm / eme) \\
\hline
\hline
\vspace{0.05em}\\
\multicolumn{7}{c}{\textbf{2020-06-19 -- Australia}}\\
\hline
\hline
Glenlee    & MB & 15:51:51.770 (0.089) & 15:51:59.334 (0.108) & 07.564 (0.197) & 1.000     & 0.978 \\
& C1R 1$^{st}$      & 15:51:37.601 (0.212) & 15:51:39.584 (0.295) & 01.983 (0.507) & 0.225 (0.073) & 0.920 \\
& C1R 2$^{nd}$      & 15:52:13.703 (0.337) & 15:52:14.004 (0.296) & 00.301 ($^{+0.633}_{-0.301}$) & 0.696 ($^{+0.301}_{-0.543}$) & 1.072 \\

\hline
\hline
\vspace{0.05em}\\
\multicolumn{7}{c}{\textbf{2019-08-08 -- La Réunion / Namibia}}\\
\hline
\hline
Ste. Marie - Mondon & MB & 21:38:15.627 (0.223) & 21:38:28.460 (0.140) & 12.833 (0.363) & 1.000 & 0.787 / 0.590 \\
Ste. Marie - Hesler & MB & 21:38:15.871 (0.244) & 21:38:28.447 (0.320) & 12.576 (0.564) & 1.000 & 1.250 / 0.847 \\
Maido               & MB & 21:38:16.659 (0.266) & 21:38:29.231 (0.109) & 12.572 (0.375) & 1.000 & 1.349 / 1.427 \\
Les Makes           & MB & 21:38:16.415 (0.144) & 21:38:29.092 (0.146) & 12.667 (0.290) & 1.000 & 0.998 / 0.821 \\
Langevin            & MB & 21:38:15.788 (0.094) & 21:38:28.339 (0.102) & 12.551 (0.196) & 1.000 & 0.750 / 0.816 \\

\hline
\hline
\vspace{0.05em}\\
\multicolumn{7}{c}{\textbf{2019-08-02 -- Australia}}\\
\hline
\hline
Yass & MB & 10:05:26.647 (0.371) & 10:05:35.976 (0.885) & 09.302 (1.256) & 1.000 & 0.681 /1.170 \\
Murrumbateman & MB & 10:05:26.657 (0.421) & 10:05:36.885 (0.405) & 10.229 (0.826) & 1.000 & 0.288 / 0.273 \\

\hline
\hline
\vspace{0.05em}\\
\multicolumn{7}{c}{\textbf{2017-08-24 -- Brazil}}\\
\hline
\hline
OPD   & MB & 02:54:17.122 (0.029) & 02:54:35.957 (0.039) & 18.834 (0.068) & 1.000 & 0.795 \\
& C1R 1$^{st}$ & 02:54:03.343 (0.182) & 02:54:04.058 (0.175) & 00.715 (0.357) & 0.782 ($^{+0.217}_{-0.548}$) & 0.910 \\
& C1R 2$^{nd}$ & 02:54:48.556 (0.153) & 02:54:49.019 (0.071) & 00.462 (0.224) & 0.286 ($^{+0.400}_{-0.056}$) & 1.141 \\

\hline
\hline
\vspace{0.05em}\\
\multicolumn{7}{c}{\textbf{2017-07-23 -- South America }}\\
\hline
\hline
S. P. A. - Fabrega   &  C1R 2$^{nd}$ & 05:55:59.034 (1.290) & 05:56:01.222 (1.372) & 02.188 ($^{+2.662}_{-2.188}$) & 0.307 ($^{+0.693}_{-0.166}$) & 1.385 \\

VLT   & C1R 1$^{st}$ & 05:55:51.104 (0.004) & 05:55:51.406 (0.008) & 00.302 (0.012) & 0.390 (0.014) & 1.403 \\
& C1R 2$^{nd}$       & 05:56:14.242 (0.011) & 05:56:14.532 (0.006) & 00.290 (0.016) & 0.429 (0.026) & 1.074 \\
& C2R 1$^{st}$       & 05:55:50.568 (0.016) & 05:55:50.740 (0.012) & 00.172 (0.028) & 0.057 (0.007) & 1.873 \\
& C2R 2$^{nd}$       & 05:56:14.956 (0.055) & 05:56:15.011 (0.055) & 00.055 ($^{+0.110}_{-0.055}$) & 0.803 ($^{+0.196}_{-0.760}$) & 0.968 \\

SSO   & C1R 1$^{st}$ & 05:55:50.836 (0.263) & 05:55:51.474 (0.243) & 00.637 (0.506) & 0.949 ($^{+0.048}_{-0.847}$) & 1.761 \\
& C1R 2$^{nd}$       & 05:56:14.242 (0.215) & 05:56:14.731 (0.068) & 00.288 (0.022) & 0.755 ($^{+0.187}_{-0.597}$) & 1.197 \\

Tolar Grande   & C1R 1$^{st}$ & 05:55:37.371 (0.026) & 05:55:37.721 (0.040) & 00.350 (0.066) & 0.271 (0.057) & 1.009 \\
& C1R 2$^{nd}$                & 05:56:03.775 (0.022) & 05:56:04.017 (0.026) & 00.242 (0.049) & 0.444 (0.090) & 1.132 \\

Cono de Arita   & MB & 05:55:48.909 (0.070) & 05:55:56.175 (0.074) & 06.955 (0.160) & 1.000 & 0.826 \\

El Rodeo   & MB & 05:55:41.690 (0.076) & 05:55:50.631 (0.091) & 08.941 (0.167) & 1.000 & 0.911 \\
& C1R 1$^{st}$  & 05:55:31.878 (0.288) & 05:55:32.374 (0.069) & 00.496 (0.357) & 0.442 ($^{+0.533}_{-0.236}$) & 0.955 \\
& C1R 2$^{nd}$  & 05:55:59.265 (0.107) & 05:55:59.819 (0.107) & 00.554 (0.214) & 0.228 ($^{+0.178}_{-0.096}$) & 0.888 \\

Cachi Adentro & MB & 05:55:41.300 (0.062) & 05:55:51.360 (0.052) & 10.060 (0.114) & 1.000 & 0.882 \\

Augusto de Lima   & MB & 05:54:24.294 (0.270) & 05:54:35.248 (0.215)& 10.955 (0.485) & 1.000 & 0.912 \\

El Salvador   & MB  & 05:55:54.242 (0.027) & 05:56:06.002 (0.021) & 11.760 (0.049) & 1.000 & 0.826 \\
& C1R 1$^{st}$      & 05:55:45.710 (0.033) & 05:55:46.009 (0.017) & 00.299 (0.049) & 0.590 (0.185) & 0.954 \\
& C1R 2$^{nd}$      & 05:56:14.349 (0.056) & 05:56:14.660 (0.140) & 00.311 (0.196) & 0.324 (0.182) & 0.944 \\

São J. Rio Preto   & MB & 05:54:39.593 (0.202) & 05:54:51.875 (0.201) & 12.282 (0.403) & 1.000 & 1.123 \\

Inca de Oro   & MB  & 05:55:54.765 (0.026) & 05:56:06.951 (0.026) & 12.186 (0.052) & 1.000 & 0.838 \\
& C1R 1$^{st}$      & 05:55:46.430 (0.043) & 05:55:46.651 (0.121) & 00.220 (0.163) & 0.396 ($^{+0.437}_{-0.242}$) & 1.296 \\
& C1R 2$^{nd}$      & 05:56:15.079 (0.027) & 05:56:15.210 (0.025) & 00.131 (0.053) & 0.559 ($^{+0.346}_{-0.192}$) & 0.903 \\

Bilac   & MB & 05:54:46.233 (3.876) & 05:54:56.465 (2.694) & 10.232 (6.569) & 1.000 & 0.552 \\

Danish Visual & C1R 1$^{st}$ & 05:55:54.536 (0.004) & 05:55:54.874 (0.006) & 00.338 (0.010) & 0.423 (0.025) & 1.123 \\
& C1R 2$^{nd}$               & 05:56:13.750 (0.011) & 05:56:14.183 (0.008) & 00.433 (0.020) & 0.458 (0.029) & 1.433 \\
& C2R 1$^{st}$               & 05:55:53.887 (0.010) & 05:55:53.914 (0.012) & 00.027 (0.022) & 0.588 (0.362) & 0.994 \\
& C2R 2$^{nd}$               & 05:56:14.868 (0.005) & 05:56:14.883 (0.006) & 00.015 (0.011) & 0.723 (0.266) & 1.101 \\

\hline
\hline
\label{tb:times_param_1}
\end{tabular}
\end{center}
\end{table*}

\setcounter{table}{0}

\begin{table*}
\begin{center}
\caption{\textbf{[Cont.]} Fitted parameters obtained for each light curve with a positive detection.}
\begin{tabular}{lcccccc}
\hline
\hline
Site   & Detection  & Immersion time & Emersion time  & Duration & Apparent & $\chi^2_{pdf}$\\
       &            & (hh:mm:ss.sss) & (hh:mm:ss.sss) & (s)      & opacity  & (imm / eme) \\
\hline
\hline
\vspace{0.05em}\\
\multicolumn{7}{c}{\textbf{2017-07-23 -- South America [Cont.] }}\\
\hline
\hline

Danish Red & C1R 1$^{st}$ & 05:55:54.539 (0.004) & 05:55:54.861 (0.005) & 00.322 (0.009) & 0.377 (0.018) & 1.077 \\
& C1R 2$^{nd}$            & 05:56:13.752 (0.008) & 05:56:14.193 (0.005) & 00.440 (0.013) & 0.365 (0.018) & 1.436 \\
& C2R 1$^{st}$            & 05:55:53.838 (0.013) & 05:55:53.924 (0.015) & 00.086 (0.028) & 0.115 (0.039) & 0.957 \\
& C2R 2$^{nd}$            & 05:56:14.822 (0.006) & 05:56:14.862 (0.008) & 00.040 (0.014) & 0.213 (0.036) & 0.904 \\

1m -- La Silla   & C1R 1$^{st}$   & 05:55:54.519 (0.013) & 05:55:54.855 (0.018) & 00.336 (0.031) & 0.394 (0.037) & 1.096 \\
& C1R 2$^{nd}$                    & 05:56:13.745 (0.010) & 05:56:14.226 (0.011) & 00.481 (0.021) & 0.386 (0.028) & 1.398 \\

TRAPPIST-South & C1R 1$^{st}$ & 05:55:54.498 (1.366) & 05:55:55.994 (1.743) & 01.495 ($^{+3.110}_{-1.495}$) & 0.067 ($^{+0.890}_{-0.018}$) & 1.131 \\
& C1R 2$^{nd}$          & 05:56:12.782 (1.403) & 05:56:14.072 (1.498) & 01.291 ($^{+2.902}_{-1.291}$) & 0.506 ($^{+0.471}_{-0.448}$) & 1.103 \\

OPD   & C1R 1$^{st}$ & 05:54:29.963 (0.009) & 05:54:30.731 (0.007) & 00.769 (0.016) & 0.381 (0.016) & 2.155 \\
& C1R 2$^{nd}$       & 05:54:39.651 (0.006) & 05:54:40.534 (0.008) & 00.883 (0.014) & 0.386 (0.012) & 2.909 \\
& C2R 1$^{st}$       & 05:54:28.778 (0.019) & 05:54:28.872 (0.027) & 00.094 (0.045) & 0.214 (0.104) & 1.243 \\
& C2R 2$^{nd}$       & 05:54:41.699 (0.004) & 05:54:41.710 (0.004) & 00.011 (0.008) & 0.978 ($^{+0.020}_{-0.115}$) & 2.588 \\

SARA   & C1R$^{c}$ & 05:56:02.499 (0.205) & 05:56:06.988 (0.041) & 04.488 (0.246) & 0.379 (0.014) & 1.113 \\

PROMPT   & C1R$^{c}$ & 05:56:02.260 (0.043) & 05:56:07.228 (0.107) & 04.968 (0.150) & 0.373 (0.016) & 2.205 \\

\hline
\hline
\vspace{0.05em}\\
\multicolumn{7}{c}{\textbf{2017-06-22 -- Namibia}}\\
\hline
\hline

Outeniqua Lodge & MB   & 21:21:20.179 (0.032) & 21:21:30.193 (0.034) & 10.013 (0.066) & 1.000 & 0.897 / 0.905 \\
& C1R 1$^{st}$         & 21:21:11.469 (0.048) & 21:21:11.768 (0.053) & 00.299 (0.101) & 0.527 ($^{+0.472}_{-0.170}$) & 0.781 \\
& C1R 2$^{nd}$         & 21:21:39.450 (0.044) & 21:21:39.745 (0.037) & 00.295 (0.081) & 0.633 ($^{+0.367}_{-0.227})$ & 0.979 \\

Onduruquea G. F. & MB  & 21:21:22.023 (0.010) & 21:21:33.634 (0.011) & 11.610 (0.021) & 1.000     & 0.809 / 0.593 \\
& C1R 1$^{st}$         & 21:21:13.486 (0.012) & 21:21:13.702 (0.015) & 00.216 (0.027) & 0.453 (0.063) & 1.058 \\
& C1R 2$^{nd}$         & 21:21:42.047 (0.014) & 21:21:42.251 (0.012) & 00.204 (0.026) & 0.403 (0.059) & 0.946 \\

Windhoek - Backes      & MB & 21:21:17.234 (0.024) & 21:21:27.189 (0.026) & 09.955 (0.050) & 1.000     & 0.897 / 1.124 \\
& C1R 1$^{st}$         & 21:21:07.646 (0.067) & 21:21:07.898 (0.069) & 00.252 (0.136) & 0.352 ($^{+0.647}_{-0.080}$) & 0.884 \\
& C1R 2$^{nd}$         & 21:21:35.709 (0.052) & 21:21:36.082 (0.076) & 00.376 (0.128) & 0.225 ($^{+0.118}_{-0.067}$) & 1.552 \\

Windhoek - Meza & MB   & 21:21:17.288 (0.028) & 21:21:27.228 (0.034) & 09.940 (0.062) & 1.000     & 1.015 / 1.187 \\
& C1R 1$^{st}$         & 21:21:07.718 (0.076) & 21:21:07.976 (0.122) & 00.258 (0.198) & 0.247 ($^{+0.752}_{-0.083}$) & 0.988 \\
& C1R 2$^{nd}$         & 21:21:35.825 (0.039) & 21:21:35.968 (0.046) & 00.143 (0.085) & 0.970 ($^{+0.030}_{-0.528}$) & 1.430 \\

Tivoli Lodge & MB      & 21:21:15.478 (0.011) & 21:21:19.838 (0.038) & 04.361 (0.049) & 1.000     & 0.522 / 0.817 \\
& C1R 1$^{st}$         & 21:21:03.630 (0.031) & 21:21:03.834 (0.038) & 00.139 (0.069) & 0.412 ($^{+0.581}_{-0.122}$) & 0.821 \\
& C1R 2$^{nd}$         & 21:21:30.580 (0.036) & 21:21:30.724 (0.018) & 00.144 (0.054) & 0.517 ($^{+0.383}_{-0.140}$) & 0.920 \\

Hakos & C1R 1$^{st}$   & 21:21:10.218 (0.007) & 21:21:10.426 (0.007) & 00.208 (0.014) & 0.433 (0.049) & 1.136 \\
& C1R 2$^{nd}$         & 21:21:36.347 (0.011) & 21:21:36.570 (0.007) & 00.223 (0.018) & 0.390 (0.042) & 0.780 \\

\hline
\hline
\vspace{0.05em}\\
\multicolumn{7}{c}{\textbf{2017-04-09 -- Namibia}}\\
\hline
\hline
Wabi Lodge & MB      & 02:11:36.536 (0.023) & 02:12:16.672 (0.021) & 40.154 (0.044) & 1.000     & 0.640 / 0.593 \\
& C1R 1$^{st}$       & 02:10:45.861 (0.045) & 02:10:47.259 (0.087) & 01.398 (0.132) & 0.304 (0.044) & 1.085 \\
& C1R 2$^{nd}$       & 02:13:00.816 (0.054) & 02:13:01.868 (0.053) & 01.052 (0.107) & 0.410 (0.056) & 0.975 \\

Weaver R. L. & MB    & 02:11:20.357 (0.010) & 02:12:15.199 (0.013) & 54.850 (0.023) & 1.000    & 1.002 / 0.968 \\
& C1R 1$^{st}$       & 02:10:38.477 (0.022) & 02:10:39.737 (0.024) & 01.260 (0.046) & 0.361 (0.024) & 1.078 \\
& C1R 2$^{nd}$       & 02:12:56.960 (0.023) & 02:12:57.986 (0.007) & 01.026 (0.030) & 0.511 (0.029) & 0.909 \\

Hakos & C1R 2$^{nd}$ & 02:13:09.054 (0.026) & 02:13:10.345 (0.020) & 01.291 (0.047)  & 0.518 (0.032) & 1.246 \\

\hline
\hline
\multicolumn{7}{l}{\rule{0pt}{3.0ex} $^{c}$ The observations at Cerro Tololo (SARA and PROMPT) were grazing over C1R, meaning that only one detection were made.}\\

\label{tb:times_param_2}
\end{tabular}
\end{center}
\emph{Note}: MB stands for the Main Body detections, C1R and C2R for the rings detections. As each transit of the ring in front of the star is considered as a detection, a given light curve may detect twice that ring (SNR allowing) during a given event. For the main body occultations the apparent opacity was not fitted and fixed as 1 (opaque body).

\end{table*}

\begin{table*}
\begin{center}
\caption{Fitted ring parameters in the ring plane obtained for each light curve.}
\begin{tabular}{lccccccc}
\hline
\hline
Site   & Detec.  & r & $W_r$  & $p_N$ & $E_p$ & $\#$ & \multirow{2}{*}{$\frac{(1-\phi_0)}{RMS}$} \\
       &         & (km) & (km) &      & (km)  & & \\
\hline
\hline
\multicolumn{8}{c}{\textbf{C1R}}\\
\hline
\hline
\vspace{0.05em}\\
\multicolumn{8}{c}{\textbf{2020-06-19 -- Australia}}\\
\hline
\hline
Glenlee    & 1$^{st}$ &  392.402 (10.723) & 40.396 (8.286) & 0.177 ($^{+0.076}_{-0.048}$) & 7.192 ($^{+2.381}_{-1.602}$) & 03 & 02.83 \\
           & 2$^{nd}$ &  384.352 (11.825) & 6.465 ($^{+10.634}_{-02.946}$) & 0.590 ($^{+0.255}_{-0.460}$) & 3.814 ($^{+1.956}_{-2.148}$) & 01 & 03.22 \\

\hline
\hline
\vspace{0.05em}\\
\multicolumn{8}{c}{\textbf{2017-08-24 -- Brazil}}\\
\hline
\hline
OPD   & 1$^{st}$ &  391.700 (6.109) & 9.807 ($^{+1.724}_{-5.492}$) & 0.214 ($^{+0.534}_{-0.042}$) & 2.103 ($^{+1.769}_{-0.296}$) & 02 & 02.98 \\
      & 2$^{nd}$ &  384.245 (3.826) & 6.465 ($^{+10.634}_{-2.946}$) & 0.590 ($^{+0.255}_{-0.460}$) & 3.814 ($^{+1.956}_{-2.148}$) & 01 & 02.88 \\

\hline
\hline
\vspace{0.05em}\\
\multicolumn{8}{c}{\textbf{2017-07-23 -- South America }}\\
\hline
\hline
SPA - Fabrega    & 2$^{nd}$ &  394.185 ($^{+36.059}_{-25.198}$) & 27.151 ($^{+33.033}_{-27.151}$) & 0.234 ($^{+0.529}_{-0,127}$) & 18.137 ($^{+2.611}_{-16.683}$) & 01 & 04.01 \\

VLT              & 1$^{st}$ &  386.211 (0.267)  & 6.596 (0.262)  & 0.297 (0.009) & 1.959 (0.140) & 04 & 23.52 \\
                 & 2$^{nd}$ &  386.274 (0.355)  & 6.524 (0.360)  & 0.327 (0.020) & 2.133 (0.255) & 03 & 24.44 \\

SSO              & 1$^{st}$ &  388.432 (11.337) & 7.920 ($^{+14.125}_{-3.701}$)  & 0.394 ($^{+0.362}_{-0.267}$) & 3.118 ($^{+0.861}_{-1.287}$) & 02 & 07.65 \\
                 & 2$^{nd}$ &  388.507 (0.491)  & 4.760 ($^{+12.671}_{-0.122}$)  & 0.749 ($^{+0.012}_{-0.637}$) & 3.565 ($^{+0.090}_{-1.658}$) & 02 & 08.90 \\

Tolar Grande     & 1$^{st}$ &  391.109 (1.646)  & 8.670 (1.635)  & 0.207 (0.043) & 1.795 (0.782) & 03 & 04.74 \\
                 & 2$^{nd}$ &  390.969 (1.222)  & 6.068 (1.229)  & 0.339 (0.069) & 2.057 (0.920) & 03 & 07.33 \\

El Rodeo         & 1$^{st}$ &  384.165 (9.431)  & 7.920 ($^{+14.125}_{-3.701}$)  & 0.394 ($^{+0.362}_{-0.267}$) & 3.118 ($^{+0.861}_{-1.287}$) & 03 & 02.61 \\
                 & 2$^{nd}$ &  384.024 (5.652)  & 15.103 ($^{+4.076}_{-3.993}$)  & 0.199 ($^{+0.108}_{-0.100}$) & 3.004 ($^{+1.295}_{-1.346}$) & 04 & 01.75 \\

El Salvador      & 1$^{st}$ &  390.513 (1.331)  & 7.811 ($^{+1.367}_{-0.824}$)  & 0.388 ($^{+0.187}_{-0.088}$) & 3.033 ($^{+1.428}_{-0.570}$) & 07 & 02.62 \\
                 & 2$^{nd}$ &  390.095 (5.325)  & 5.390 ($^{+7.818}_{-1.695}$)  & 0.243 ($^{+0.159}_{-0.137}$) & 1.307 ($^{+0.854}_{-0.530}$) & 06 & 01.72 \\

Inca de Oro      & 1$^{st}$ &  390.145 (4.428)  & 6.614 ($^{+2.553}_{-4.898}$)  & 0.228 ($^{+0.512}_{-0.094}$) & 1.510 ($^{+0.621}_{-0.714}$) & 03 & 02.35 \\
                 & 2$^{nd}$ &  389.436 (1.440)  & 3.915 ($^{+1.020}_{-1.561}$)  & 0.394 ($^{+0.366}_{-0.126}$) & 1.542 ($^{+1.077}_{-0.423}$) & 03 & 03.29 \\

Danish Visual    & 1$^{st}$ &  386.747 (0.185)  & 6.064 (0.179)  & 0.323 (0.019) & 1.959 (0.176) & 11 & 08.22 \\
                 & 2$^{nd}$ &  385.576 (0.369)  & 8.191 (0.378)  & 0.349 (0.022) & 2.859 (0.320) & 14 & 08.60 \\

Danish Red       & 1$^{st}$ &  386.840 (0.167)  & 5.779 (0.162)  & 0.288 (0.014) & 1.664 (0.130) & 12 & 09.75 \\
                 & 2$^{nd}$ &  385.686 (0.240)  & 8.326 (0.246)  & 0.278 (0.014) & 2.315 (0.188) & 15 & 09.62 \\
1m -- La Silla   & 1$^{st}$ &  387.081 (0.573)  & 6.035 (0.557)  & 0.300 (0.028) & 1.811 (0.352) & 03 & 08.59 \\
                 & 2$^{nd}$ &  385.908 (0.388)  & 9.107 (0.398)  & 0.294 (0.021) & 2.677 (0.317) & 05 & 08.42 \\

TRAPPIST-South & 1$^{st}$ &  376.900 ($^{+60.358}_{-49.445}$) & 25.448 ($^{+57.334}_{-22.207}$) & 0.475 ($^{+ 0.287}_{ -0.440}$) & 12.099 ($^{+ 7.152}_{-10.784}$) & 01 & 02.95 \\
         & 2$^{nd}$ &  375.774 ($^{+55.900}_{-46.330}$) & 19.586 ($^{+57.596}_{-16.672}$) & 0.608 ($^{+ 0.154}_{ -0.575}$) & 11.918 ($^{+ 5.874}_{-10.760}$) & 01 & 02.85 \\

OPD   & 1$^{st}$ &  385.321 (0.149) & 6.456 (0.134) & 0.291 (0.012) &1.879 (0.118) & 07 & 21.10 \\
      & 2$^{nd}$ &  385.441 (0.131) & 8.920 (0.141) & 0.294 (0.009) &2.622 (0.123) & 08 & 21.41 \\

SARA   & $^a$ &  388.928 ($^{+0.931}_{-0.852}$) & >4.443 (0.244) & 0.289 (0.011) & >1.284 (0.122) & 03 & 06.58 \\
PROMPT & $^a$ &  389.823 ($^{+0.682}_{-0.405}$) & >4.819 (0.146) & 0.284 (0.012) & >1.369 (0.101) & 09 & 16.38 \\

\hline
\hline
\vspace{0.05em}\\
\multicolumn{8}{c}{\textbf{2017-06-22 -- Namibia}}\\
\hline
\hline

Outeniqua     & 1$^{st}$ &  388.073 (2.716) & 8.266 ($^{+1.972}_{-2.750}$) & 0.410 ($^{+0.367}_{-0.132}$) & 3.391 ($^{+2.654}_{-1.081}$) & 04 & 02.45 \\
              & 2$^{nd}$ &  390.936 (2.179) & 8.104 ($^{+1.706}_{-1.880}$) & 0.492 ($^{+0.285}_{-0.177}$) & 3.989 ($^{+2.919}_{-1.370}$) & 04 & 02.65 \\
Onduruquea    & 1$^{st}$ &  391.014 (0.739) & 5.891 ($^{+0.441}_{-0.636}$) & 0.360 ($^{+0.042}_{-0.057}$) & 2.121 ($^{+0.179}_{-0.351}$) & 03 & 06.09 \\
              & 2$^{nd}$ &  390.503 (0.711) & 5.550 ($^{+0.532}_{-0.486}$) & 0.314 ($^{+0.045}_{-0.047}$) & 1.744 ($^{+0.239}_{-0.260}$) & 03 & 05.28 \\
Wdh - Backes  & 1$^{st}$ &  393.714 (3.655) & 7.459 ($^{+2.267}_{-3.973}$) & 0.274 ($^{+0.504}_{-0.062}$) & 2.045 ($^{+1.514}_{-0.381}$) & 02 & 05.51 \\
              & 2$^{nd}$ &  389.827 (3.439) & 9.987 ($^{+2.967}_{-2.690}$) & 0.175 ($^{+0.092}_{-0.052}$) & 1.746 ($^{+0.560}_{-0.416}$) & 03 & 03.90 \\
Wdh - Meza    & 1$^{st}$ &  391.699 (5.323) & 8.189 ($^{+3.029}_{-6.027}$) & 0.193 ($^{+0.585}_{-0.064}$) & 1.577 ($^{+1.340}_{-0.590}$) & 02 & 03.01 \\
              & 2$^{nd}$ &  389.881 (2.284) & 3.731 ($^{+2.011}_{-0.731}$) & 0.755 ($^{+0.023}_{-0.411}$) & 2.816 ($^{+0.810}_{-1.172}$) & 01 & 04.82 \\
Tivoli        & 1$^{st}$ &  391.348 (1.781) & 3.872 ($^{+1.269}_{-2.034}$) & 0.320 ($^{+0.452}_{-0.095}$) & 1.240 ($^{+0.388}_{-0.334}$) & 02 & 03.43 \\
              & 2$^{nd}$ &  390.145 (1.394) & 3.909 ($^{+1.008}_{-1.561}$) & 0.402 ($^{+0.298}_{-0.109}$) & 1.571 ($^{+0.489}_{-0.382}$) & 03 & 04.04 \\
Hakos         & 1$^{st}$ &  387.490 (0.350) & 5.095 ($^{+0.273}_{-0.197}$) & 0.342 ($^{+0.033}_{-0.043}$) & 1.740 ($^{+0.202}_{-0.176}$) & 04 & 05.57 \\
              & 2$^{nd}$ &  393.668 (0.452) & 5.558 ($^{+0.365}_{-0.426}$) & 0.313 ($^{+0.031}_{-0.043}$) & 1.739 ($^{+0.125}_{-0.250}$) & 04 & 05.11 \\

\hline
\hline
\multicolumn{8}{l}{\rule{0pt}{1.0ex} $^{a}$ As we have a grazing detection at SARA and PROMPT, we give a lower limit to the radial width.}\\

\label{tb:ring_parameters_1}
\end{tabular}
\end{center}
\end{table*}

\setcounter{table}{1}

\begin{table*}
\begin{center}
\caption{\textbf{[Cont.]} Fitted ring parameters in the ring plane obtained for each light curve.}
\begin{tabular}{lccccccc}
\hline
\hline
Site   & Detec.  & r & $W_r$  & $p_N$ & $E_p$ & $\#$ & \multirow{2}{*}{$\frac{(1-\phi_0)}{RMS}$} \\
       &         & (km) & (km) &      & (km)  & & \\
\hline
\hline
\multicolumn{8}{c}{\textbf{C1R}}\\
\hline
\hline
\vspace{0.05em}\\
\multicolumn{8}{c}{\textbf{2017-04-09 -- Namibia}}\\
\hline
\hline
Wabi Lodge   & 1$^{st}$ & 386.859 (0.635) & 7.056 (0.258) & 0.241 (0.035) & 1.700 (0.318) & 16 & 02.43 \\
             & 2$^{nd}$ & 389.603 (0.565) & 5.725 (0.167) & 0.326 (0.044) & 1.866 (0.314) & 12 & 03.07 \\
Weaver R. L. & 1$^{st}$ & 388.185 (0.240) & 7.591 (0.717) & 0.287 (0.019) & 2.179 (0.364) & 18 & 04.02 \\
             & 2$^{nd}$ & 387.684 (0.218) & 5.712 (0.581) & 0.406 (0.023) & 2.319 (0.381) & 15 & 05.18 \\
Hakos        & 2$^{nd}$ & 388.314 (0.200) & 5.642 (0.205) & 0.411 (0.025) & 2.319 (0.230) & 16 & 08.63 \\

\hline
\hline
\multicolumn{8}{c}{\textbf{C2R}}\\
\hline
\hline
\vspace{0.05em}\\
\multicolumn{8}{c}{\textbf{2017-07-23 -- South America }}\\
\hline
\hline
VLT           & 1$^{st}$ & 401.125 (0.630) & 3.716 ($^{+0.631}_{-0.527}$) & 0.045 ($^{+0.005}_{-0.006}$) & 0.168 ($^{+0.016}_{-0.020}$) & 02 & 04.26 \\
              & 2$^{nd}$ & 400.790 (0.248) & 0.159 ($^{+3.110}_{-0.097}$) & 0.664 ($^{+0.099}_{-0.631}$) & 0.105 ($^{+0.096}_{-0.061}$) & 01 & 04.05 \\

Danish Visual & 1$^{st}$ & 400.417 (0.425) & 0.277 ($^{+0.640}_{-0.129}$) & 0.445 ($^{+0.317}_{-0.294}$) & 0.123 ($^{+0.036}_{-0.026}$) & 04 & 02.65 \\
              & 2$^{nd}$ & 400.944 (0.213) & 0.258 ($^{+0.161}_{-0.072}$) & 0.598 ($^{+0.158}_{-0.264}$) & 0.154 ($^{+0.039}_{-0.034}$) & 04 & 04.08 \\

Danish Red    & 1$^{st}$ & 400.805 (0.543) & 0.191 ($^{+1.233}_{-0.073}$) & 0.500 ($^{+0.212}_{-0.420}$) & 0.095 ($^{+0.021}_{-0.016}$) & 04 & 02.96 \\
              & 2$^{nd}$ & 400.286 (0.504) & 0.801 ($^{+0.440}_{-0.559}$) & 0.158 ($^{+0.282}_{-0.056}$) & 0.127 ($^{+0.029}_{-0.035}$) & 04 & 02.70 \\

OPD           & 1$^{st}$ & 399.263 (0.530) & 0.631 ($^{+0.545}_{-0.185}$) & 0.152 ($^{+0.067}_{-0.073}$) & 0.096 ($^{+0.015}_{-0.012}$) & 01 & 05.75 \\
              & 2$^{nd}$ & 400.301 (0.095) & 0.095 ($^{+0.015}_{-0.010}$) & 0.759 ($^{+0.002}_{-0.091}$) & 0.072 ($^{+0.010}_{-0.009}$) & 02 & 03.78 \\

\hline
\hline
\label{tb:ring_parameters_2}
\end{tabular}
\end{center}
\emph{Note}: The radial distance ($r$) to the ring centre (assumed to be circular), the radial width ($W_r$), normal opacity ($p_N$) and equivalent width ($E_p$). There is also the number of data points within the detection. Lastly there is the ratio of bottom flux over the RMS of each light curve, that can be used to quantify the SNR of the detection.

\end{table*}

\section{Chariklo's updated positions} \label{App:update_positions}

Stellar occultations provide precise astrometric position of the occulting body relative to the occulted star. If a new and more accurate position of the star is available, we can update the objects position \citep{Rommel_2020}. Table~\ref{tb:positions} provides the solutions derived from seven stellar occultations observed between 2017 and 2020, using the Gaia Early Data Release 3 (GEDR3). Here, we also update results obtained from 12 occultations between 2013 and 2016 \citep{Braga-Ribas_2014,Berard_2017}, using again the GEDR3. 

\cite{Desmars_2017DPS} updated the positions obtained from pre-2017 events using the GDR2 catalogue. Let us consider Chariklo's previous position ($\alpha_c,\delta_c$), the GDR2 star position ($\alpha_{GDR2},\delta_{GDR2}$) and  the GEDR3 ($\alpha_{GEDR3}$, $\delta_{GEDR3}$), both at epoch. The updated Chariklo position ($\alpha'_c,\delta'_c$) is, then, obtained from
\begin{eqnarray}
\alpha'_c &=& \alpha_{c} + \alpha_{GEDR3} - \alpha_{GDR2} \label{eq:updtae_alfa}\\
\delta'_c &=& \delta_{c} + \delta_{GEDR3} - \delta_{GDR2} \label{eq:updtae_delta}
\end{eqnarray}
The updated Chariklo positions between 2013 and 2017 are given in Table \ref{tb:positions_update}.

\begin{table*}
\begin{center}
\caption{Update of astrometrical Chariklo's positions for the events observed between 2013 and 2016.}
\begin{tabular}{ccc}
\hline
\hline
Date and time UTC & Right Ascension & Declination \\
\hline
\hline
2013-06-03 06:25:30.000 & 16$^h$ 56$^m$ 06$^s$.5117522 & -40$^\circ$ 31' 30''.107042  \\ 
2014-02-16 07:45:35.000 & 17$^h$ 35$^m$ 55$^s$.2915296 & -38$^\circ$ 05' 17''.300445  \\
2014-03-16 20:31:45.000 & 17$^h$ 40$^m$ 39$^s$.8348967 & -38$^\circ$ 25' 46''.483279  \\
2014-04-29 23:14:12.000 & 17$^h$ 39$^m$ 02$^s$.1051292 & -38$^\circ$ 52' 48''.749358  \\
2014-06-28 22:24:35.000 & 17$^h$ 24$^m$ 50$^s$.3869110 & -38$^\circ$ 41' 05''.485778  \\
2015-04-26 02:11:58.000 & 18$^h$ 10$^m$ 46$^s$.1055222 & -36$^\circ$ 38' 56''.597743  \\
2015-05-12 17:55:40.000 & 18$^h$ 08$^m$ 29$^s$.3099452 & -36$^\circ$ 44' 56''.801226  \\
2016-07-26 00:00:00.000 & 18$^h$ 20$^m$ 35$^s$.3610788 & -34$^\circ$ 02' 29''.041382  \\
2016-08-10 14:23:00.000 & 18$^h$ 17$^m$ 47$^s$.3503443 & -33$^\circ$ 51' 02''.478041  \\
2016-08-10 16:43:00.000 & 18$^h$ 17$^m$ 46$^s$.4742139 & -33$^\circ$ 50' 57''.707107  \\
2016-08-15 11:38:00.000 & 18$^h$ 17$^m$ 06$^s$.2478821 & -33$^\circ$ 46' 56''.499075  \\
2016-10-01 10:10:00.000 & 18$^h$ 16$^m$ 20$^s$.0944710 & -33$^\circ$ 01' 10''.842803  \\
\hline
\hline
\label{tb:positions_update}
\end{tabular}
\end{center}

\end{table*}

\section{The case of the 2019-09-04 event in Hawaii} \label{App:Paul_Maley}

\begin{figure}[h]
\centering
\includegraphics[width=0.50\textwidth]{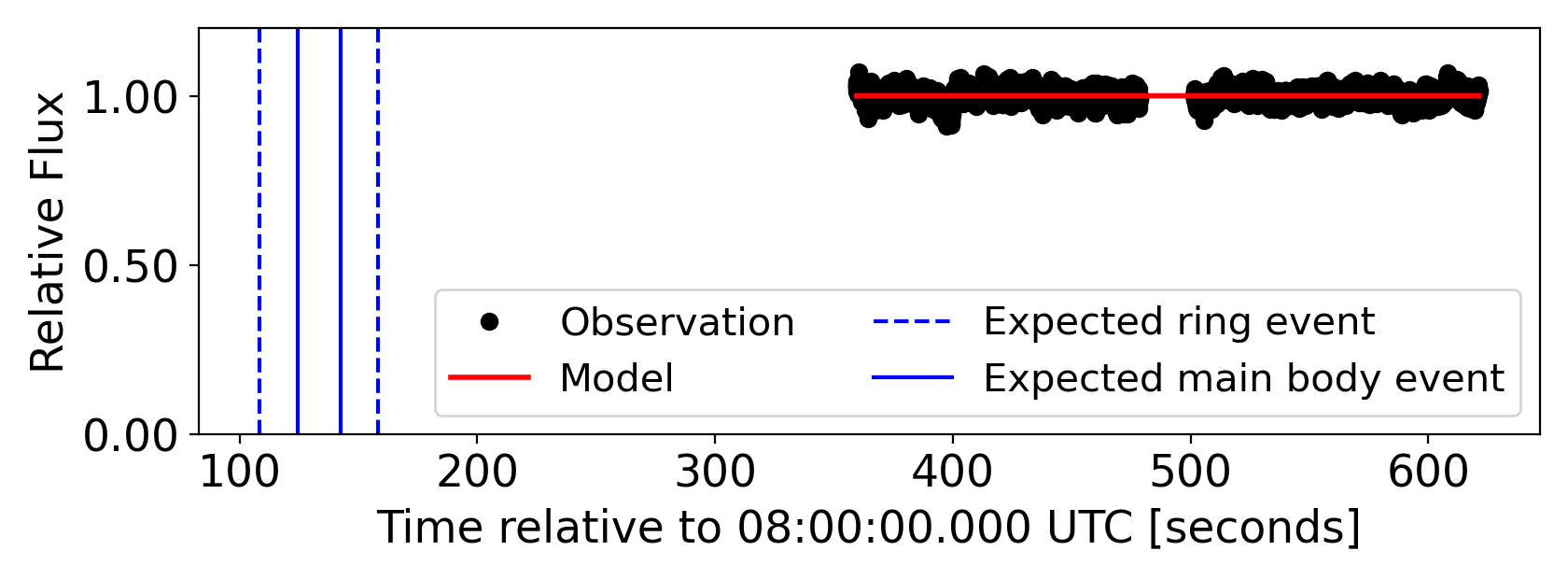}
\caption{Light curve obtained at Gemini North on 2017-07-23. The solid blue vertical lines stands for the expected times for the main body occultation and the dashed lines for the expected times for the ring. The black points are the data points, which were taken in two blocks separated by a few seconds. See text for details.}
\label{Fig:Gemini}
\end{figure}

On 2019-09-04, Chariklo's system occulted a mag G 13.586 star as seen from Hawaii. A run was triggered to observe the event from the 8.1-m Gemini North telescope in Mauna Kea. Unfortunately, the weather only allowed the observations to start a few minutes after the occultation as can be seen in Figure \ref{Fig:Gemini}. 

\begin{figure}[h]
\centering
\includegraphics[width=0.50\textwidth]{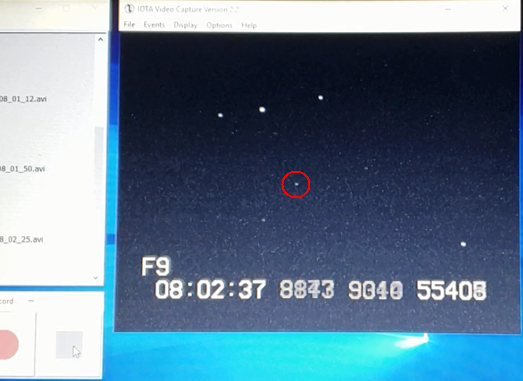}
\caption{%
One frame from the smartphone video of the notebook screen, during the 2019-09-04 event. The occulted star is highlighted by the red circle.}%
\label{Fig:PM_Screen_cell}
\end{figure}

A simultaneous observation was set up at Mauna Loa, using a 0.36-m portable telescope. Due to a technical failure, the images were not recorded, but were being displayed on screen. The observers could use a smartphone to film the notebook screen, with the GPS time inserted in each frame, see Figure \ref{Fig:PM_Screen_cell}. 

\begin{figure}[h]
\centering
\includegraphics[width=0.50\textwidth]{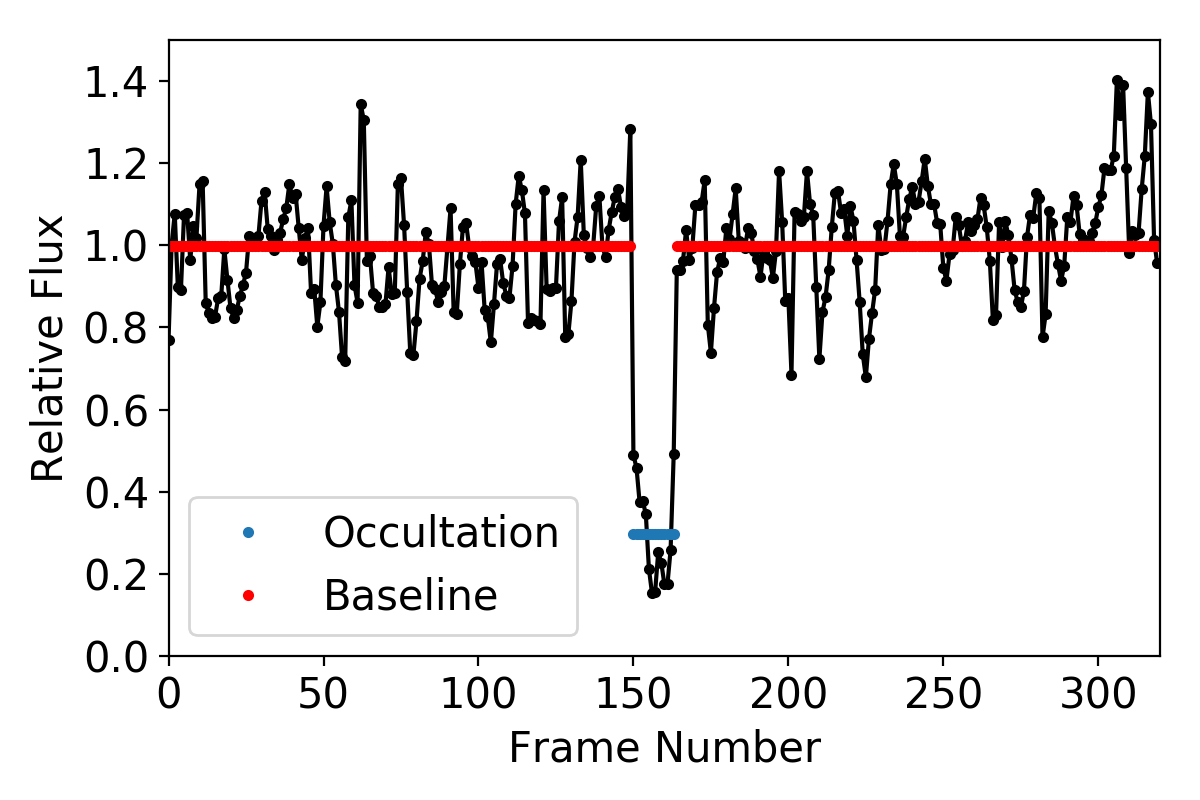}
\caption{Normalised flux vs. frame number obtained at Mauna Loa during the 2019-09-04 event, showing the occultation by C1R. The red line stands for the baseline flux (1.0).}%
\label{Fig:PM_lc}
\end{figure}

As the smartphone video recording started after the occultation by the main body, only a transit of C1R in front of the star was eventually recorded. This atypical data-set was analysed using the aperture photometry package of \texttt{PyMovie} \citep{Pymovie}, resulting in the light curve shown in Figure \ref{Fig:PM_lc}. As we could not retrieve the exact mid-time for each frame, we determined the immersion time using the time stamped in the images right before and after the star disappearance. We then consider that the mean time between the images with and without the star provides 
the immersion time, with an uncertainty of half the time between these two frames. The same protocol was followed to determine the emersion time. 

The mid-time of the C1R occultation is then evaluated at UTC 08:02:38.993$\pm$0.017~s. The \textsc{nima v.18} Chariklo's ephemeris having an uncertainty of 20 km for this event, this mid-time corresponds to a $(f,~g)$ position in the sky plane of $(250.945~\pm~20.0,~-36.121~\pm~20.0)$~km. The radial width of CR1 in the ring plane is then $6.77~\pm~0.53$~km, with normal opacity of $0.370~\pm~0.094$. These results are consistent with the obtained parameters from other occultations. If more observations of this event were to be obtained by other teams, the data presented here could be used to better constrain the geometry of this occultation. 

\section{Light curves} \label{App:Lightcurves}

This section provides the plots of the normalised light curves vs. UTC analysed in this work. There are forty light curves obtained during the seven occultations observed between 2017 and 2020 (excluding the 2019-09-04 event). The events are listed in inverse chronological order, and the light curves are plotted from the northernmost to the southernmost stations. The date, event and observer are indicated in the title and label of each figure. The upper panel contains the complete normalised light curve (black dots) and the fitted model (red line). The bottom panels contain zoom-in views of a few seconds each centred on the detections of C1R and the immersions and emersions behind the main body, when detected.

\begin{figure}[h]
\centering
\includegraphics[width=0.50\textwidth]{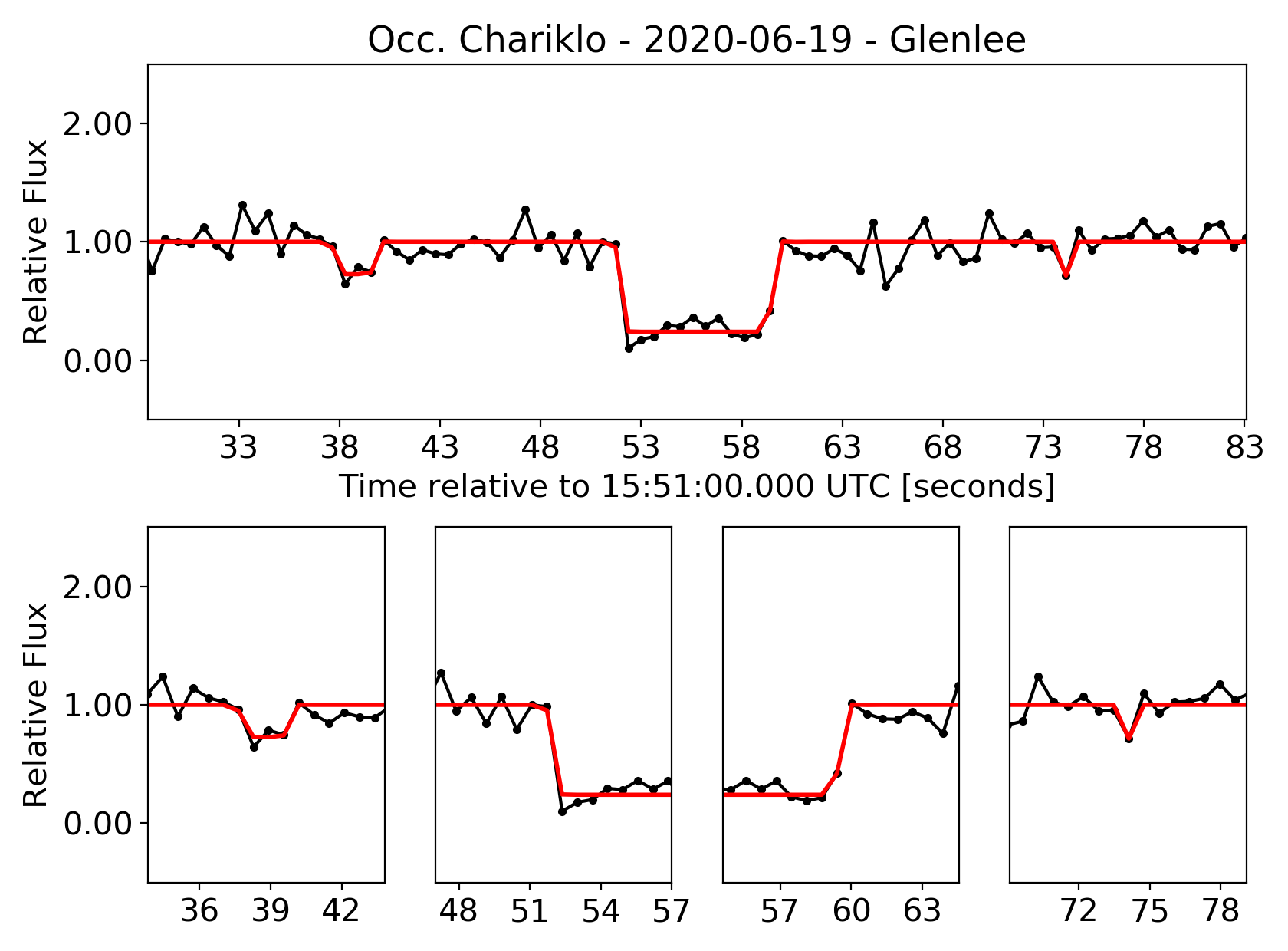}
\caption{Light curve obtained in Glenlee on 2020-06-19. The event, date and observer are indicated in the title and label. The upper panel contains the complete normalised light curve (black dots) and the fitted model (red line). The bottom panels contain zoom-in views of a few seconds each centred on the detections of C1R and the immersions and emersions behind the main body.}
\end{figure}               

\begin{figure}[h]
\centering
\includegraphics[width=0.50\textwidth]{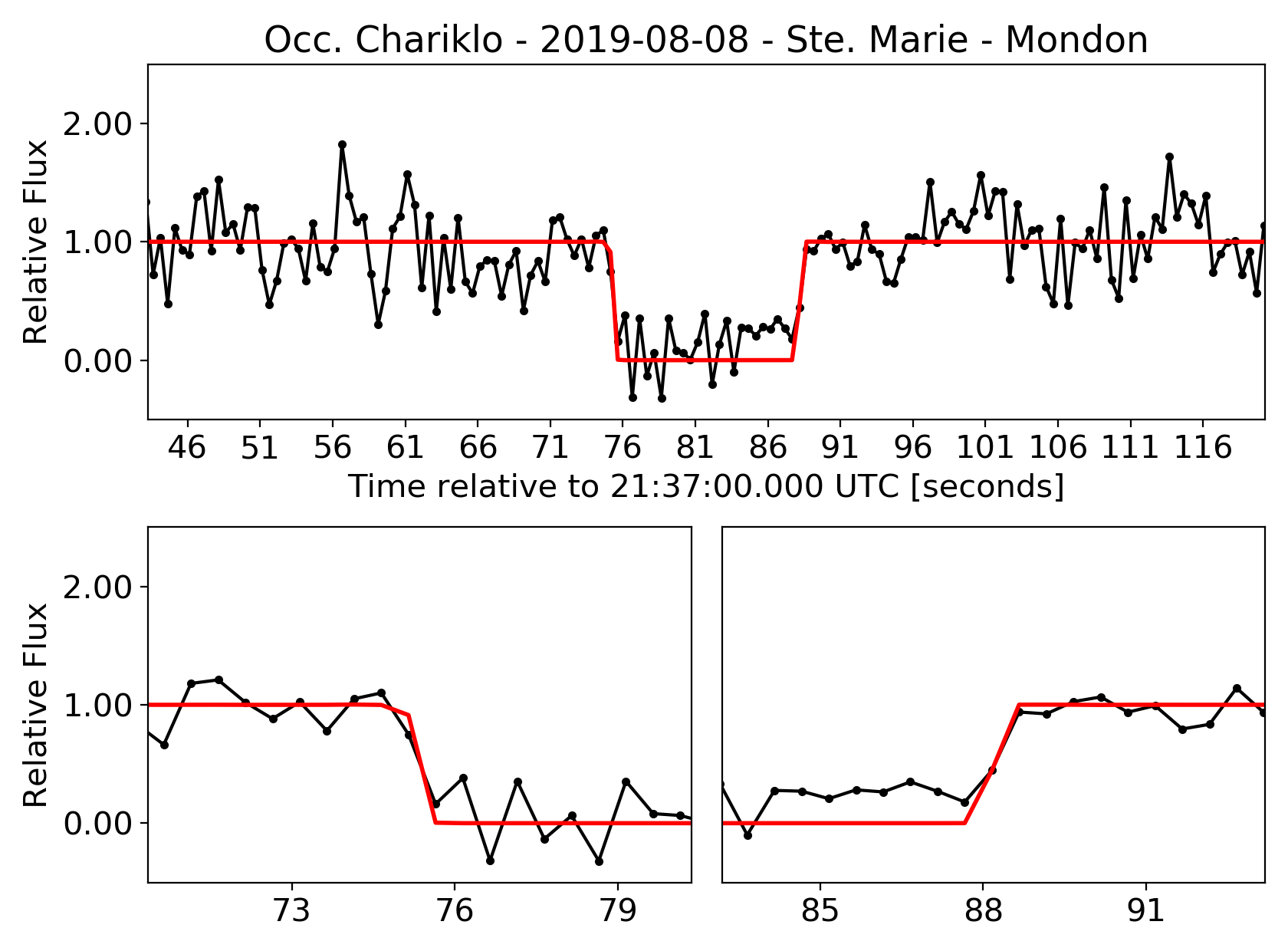}
\caption{Light curve obtained at Ste. Marie by Mondon on 2019-08-08.}
\end{figure}               

\begin{figure}[h]
\centering
\includegraphics[width=0.50\textwidth]{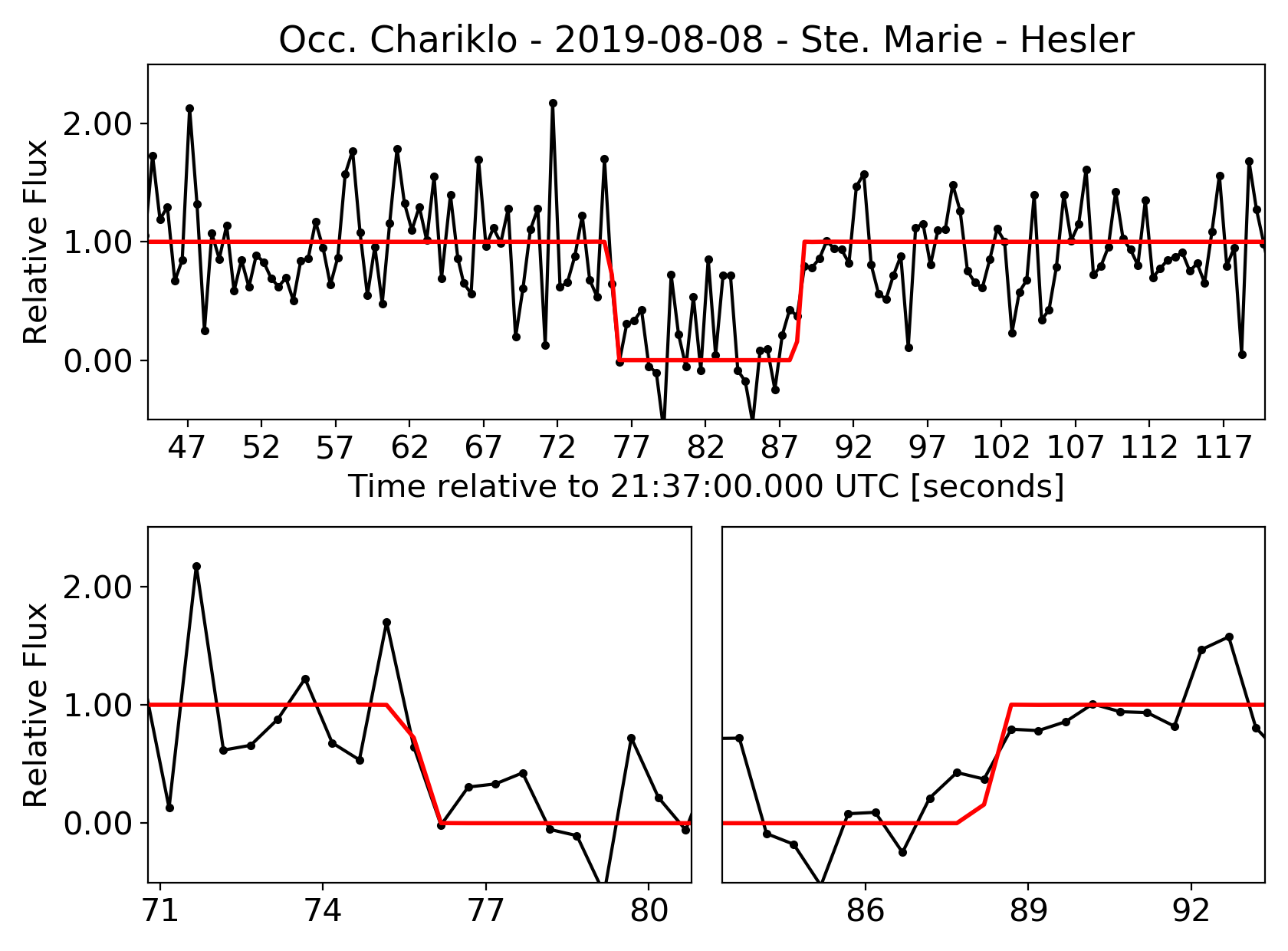}
\caption{Light curve obtained at Ste. Marie by Hesler on 2019-08-08.}
\label{Fig:lc_all_1}
\end{figure}

\begin{figure}[h]
\centering
\includegraphics[width=0.50\textwidth]{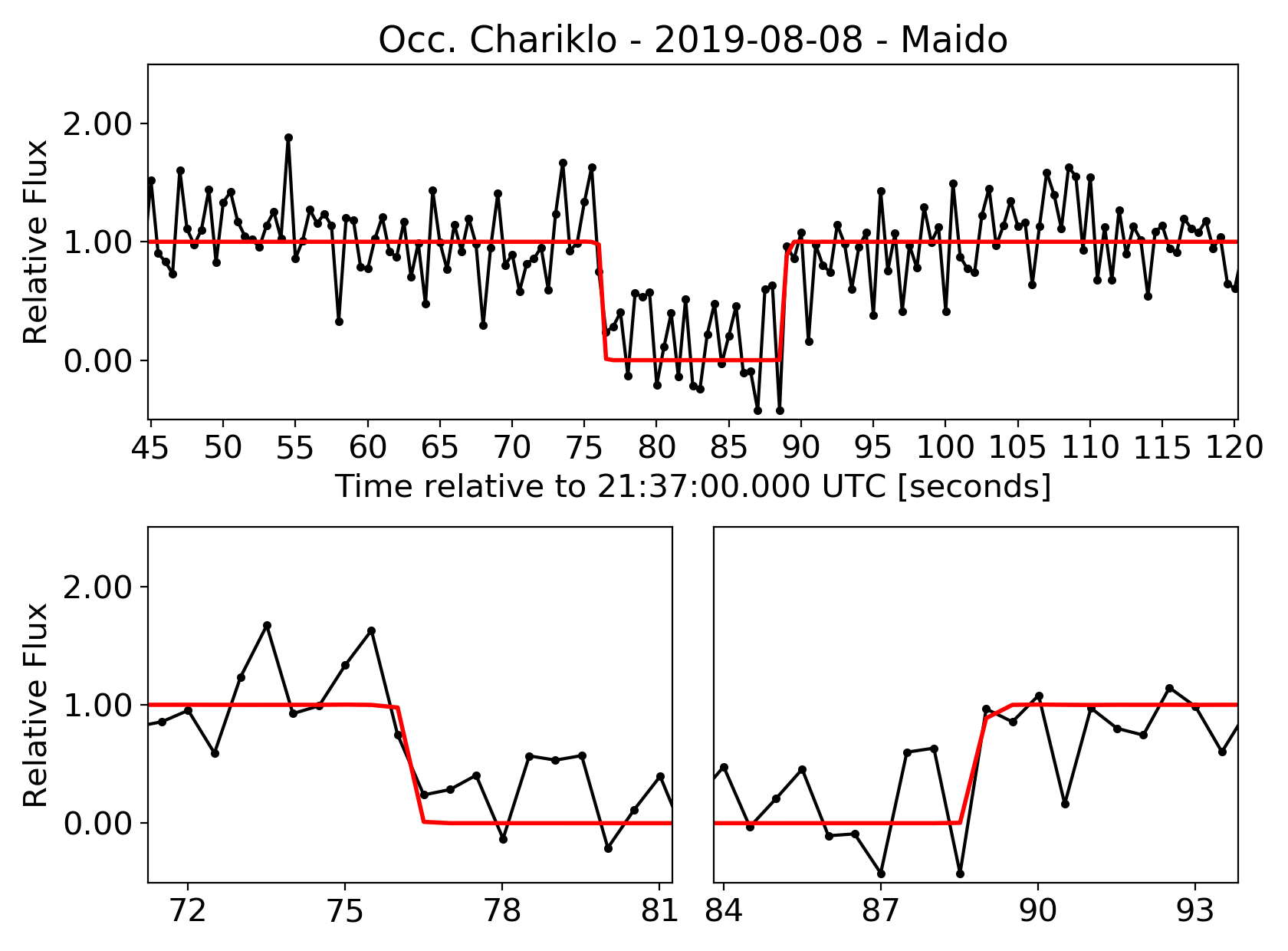}
\caption{Light curve obtained at Maido on 2019-08-08.}
\label{Fig:lc_all_1}
\end{figure}               

\begin{figure}[h]
\centering
\includegraphics[width=0.50\textwidth]{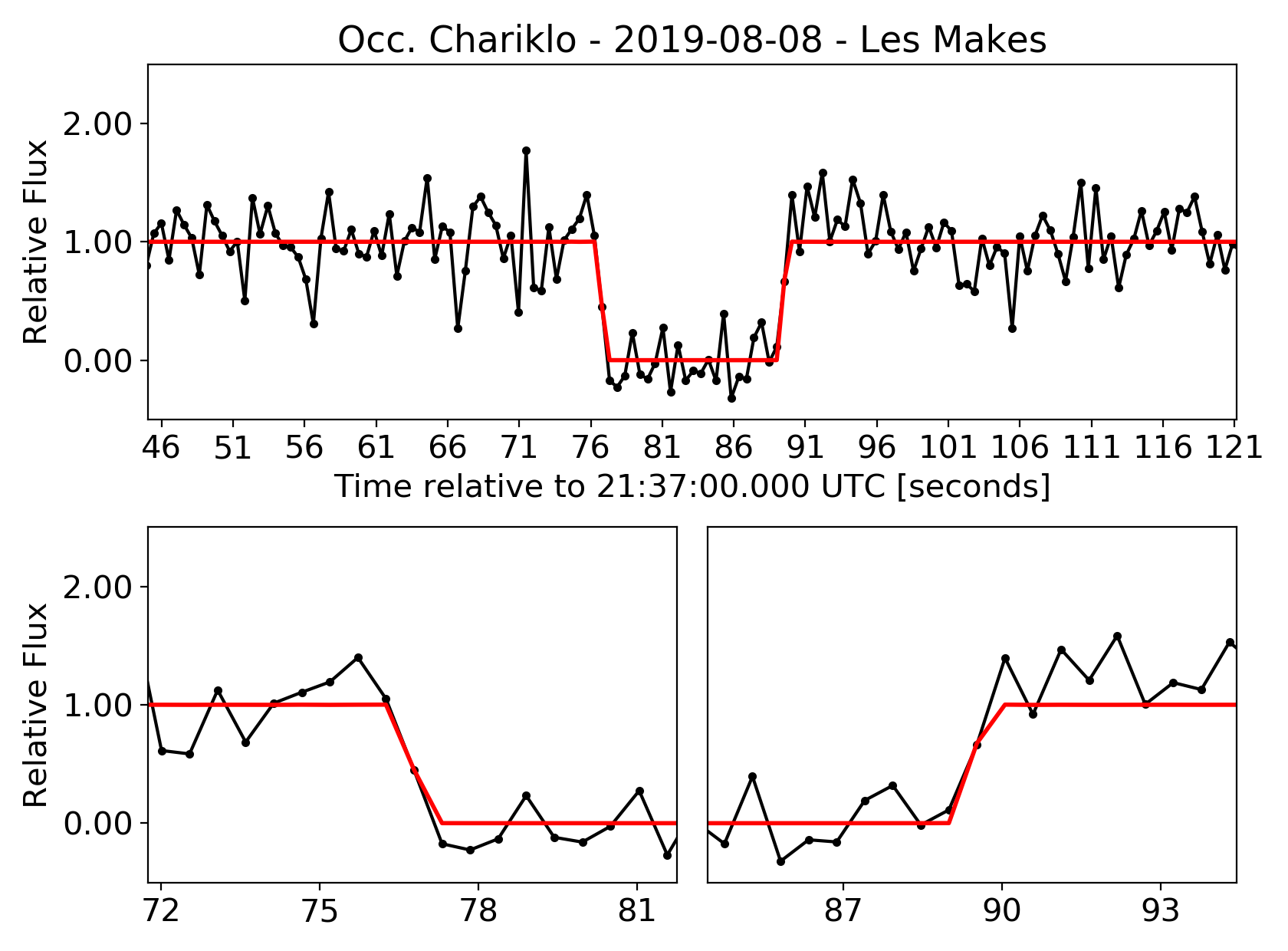}
\caption{Light curve obtained at Les Makes on 2019-08-08.}
\label{Fig:lc_all_1}
\end{figure}               

\begin{figure}[h]
\centering
\includegraphics[width=0.50\textwidth]{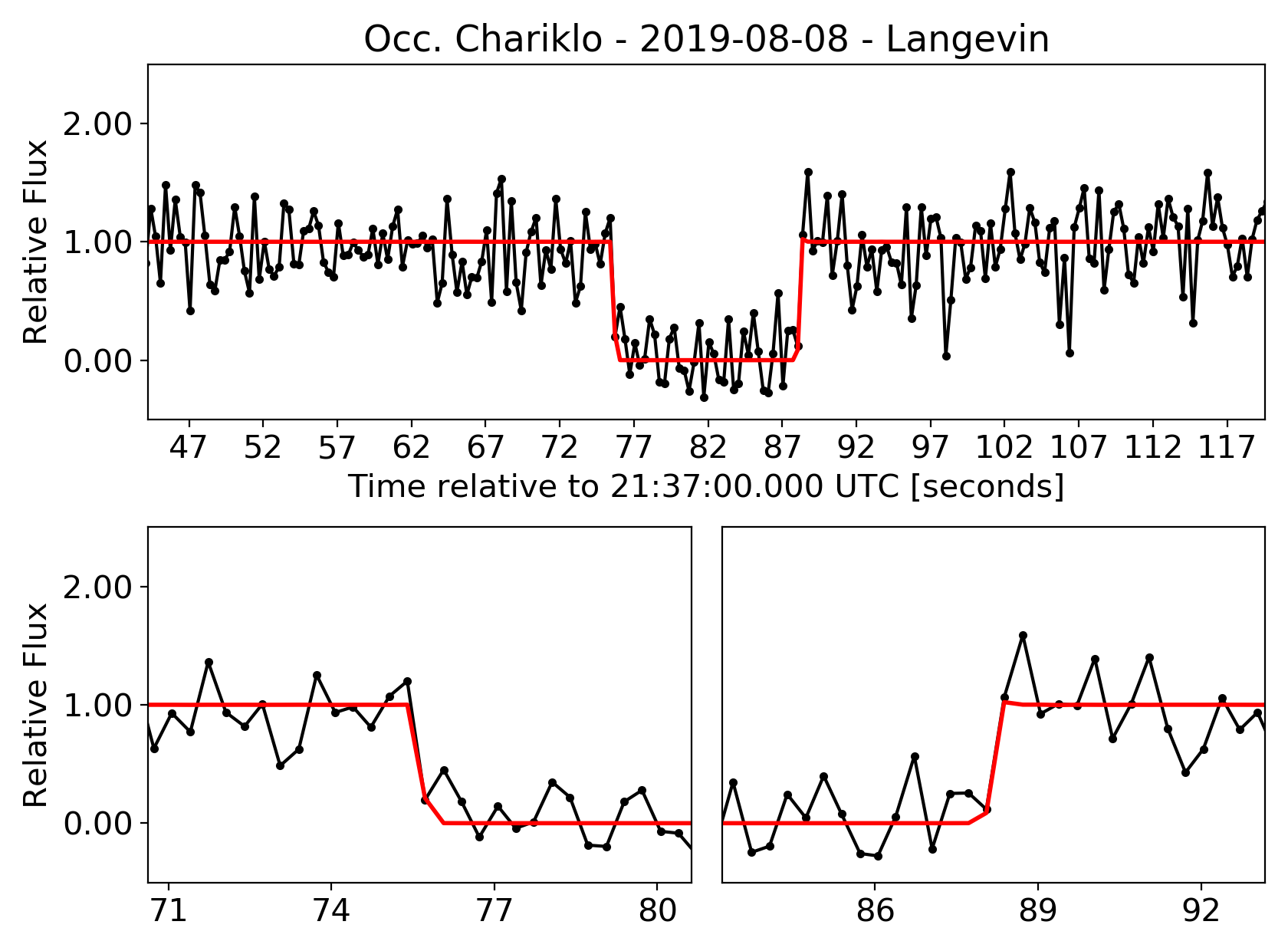}
\caption{Light curve obtained at Langevin on 2019-08-08.}
\label{Fig:lc_all_1}
\end{figure}               

\begin{figure}[h]
\centering
\includegraphics[width=0.50\textwidth]{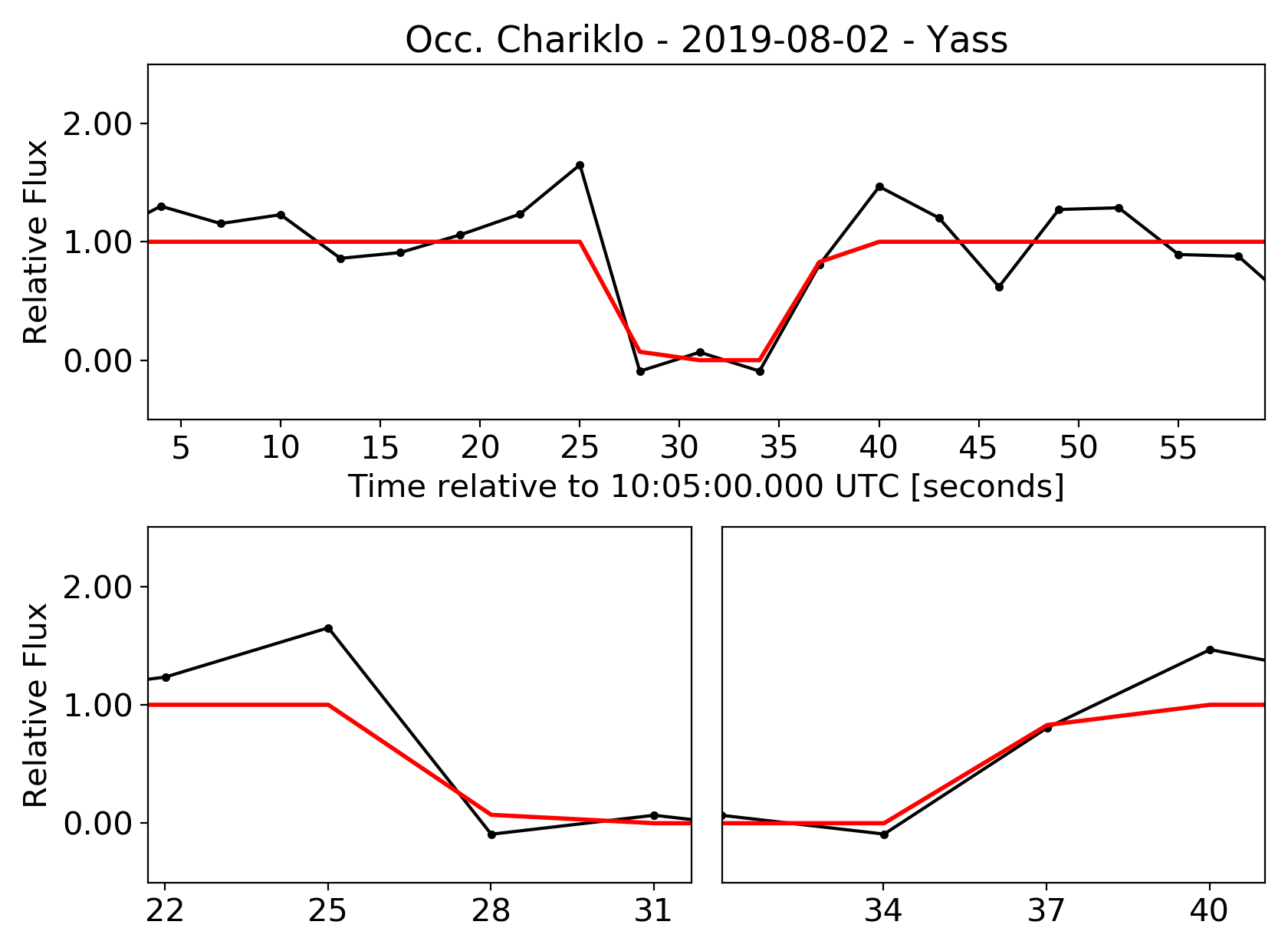}
\caption{Light curve obtained at Yass on 2019-08-02.}
\label{Fig:lc_all_1}
\end{figure}               

\begin{figure}[h]
\centering
\includegraphics[width=0.50\textwidth]{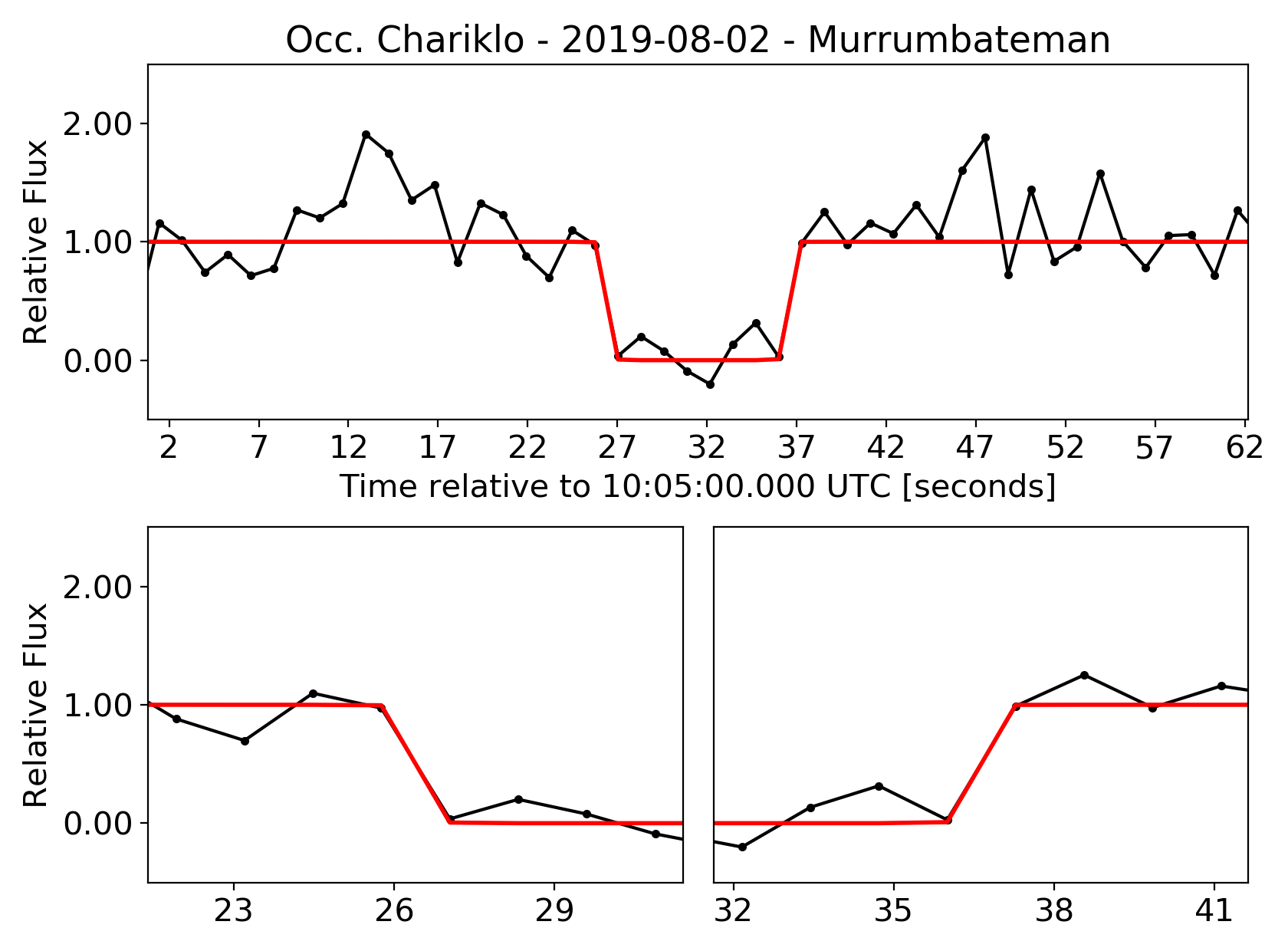}
\caption{Light curve obtained at Murrumbateman on 2019-08-02.}
\label{Fig:lc_all_1}
\end{figure}               


\begin{figure}[h]
\centering
\includegraphics[width=0.50\textwidth]{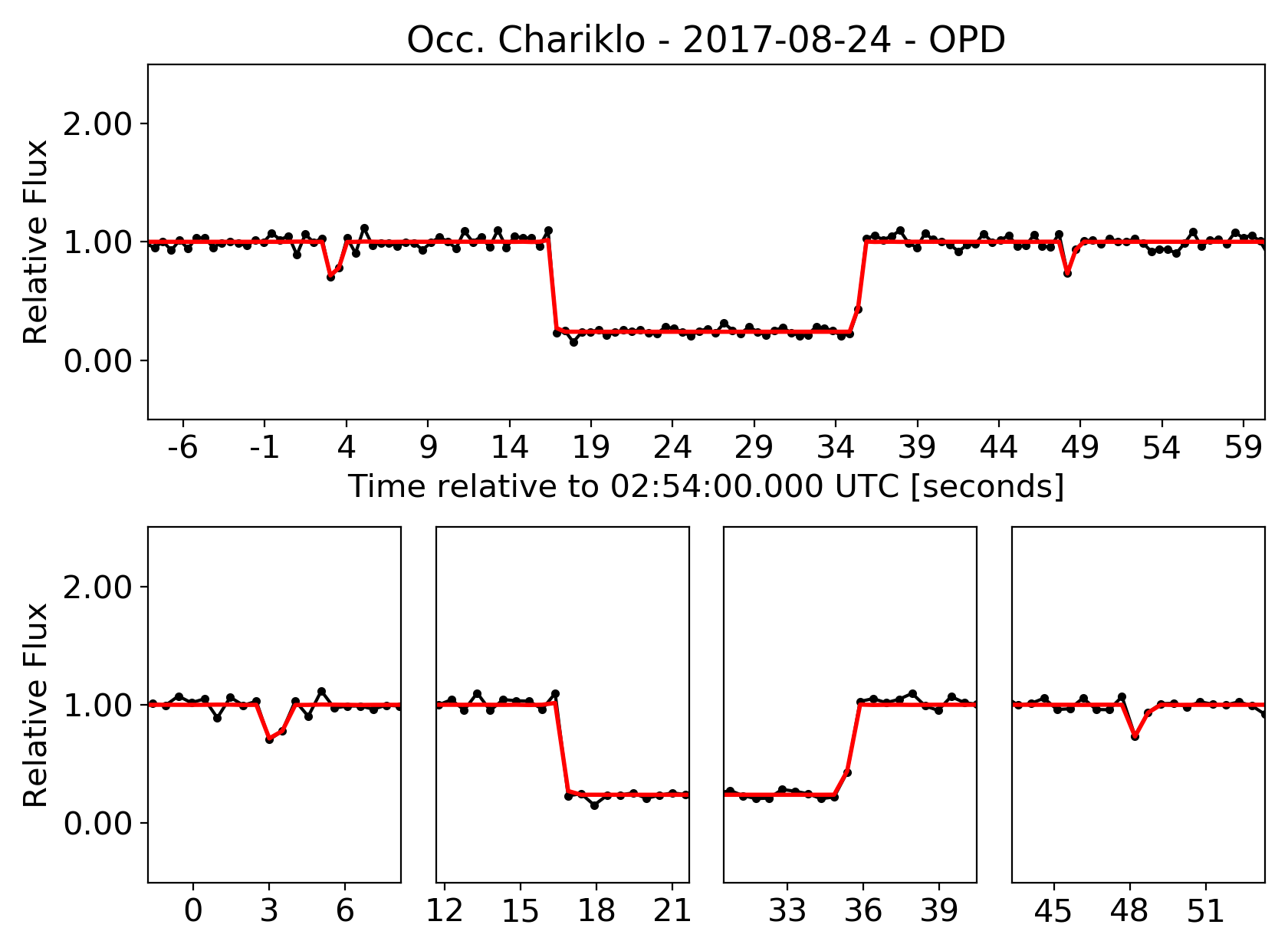}
\caption{Light curve obtained at OPD on 2017-08-24.}
\label{Fig:lc_all_1}
\end{figure}               


\begin{figure}[h]
\centering
\includegraphics[width=0.50\textwidth]{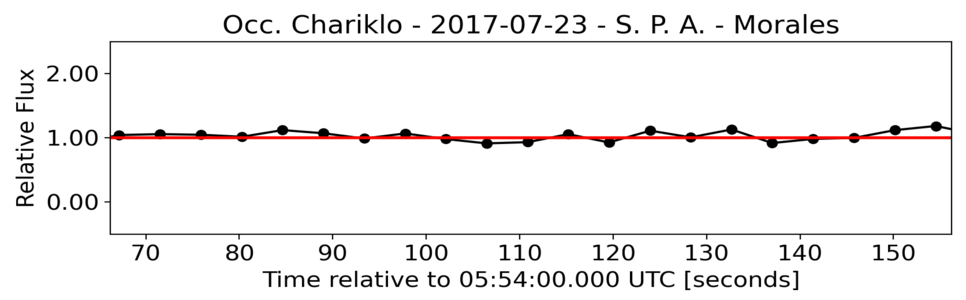}
\caption{Light curve obtained at San Pedro de Atacama by Morales on 2017-07-23.}
\label{Fig:lc_all_1}
\end{figure}

\begin{figure}[h]
\centering
\includegraphics[width=0.50\textwidth]{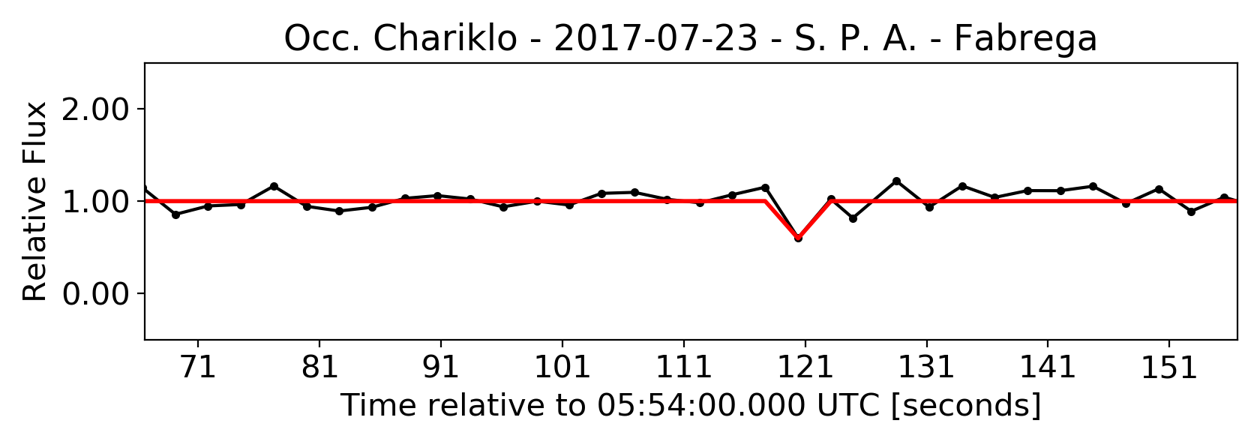}
\caption{Light curve obtained at San Pedro de Atacama by Fabrega on 2017-07-23.}
\label{Fig:lc_all_1}
\end{figure}               

\begin{figure}[h]
\centering
\includegraphics[width=0.50\textwidth]{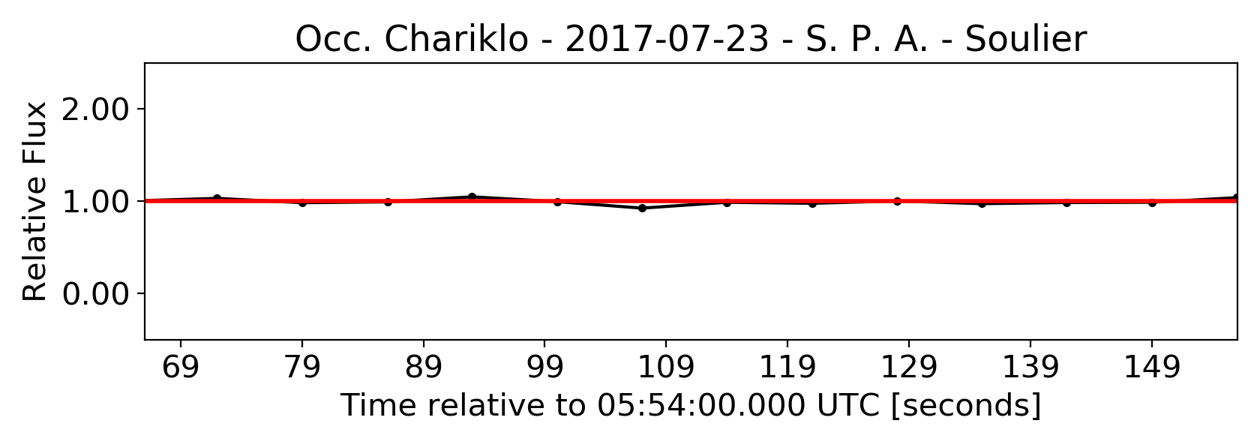}
\caption{Light curve obtained at San Pedro de Atacama by Soulier on 2017-07-23.}
\label{Fig:lc_all_1}
\end{figure}               

\begin{figure}[h]
\centering
\includegraphics[width=0.50\textwidth]{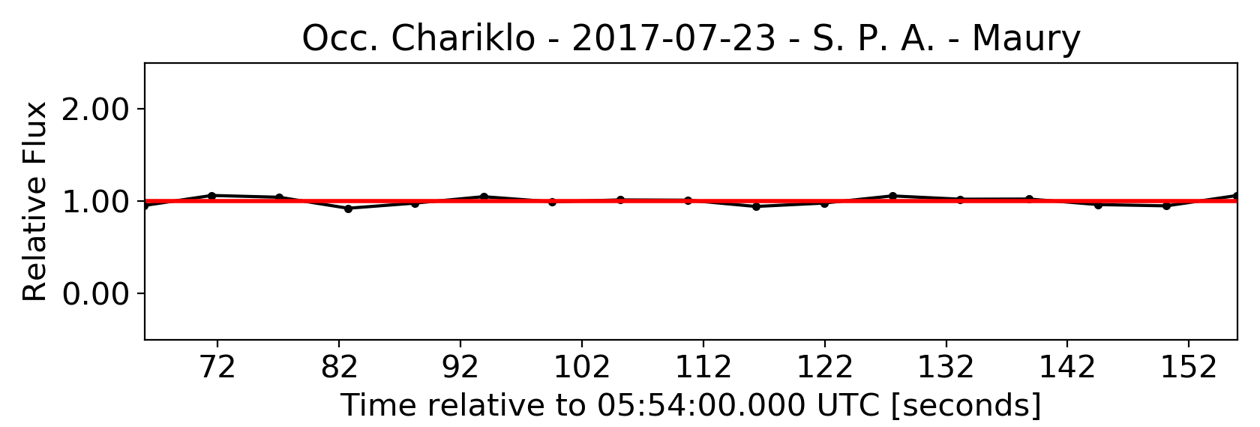}
\caption{Light curve obtained at San Pedro de Atacama by Maury on 2017-07-23.}
\label{Fig:lc_all_1}
\end{figure}               

\begin{figure}[h]
\centering
\includegraphics[width=0.50\textwidth]{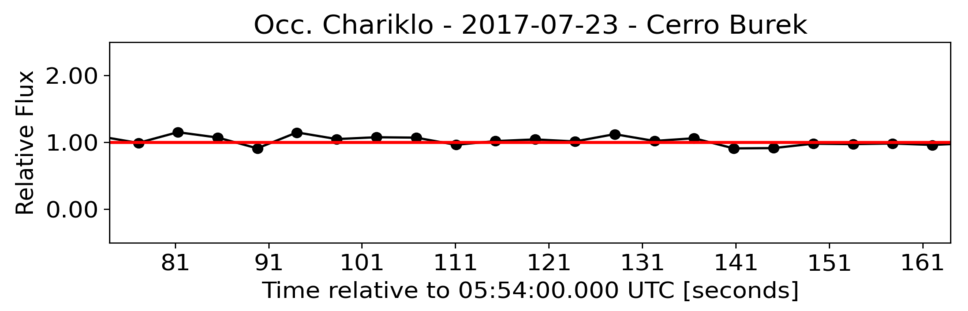}
\caption{Light curve obtained at Cerro Burek on 2017-07-23.}
\label{Fig:lc_all_1}
\end{figure}               

\begin{figure}[h]
\centering
\includegraphics[width=0.50\textwidth]{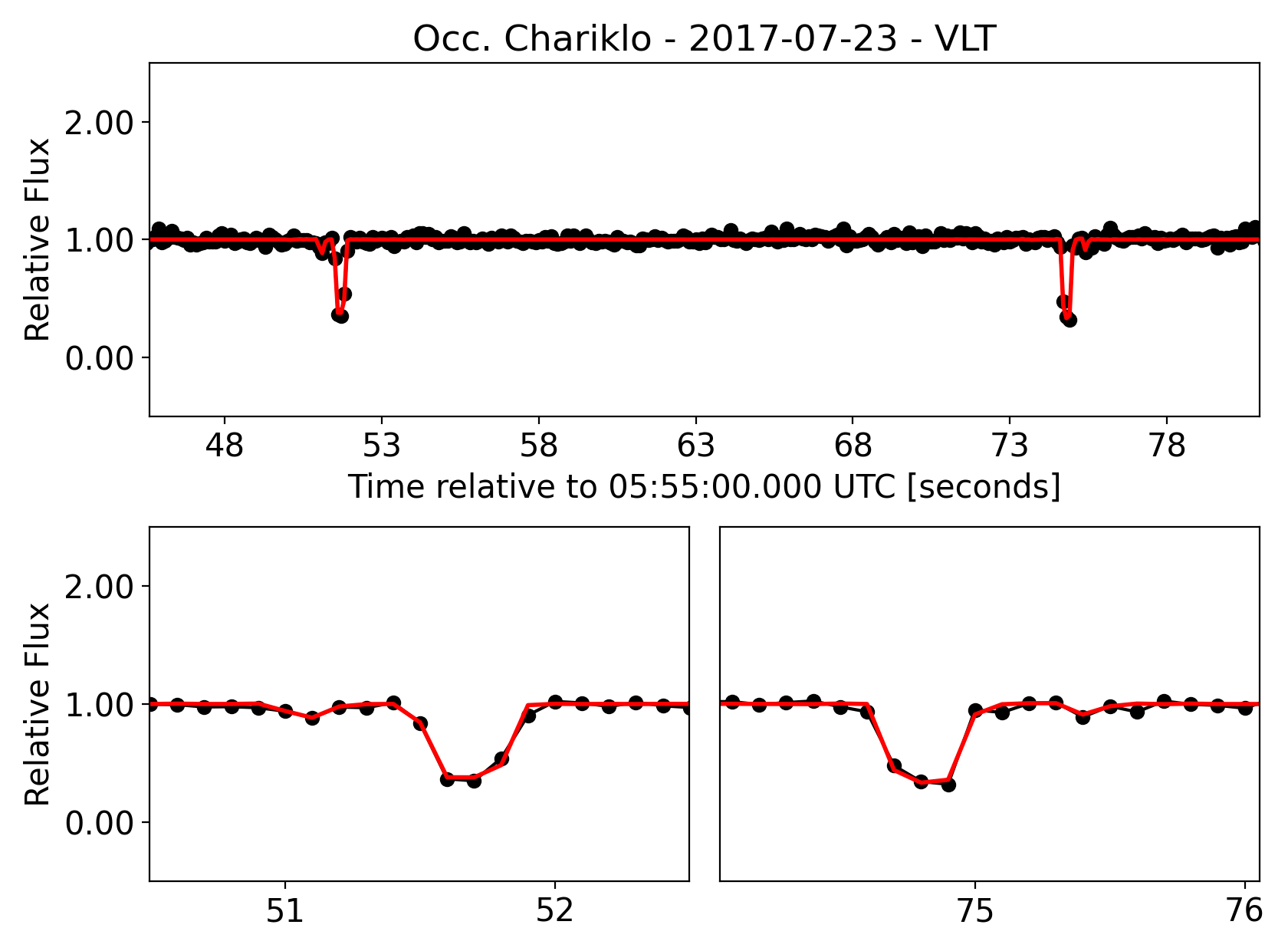}
\caption{Light curve obtained at the VLT on 2017-07-23.}
\label{Fig:lc_all_1}
\end{figure}               

\begin{figure}[h]
\centering
\includegraphics[width=0.50\textwidth]{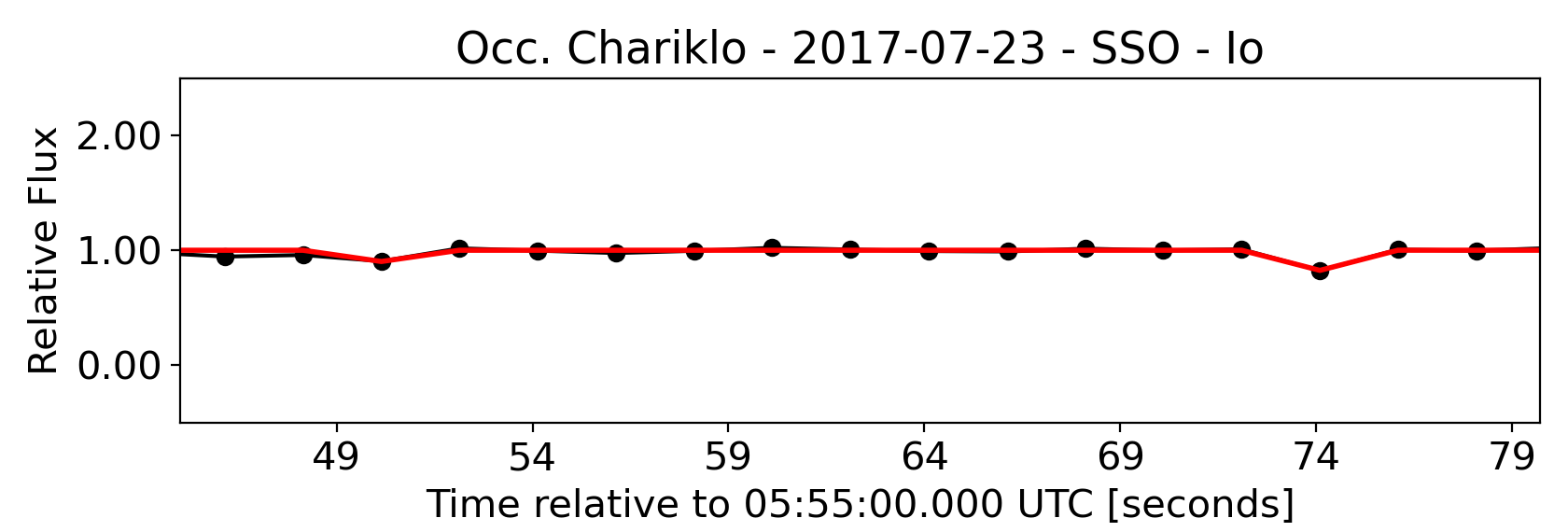}
\caption{Light curve obtained at the SSO - Io on 2017-07-23.}
\label{Fig:lc_all_1}
\end{figure}               

\begin{figure}[h]
\centering
\includegraphics[width=0.50\textwidth]{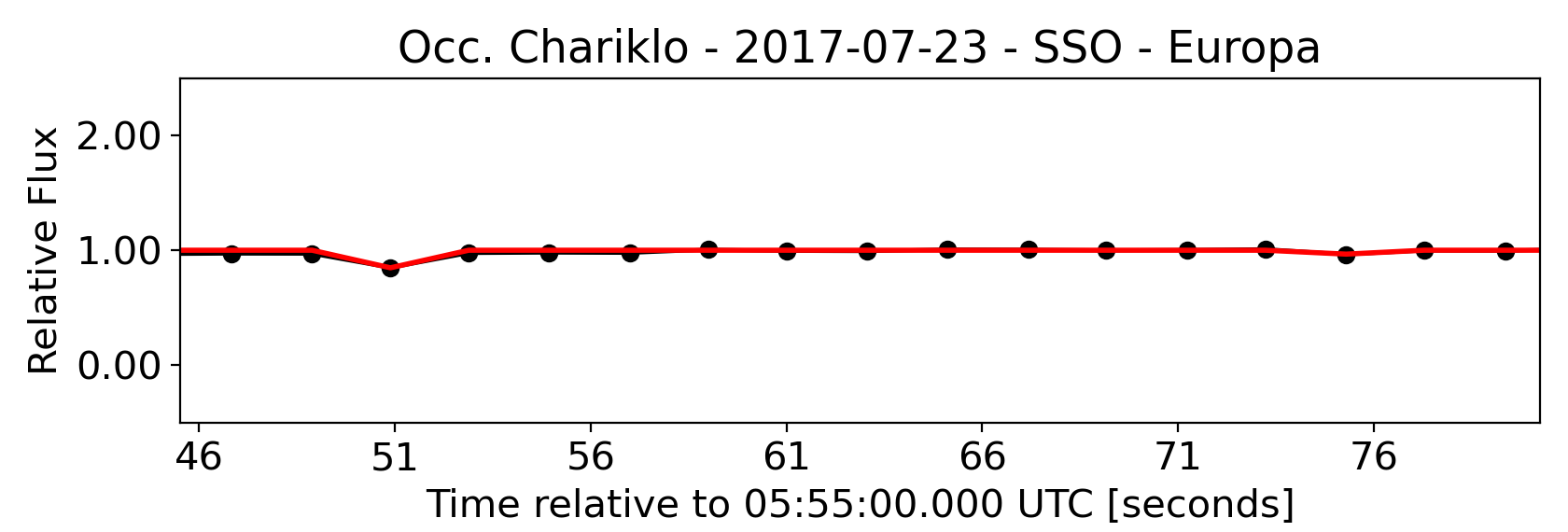}
\caption{Light curve obtained at the SSO - Europa on 2017-07-23.}
\label{Fig:lc_all_1}
\end{figure}               

\begin{figure}[h]
\centering
\includegraphics[width=0.50\textwidth]{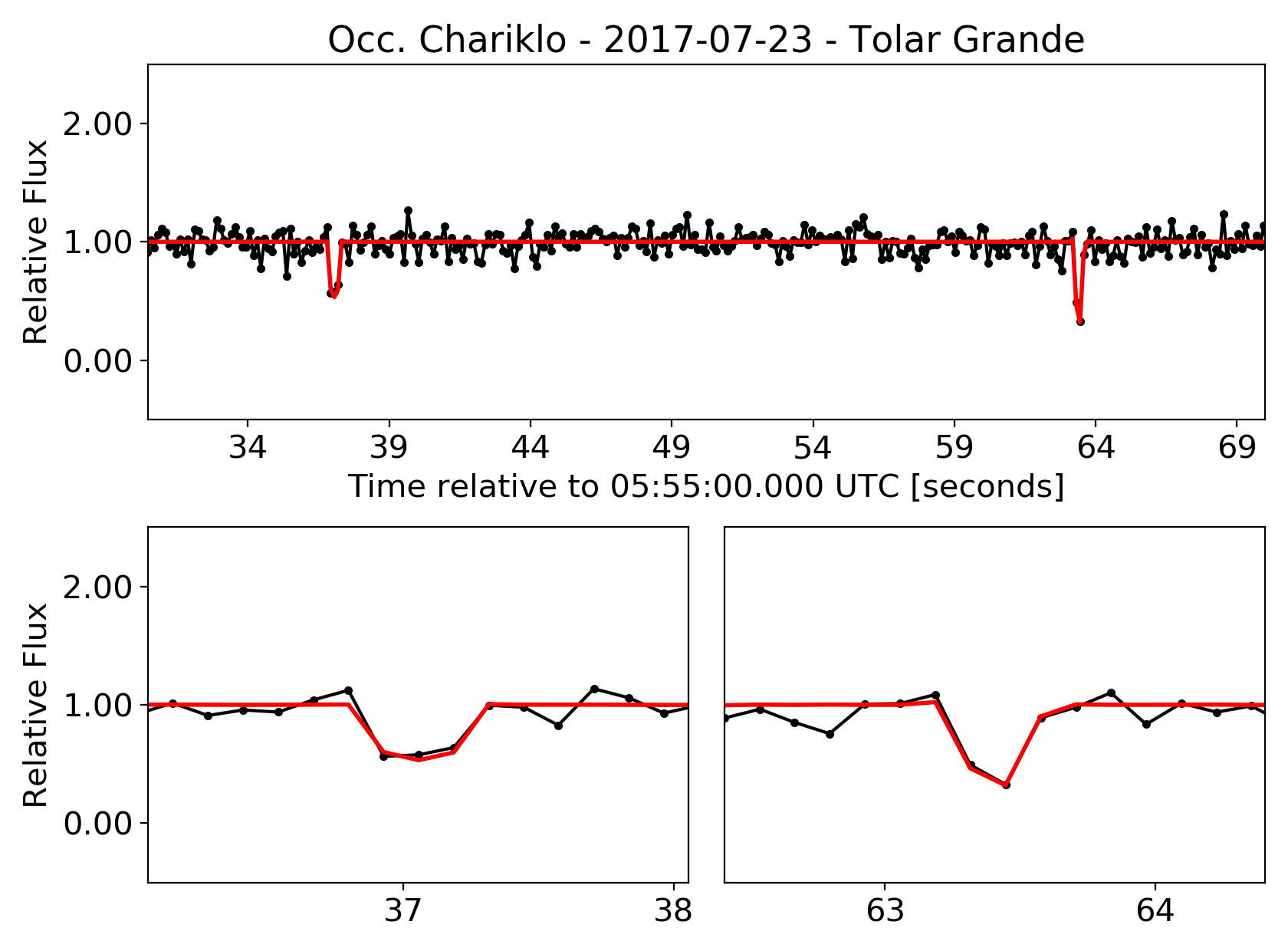}
\caption{Light curve obtained at Tolar Grande on 2017-07-23.}
\label{Fig:lc_all_1}
\end{figure}               

\begin{figure}[h]
\centering
\includegraphics[width=0.50\textwidth]{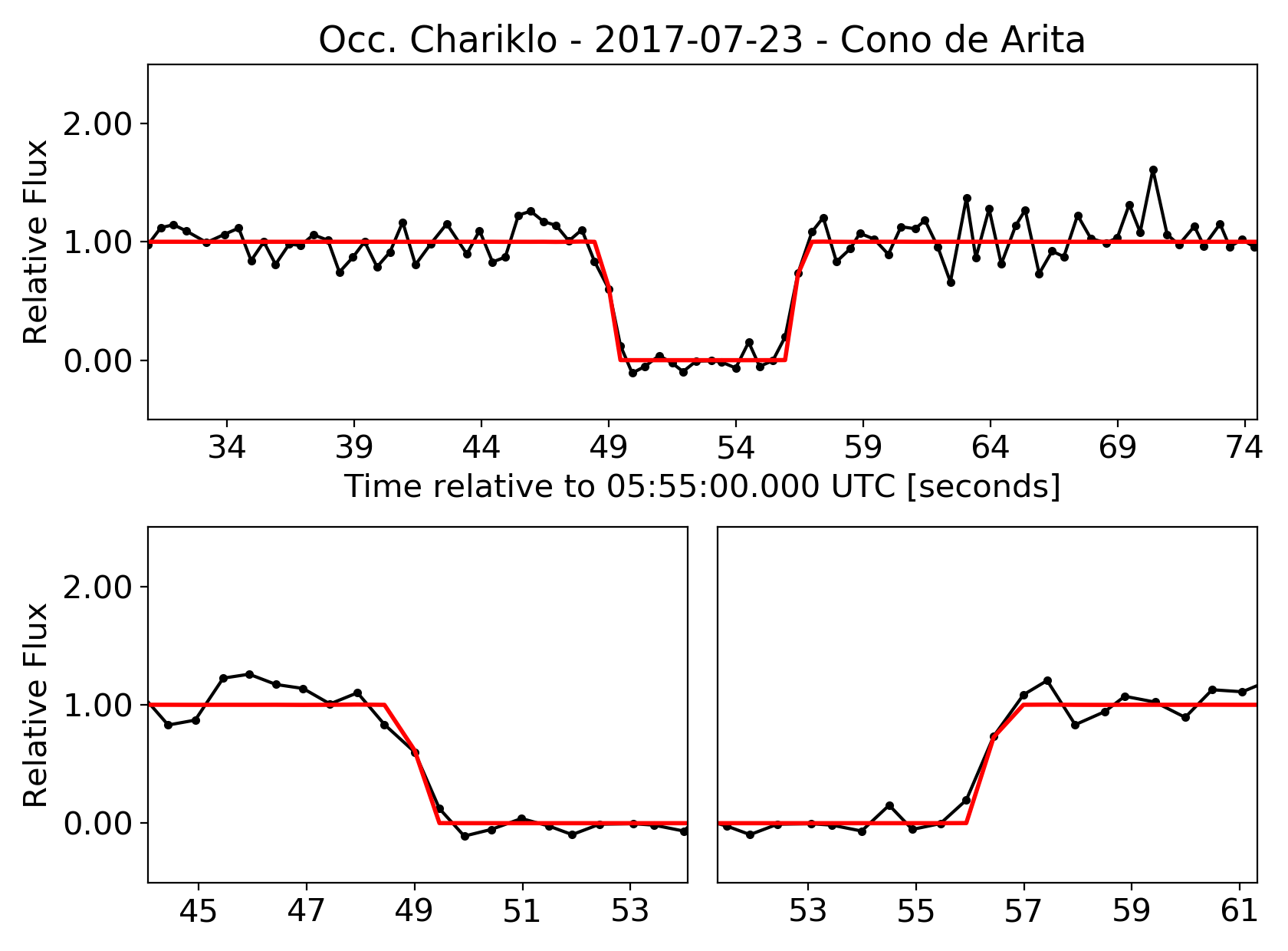}
\caption{Light curve obtained at Cono de Arita on 2017-07-23.}
\label{Fig:lc_all_1}
\end{figure}               

\begin{figure}[h]
\centering
\includegraphics[width=0.50\textwidth]{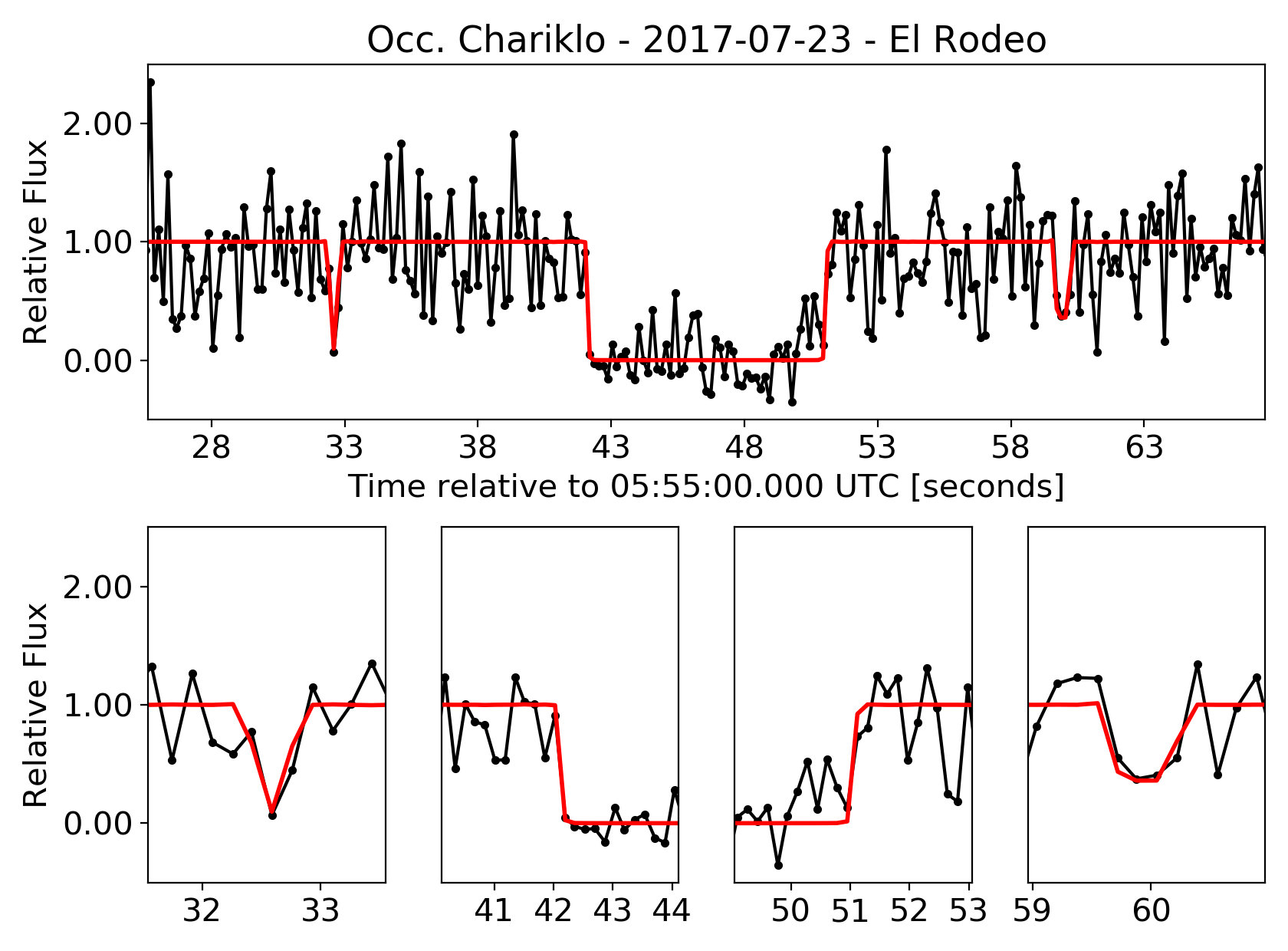}
\caption{Light curve obtained at El Rodeo on 2017-07-23.}
\label{Fig:lc_all_1}
\end{figure}               

\begin{figure}[h]
\centering
\includegraphics[width=0.50\textwidth]{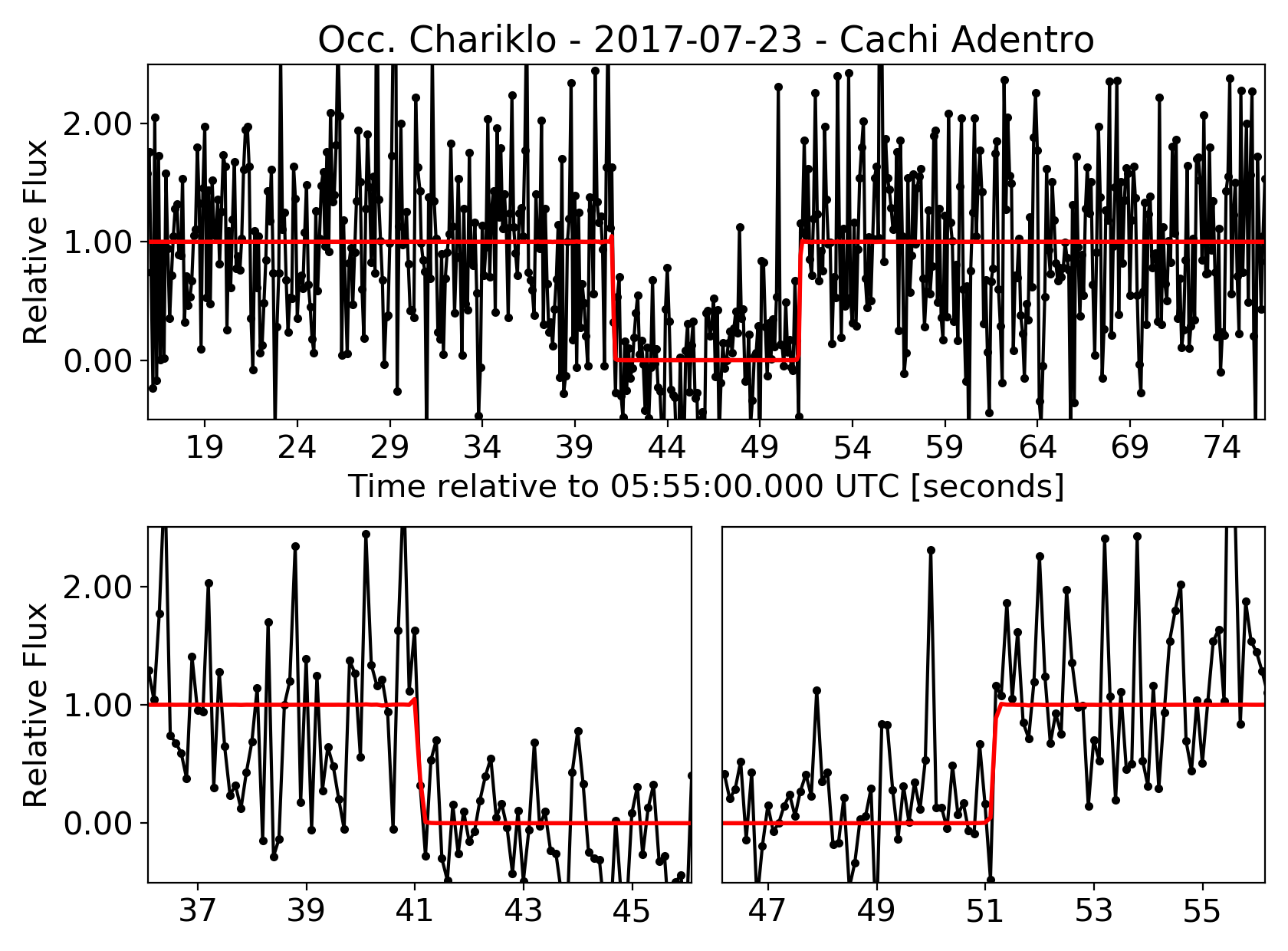}
\caption{Light curve obtained at Cachi Adentro on 2017-07-23.}
\label{Fig:lc_all_1}
\end{figure}               

\begin{figure}[h]
\centering
\includegraphics[width=0.50\textwidth]{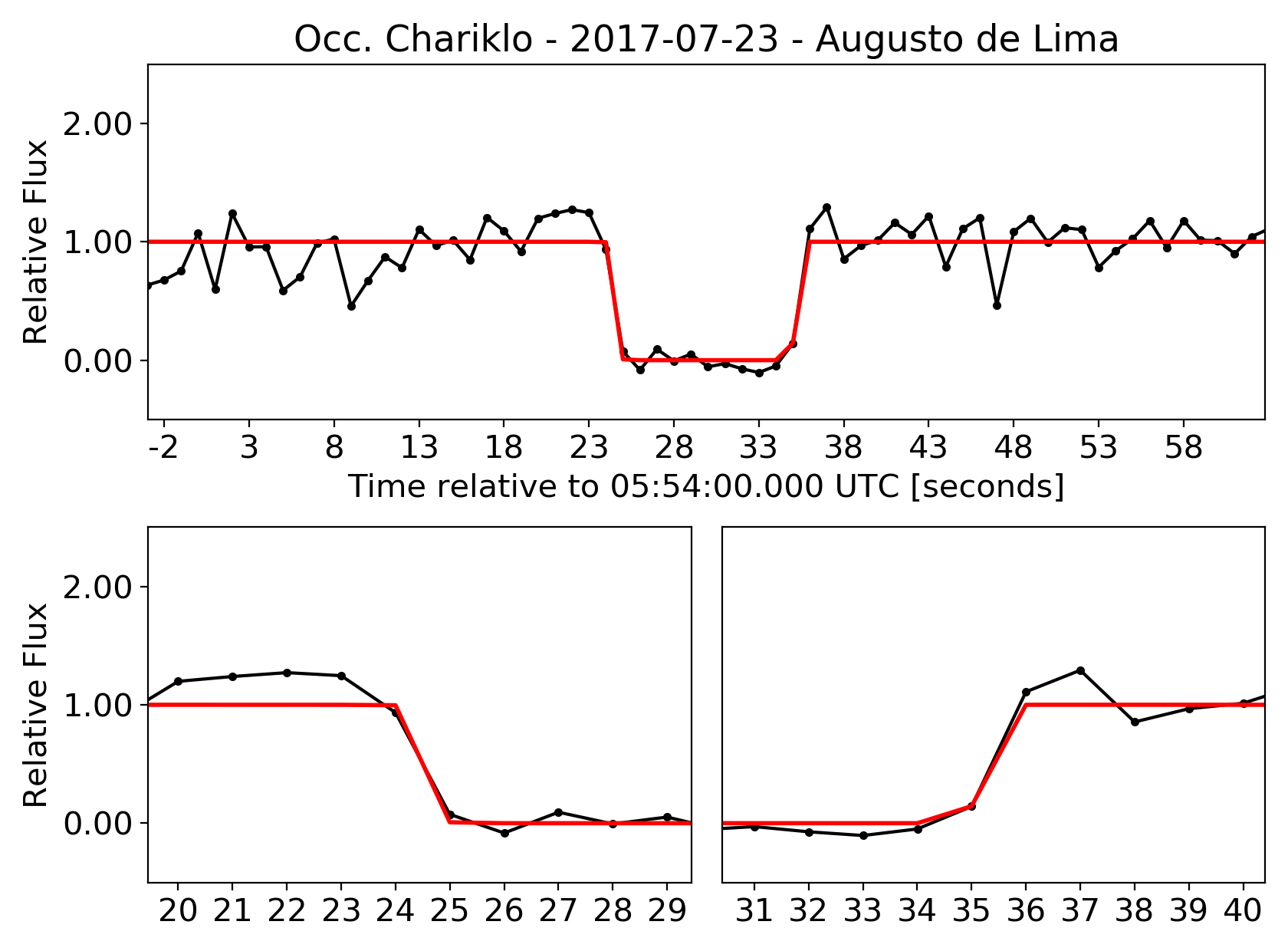}
\caption{Light curve obtained at Augusto de Lima on 2017-07-23.}
\label{Fig:lc_all_1}
\end{figure}               

\begin{figure}[h]
\centering
\includegraphics[width=0.50\textwidth]{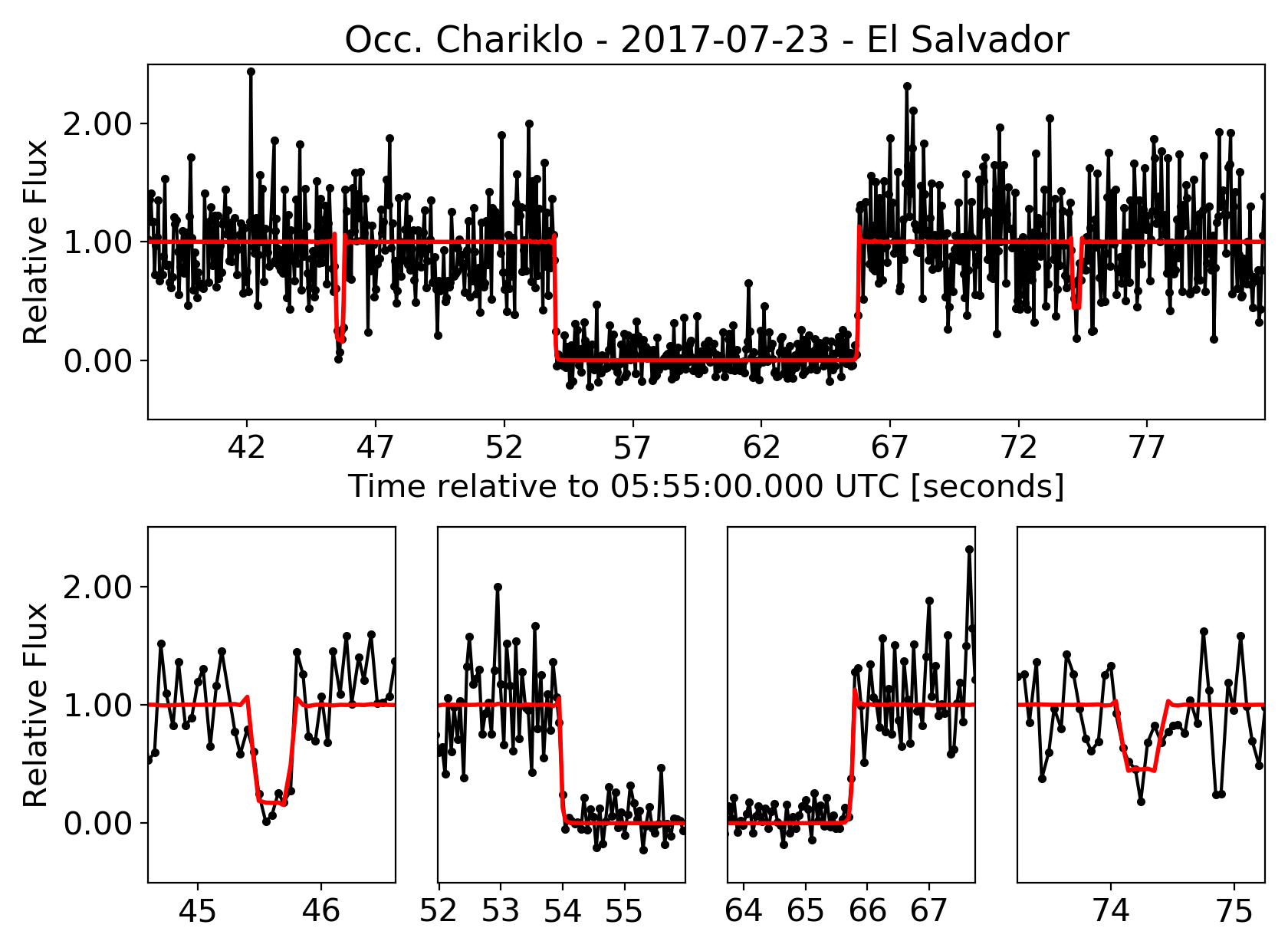}
\caption{Light curve obtained at El Salvador on 2017-07-23.}
\label{Fig:lc_all_1}
\end{figure}               

\begin{figure}[h]
\centering
\includegraphics[width=0.50\textwidth]{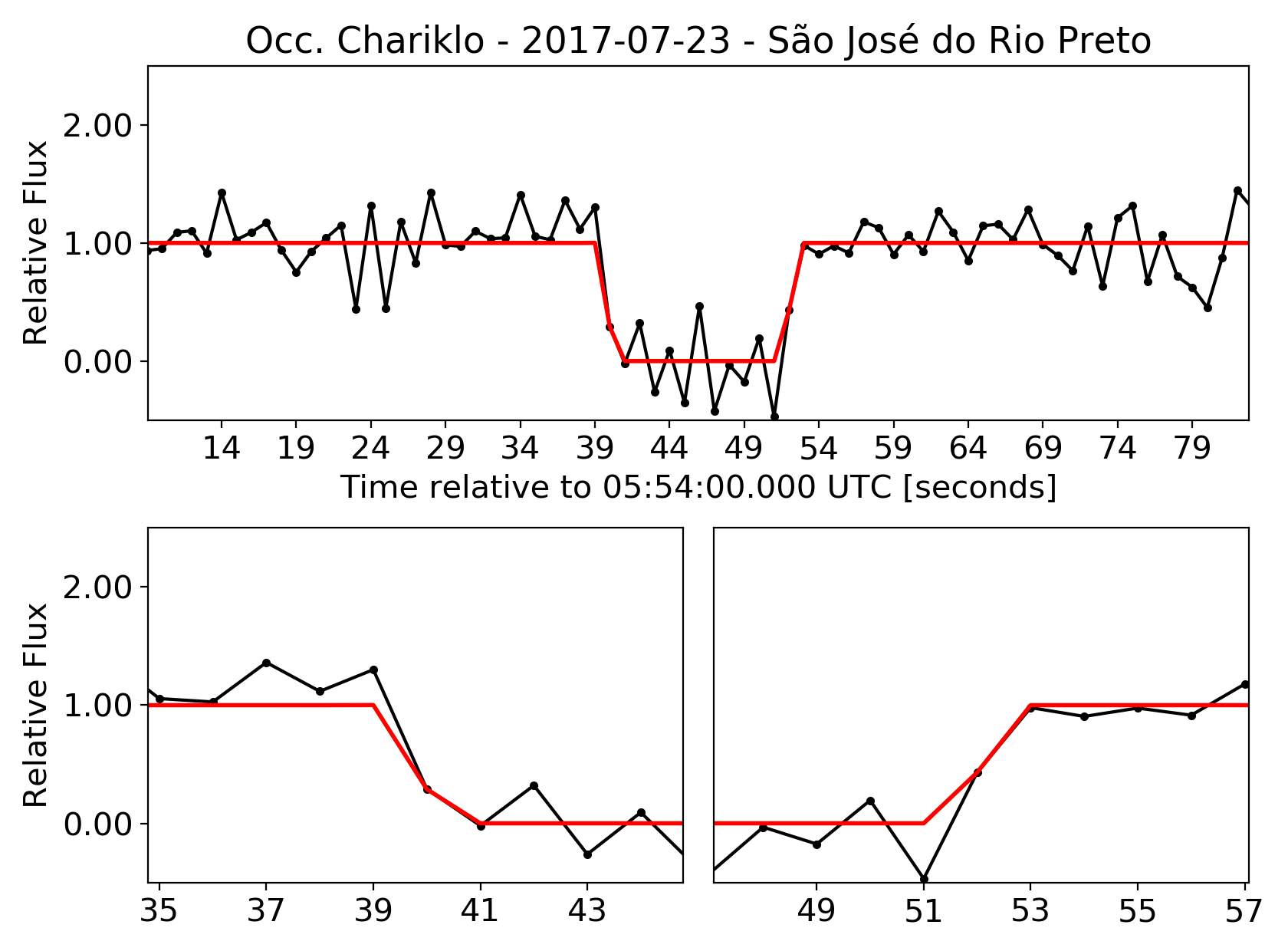}
\caption{Light curve obtained at São José do Rio Preto on 2017-07-23.}
\label{Fig:lc_all_1}
\end{figure}               

\begin{figure}[h]
\centering
\includegraphics[width=0.50\textwidth]{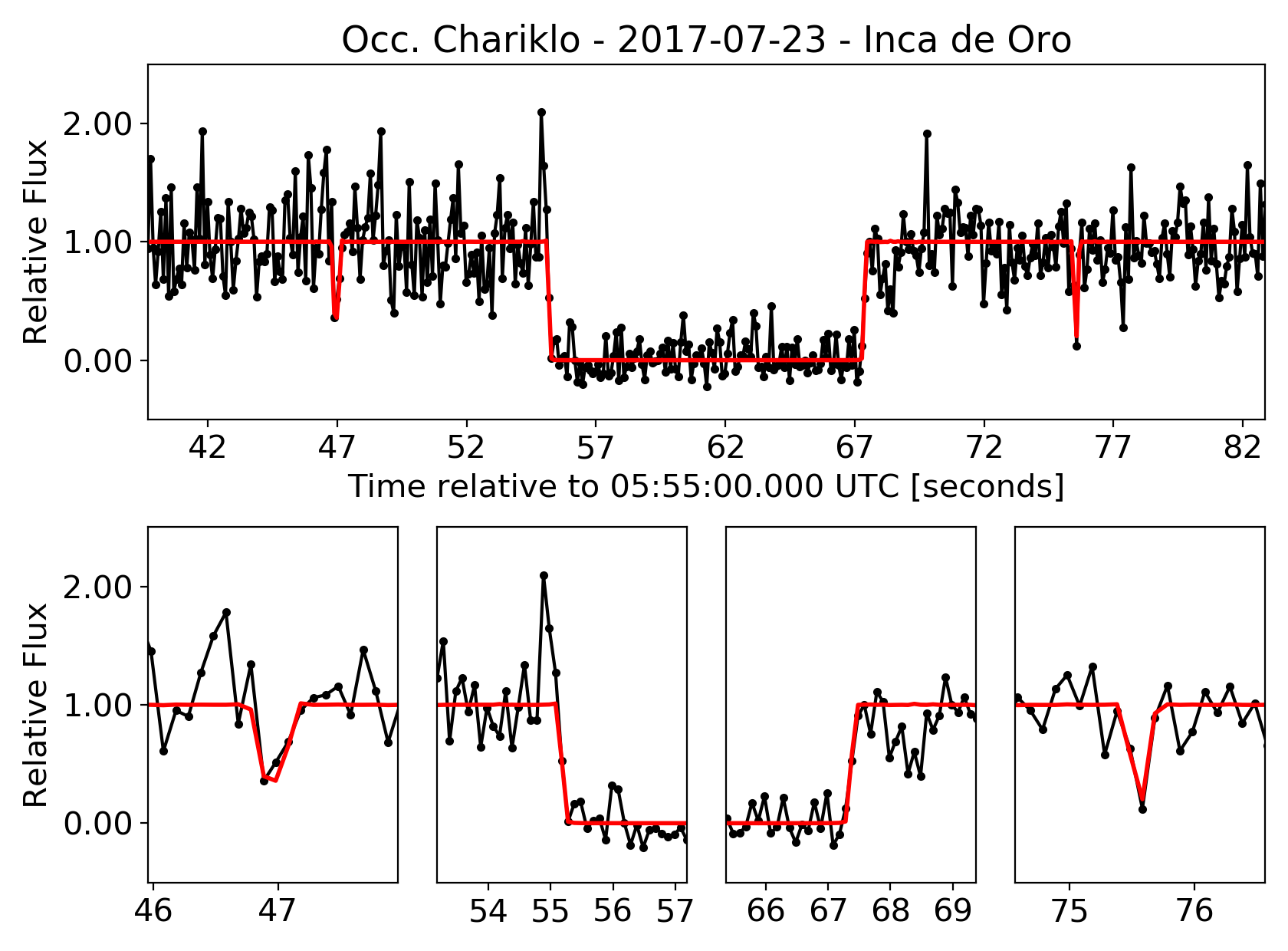}
\caption{Light curve obtained at Inca de Oro on 2017-07-23.}
\label{Fig:lc_all_1}
\end{figure}               

\begin{figure}[h]
\centering
\includegraphics[width=0.50\textwidth]{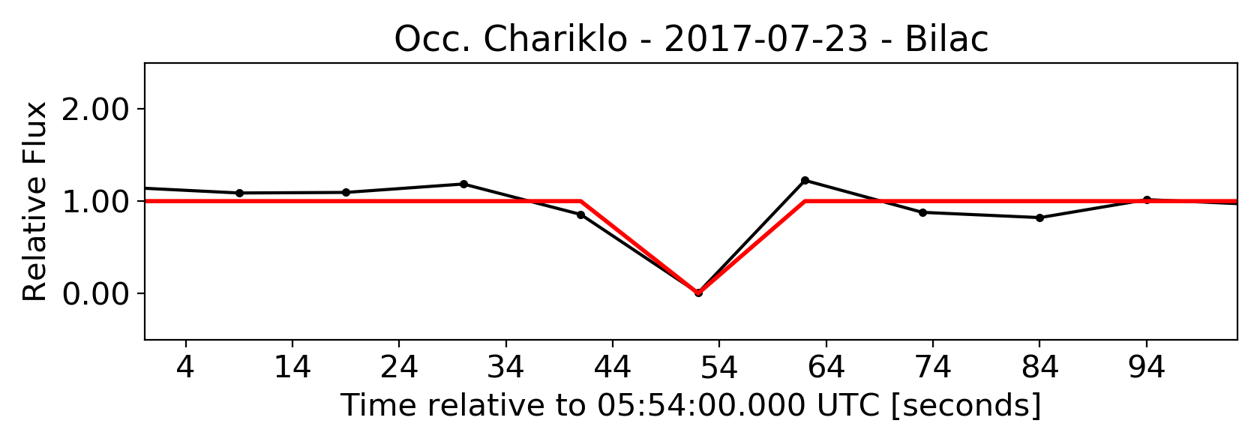}
\caption{Light curve obtained at Bilac on 2017-07-23.}
\label{Fig:lc_all_1}
\end{figure}               

\begin{figure}[h]
\centering
\includegraphics[width=0.50\textwidth]{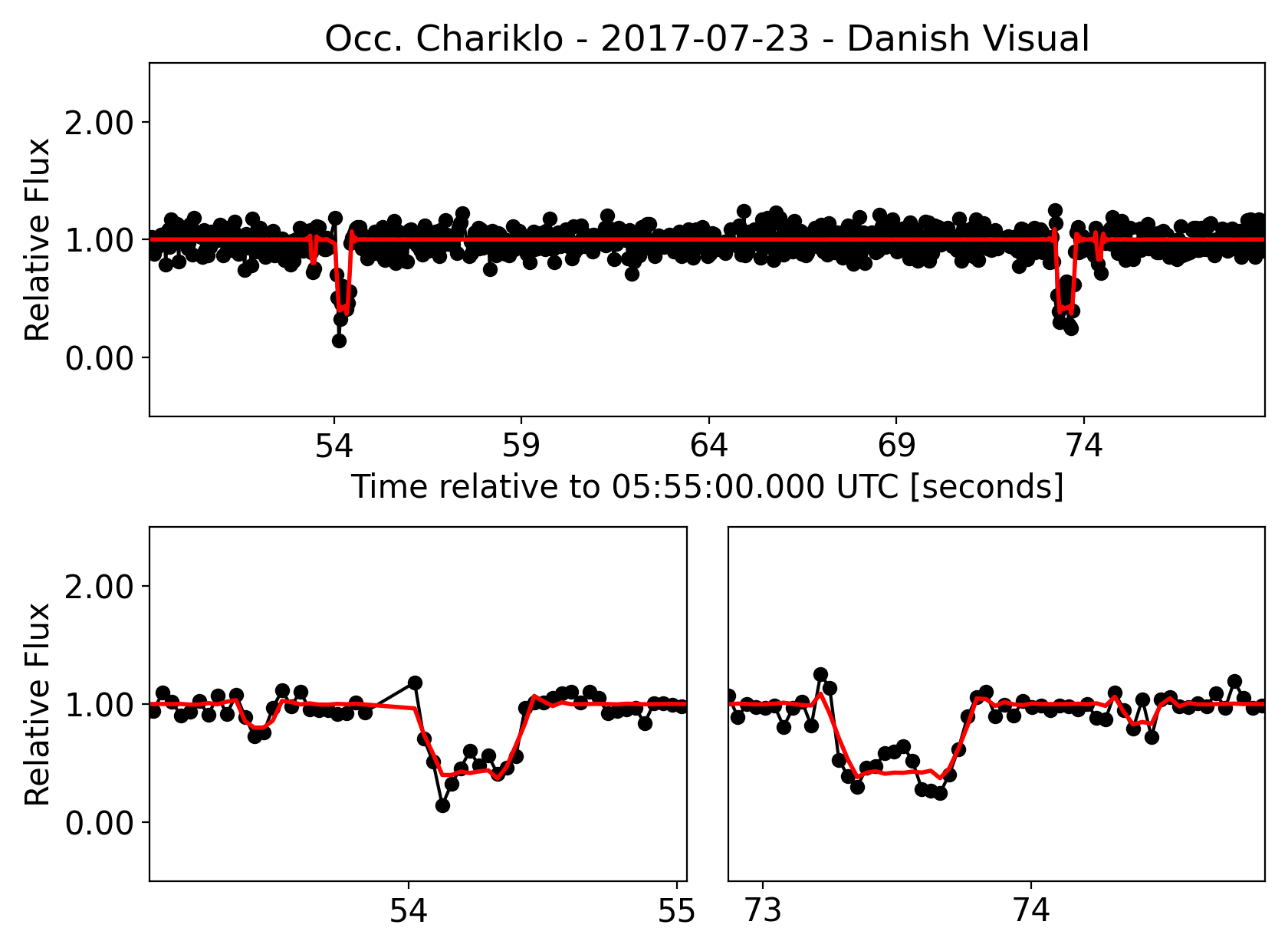}
\caption{Light curve obtained with Danish with the Visual band on 2017-07-23.}
\label{Fig:lc_all_1}
\end{figure}               

\begin{figure}[h]
\centering
\includegraphics[width=0.50\textwidth]{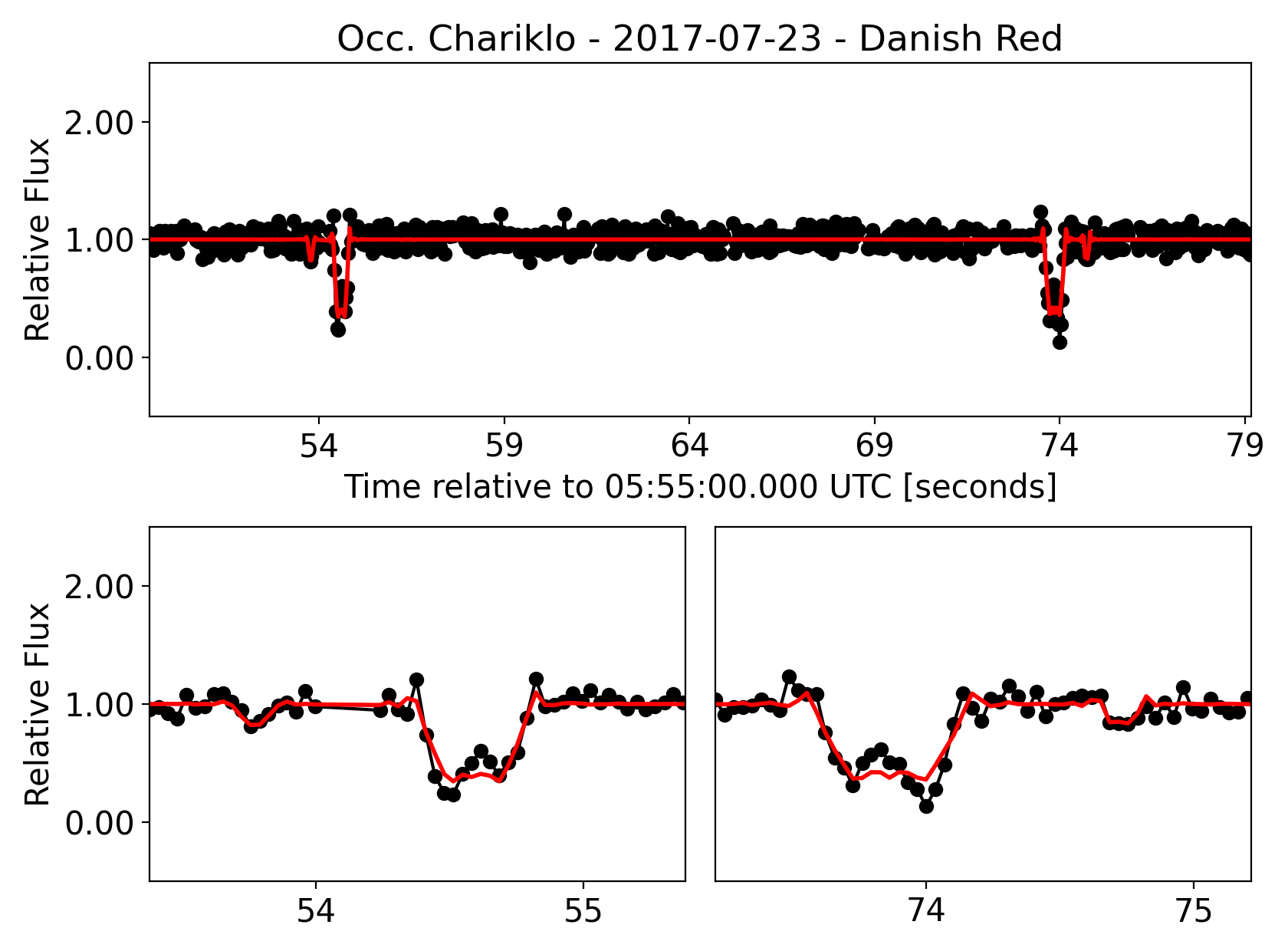}
\caption{Light curve obtained with Danish with the Red band on 2017-07-23.}
\label{Fig:lc_all_1}
\end{figure}               

\begin{figure}[h]
\centering
\includegraphics[width=0.50\textwidth]{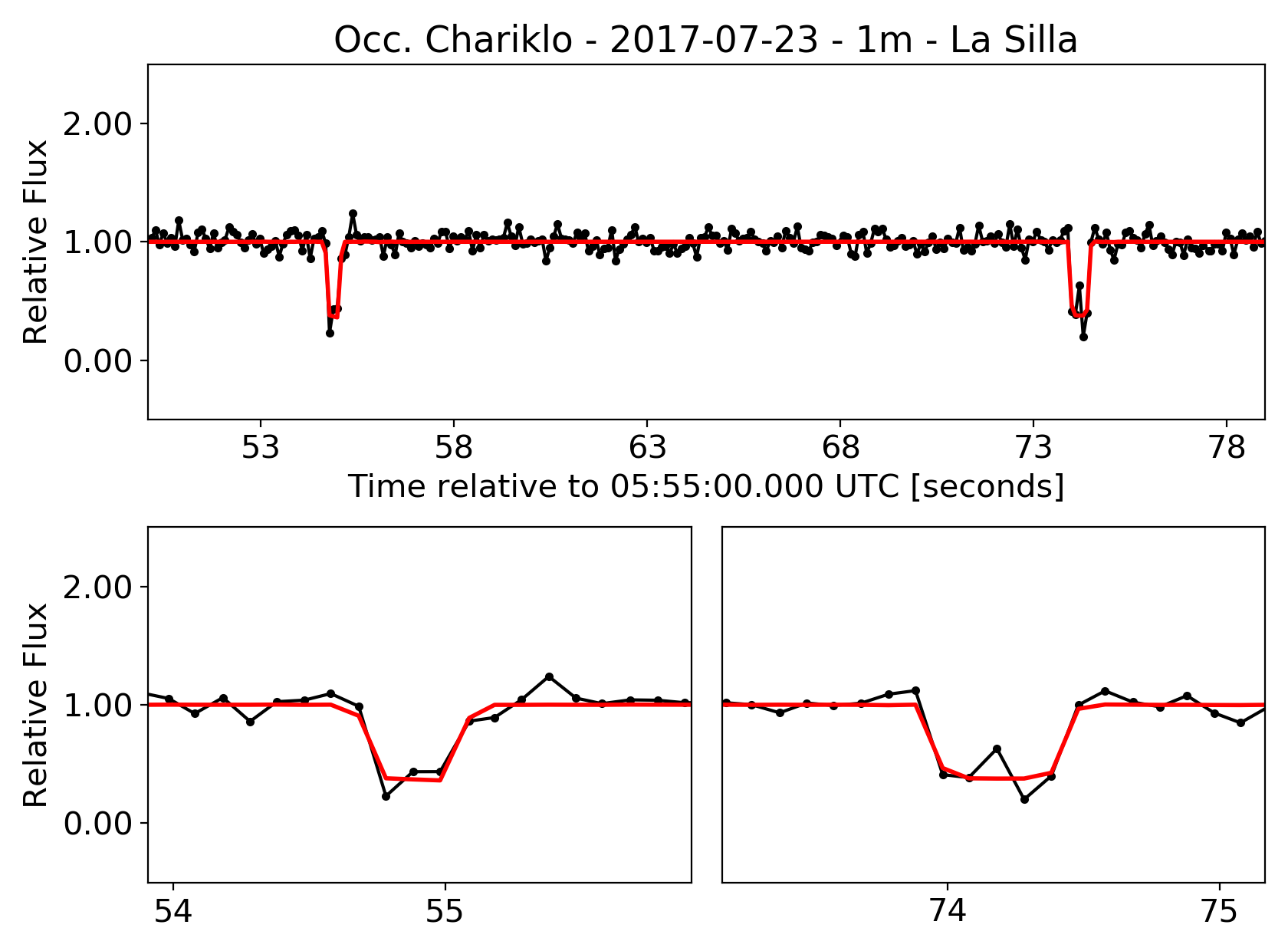}
\caption{Light curve obtained with the 1m telescope in La Silla on 2017-07-23.}
\label{Fig:lc_all_1}
\end{figure}               

\begin{figure}[h]
\centering
\includegraphics[width=0.50\textwidth]{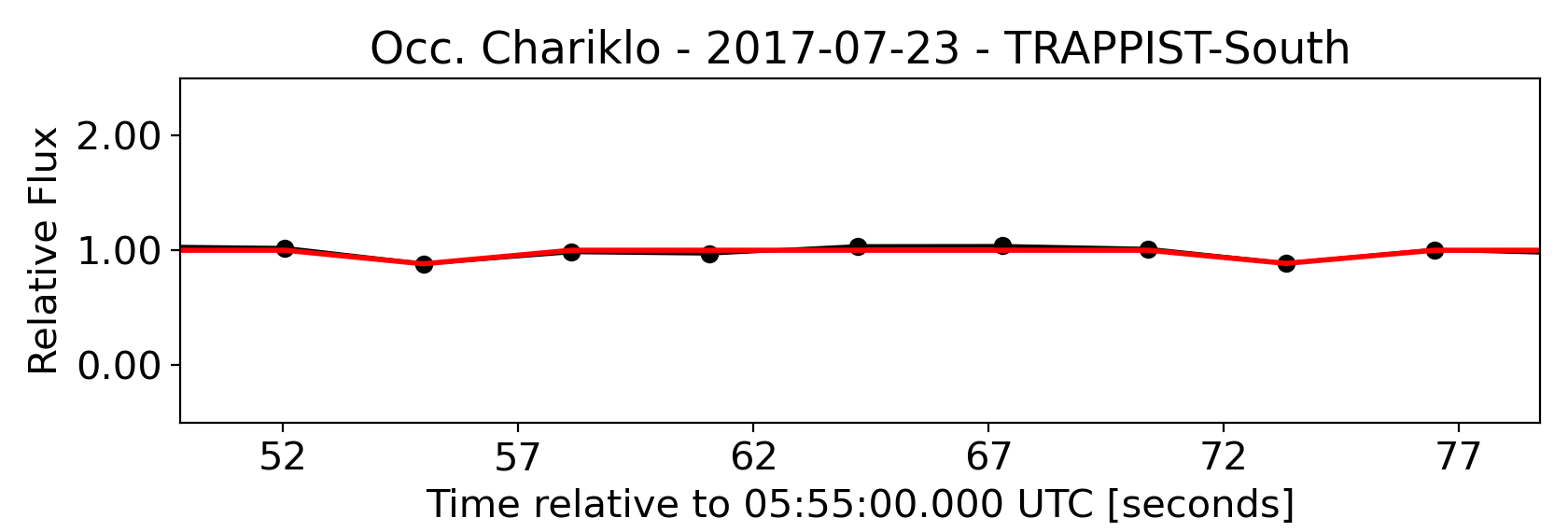}
\caption{Light curve obtained with TRAPPIST-South on 2017-07-23.}
\label{Fig:lc_all_1}
\end{figure}               

\begin{figure}[h]
\centering
\includegraphics[width=0.50\textwidth]{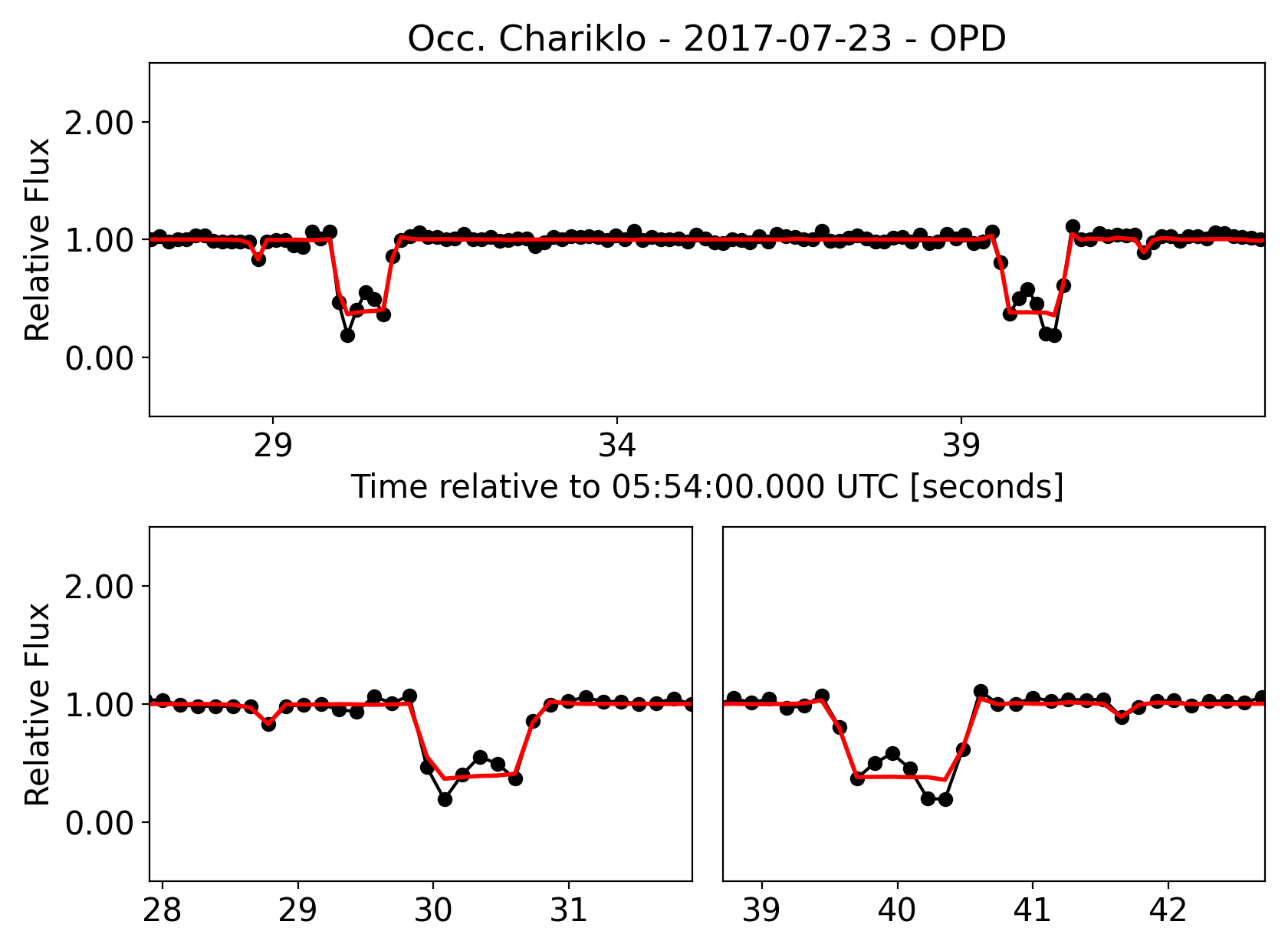}
\caption{Light curve obtained at the OPD on 2017-07-23.}
\label{Fig:lc_all_1}
\end{figure}               

\begin{figure}[h]
\centering
\includegraphics[width=0.50\textwidth]{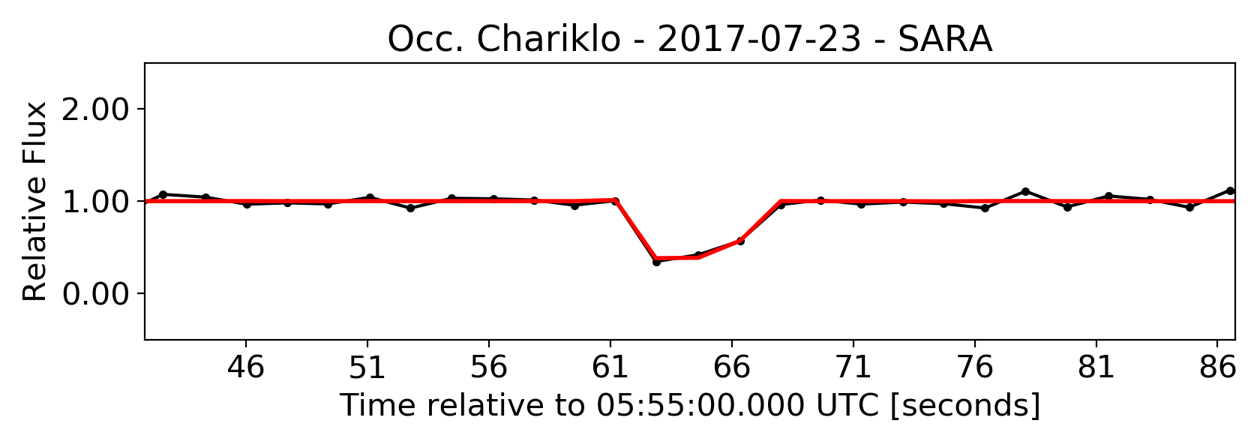}
\caption{Light curve obtained with SARA on 2017-07-23.}
\label{Fig:lc_all_1}
\end{figure}               

\begin{figure}[h]
\centering
\includegraphics[width=0.50\textwidth]{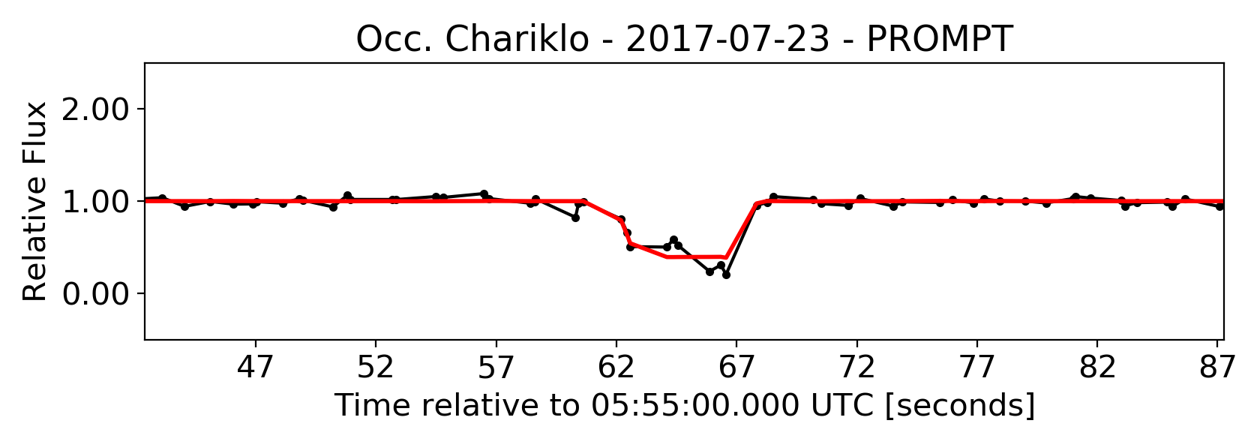}
\caption{Light curve obtained with PROMPT on 2017-07-23.}
\label{Fig:lc_all_1}
\end{figure}


\begin{figure}[h]
\centering
\includegraphics[width=0.50\textwidth]{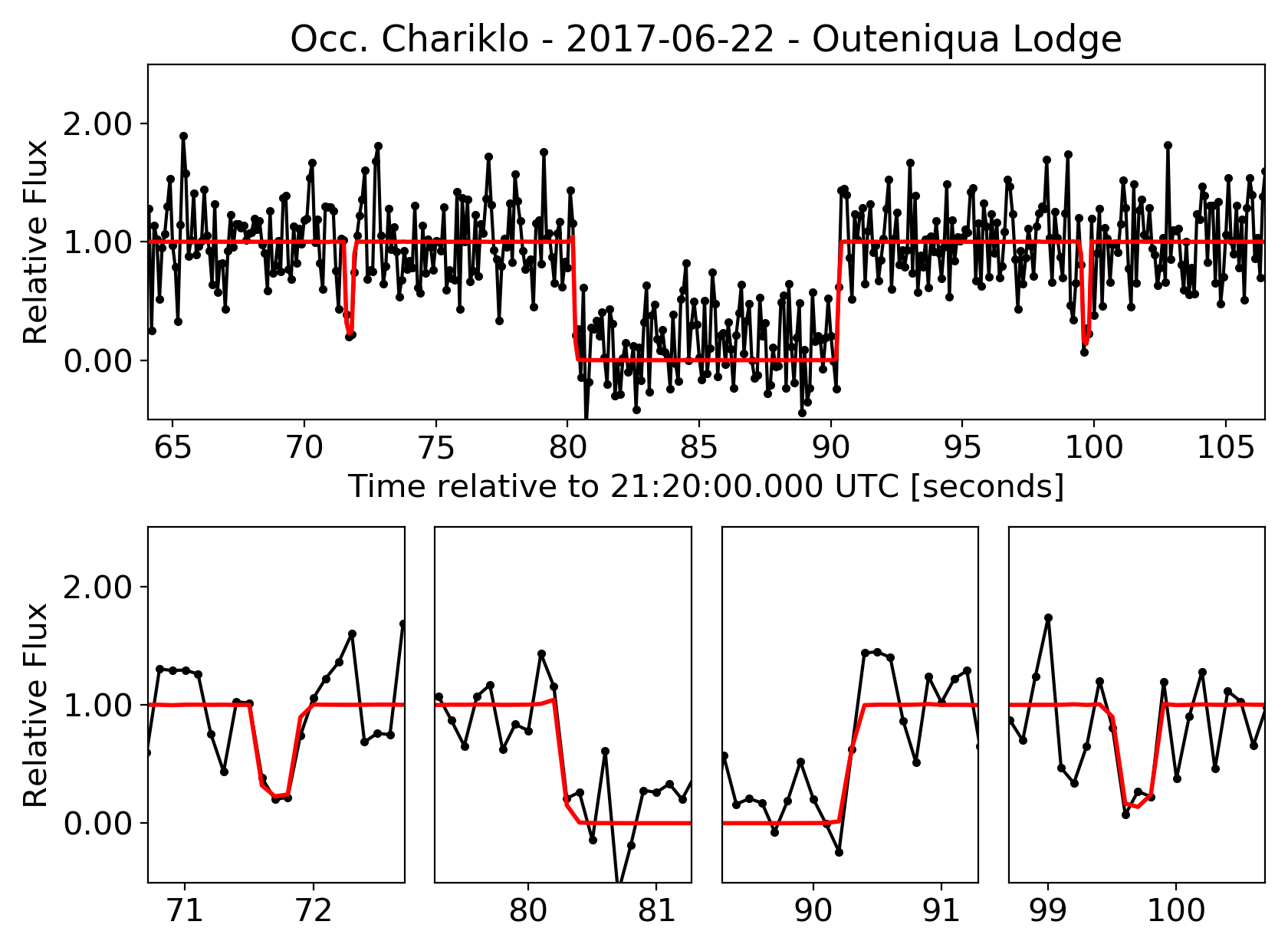}
\caption{Light curve obtained at Outeniqua Lodge on 2017-06-22.}
\label{Fig:lc_all_1}
\end{figure}               

\begin{figure}[h]
\centering
\includegraphics[width=0.50\textwidth]{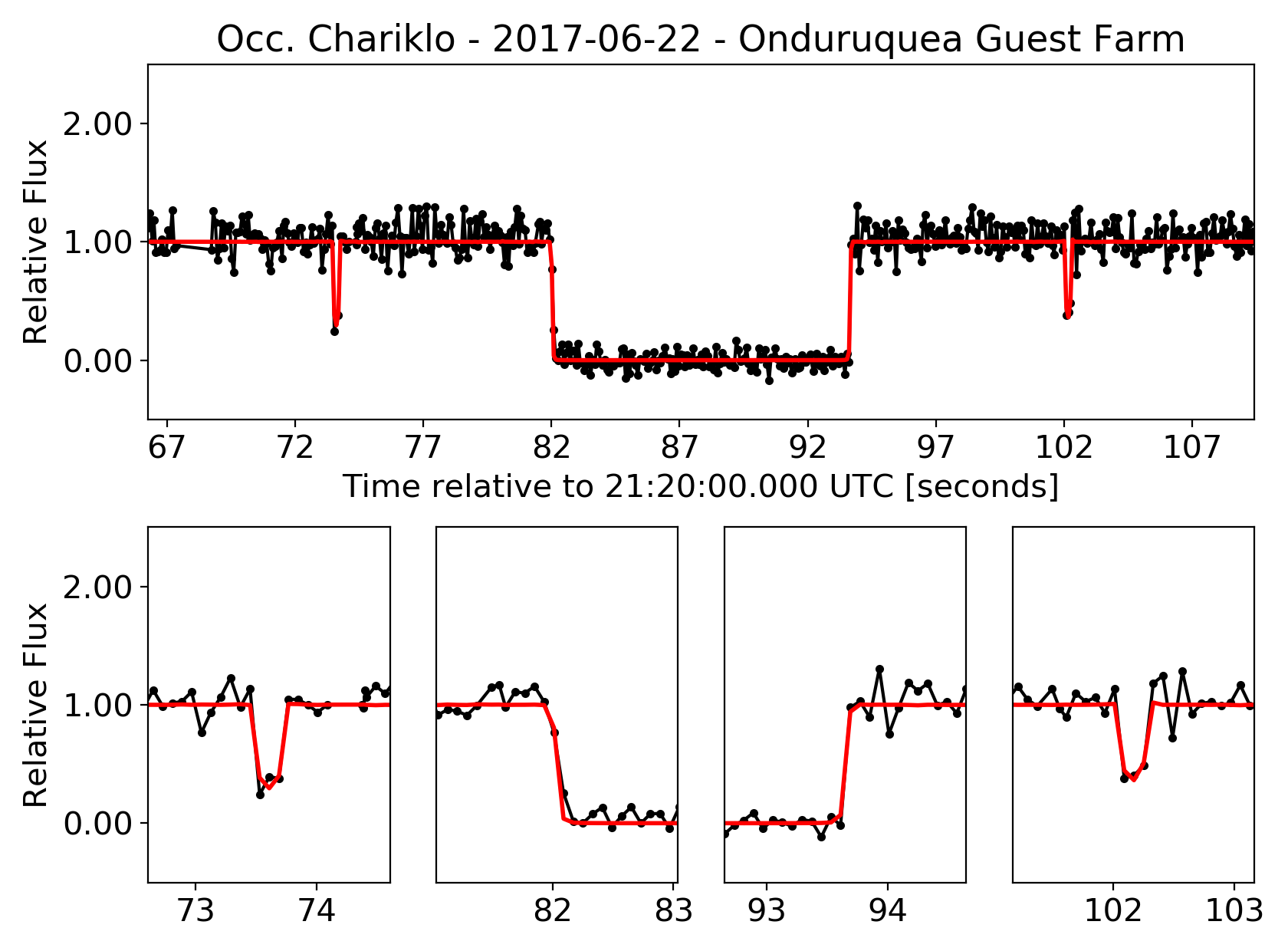}
\caption{Light curve obtained at Onduruquea Guest Farm on 2017-06-22.}
\label{Fig:lc_all_1}
\end{figure}               

\begin{figure}[h]
\centering
\includegraphics[width=0.50\textwidth]{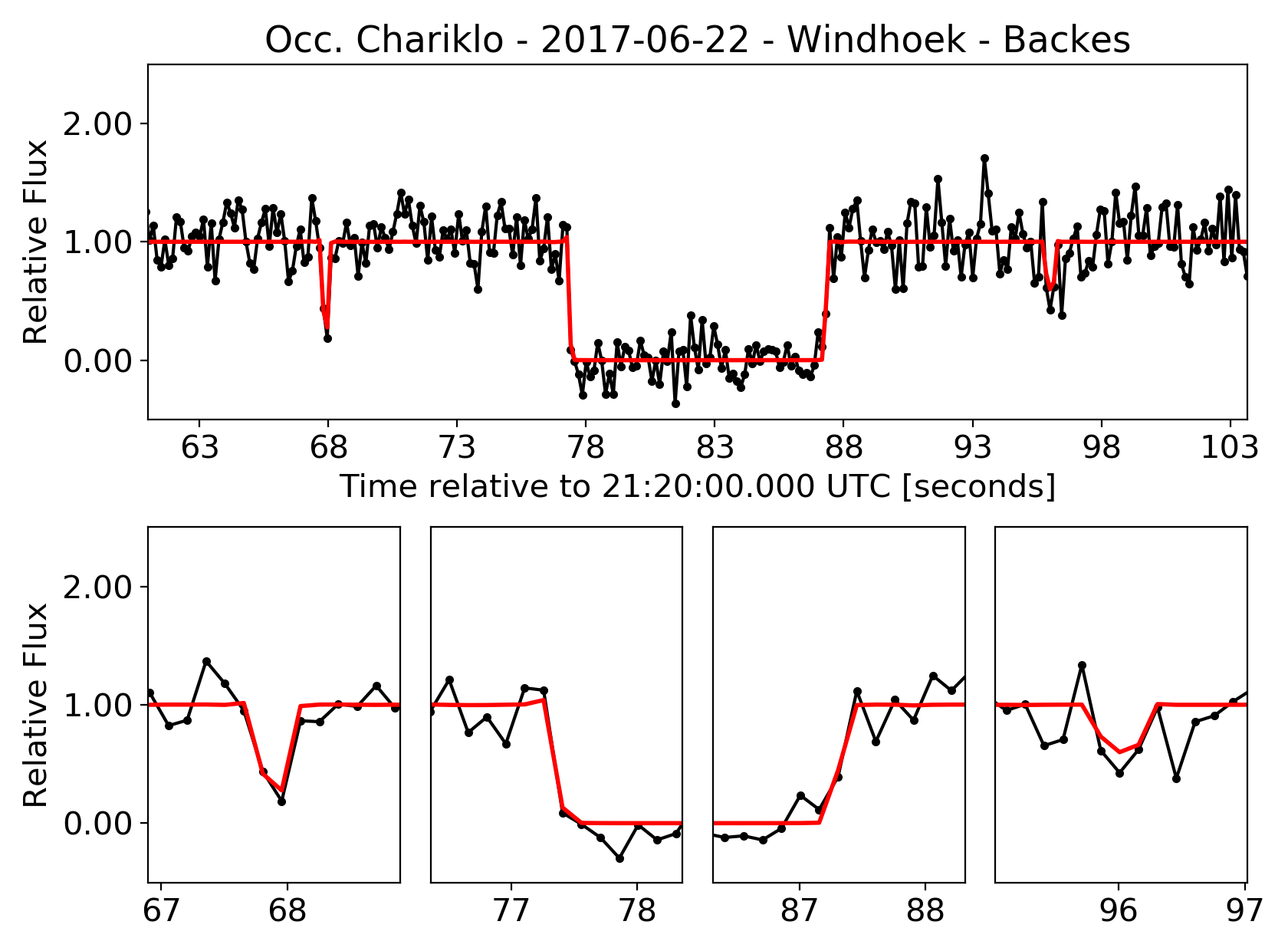}
\caption{Light curve obtained at Windhoek by Backes on 2017-06-22.}
\label{Fig:lc_all_1}
\end{figure}               

\begin{figure}[h]
\centering
\includegraphics[width=0.50\textwidth]{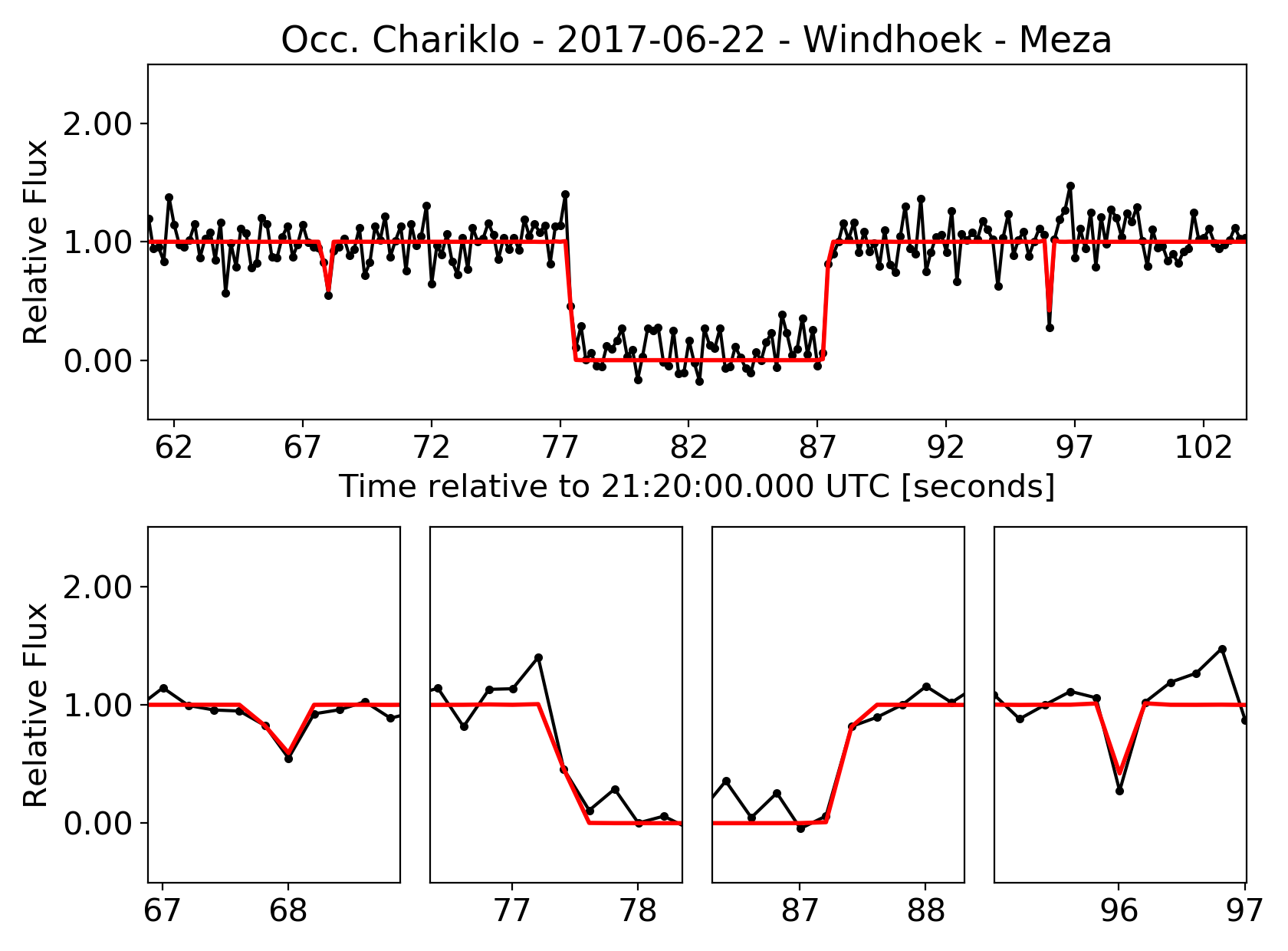}
\caption{Light curve obtained at Windhoek by Meza on 2017-06-22.}
\label{Fig:lc_all_1}
\end{figure}               

\begin{figure}[h]
\centering
\includegraphics[width=0.50\textwidth]{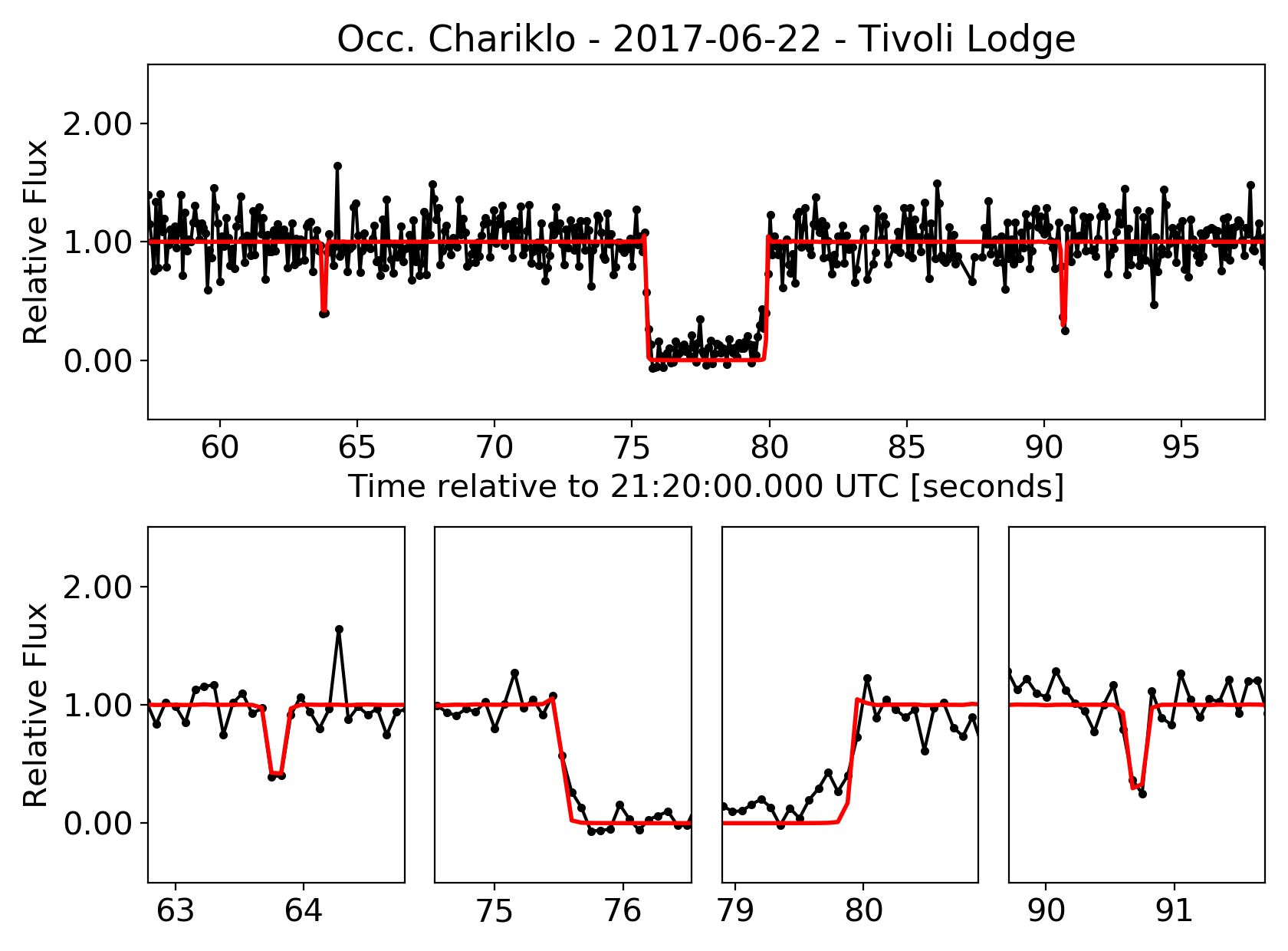}
\caption{Light curve obtained at Tivoli Lodge on 2017-06-22.}
\label{Fig:lc_all_1}
\end{figure}               

\begin{figure}[h]
\centering
\includegraphics[width=0.50\textwidth]{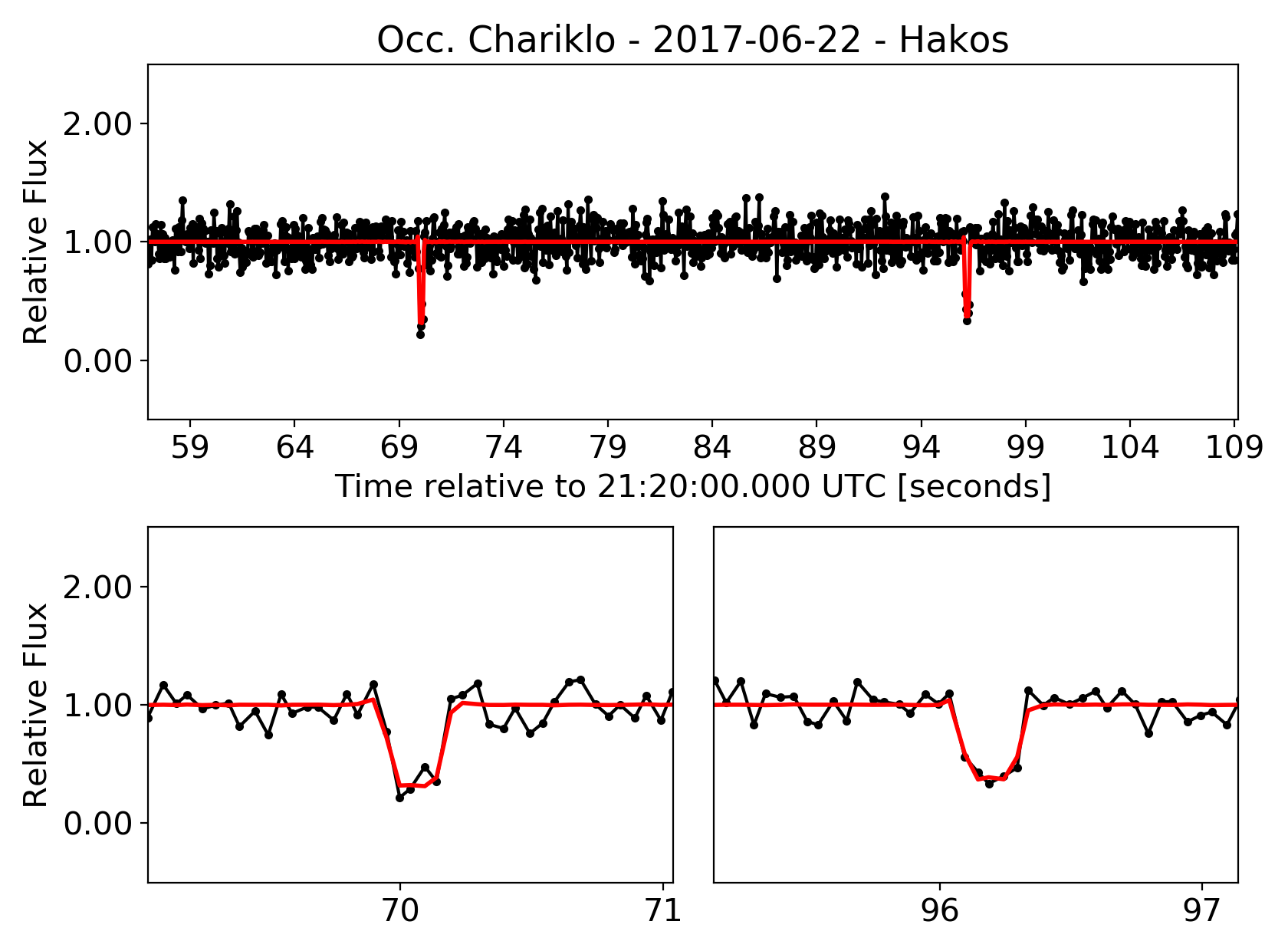}
\caption{Light curve obtained at Hakos on 2017-06-22.}
\label{Fig:lc_all_1}
\end{figure}               


\begin{figure}[h]
\centering
\includegraphics[width=0.50\textwidth]{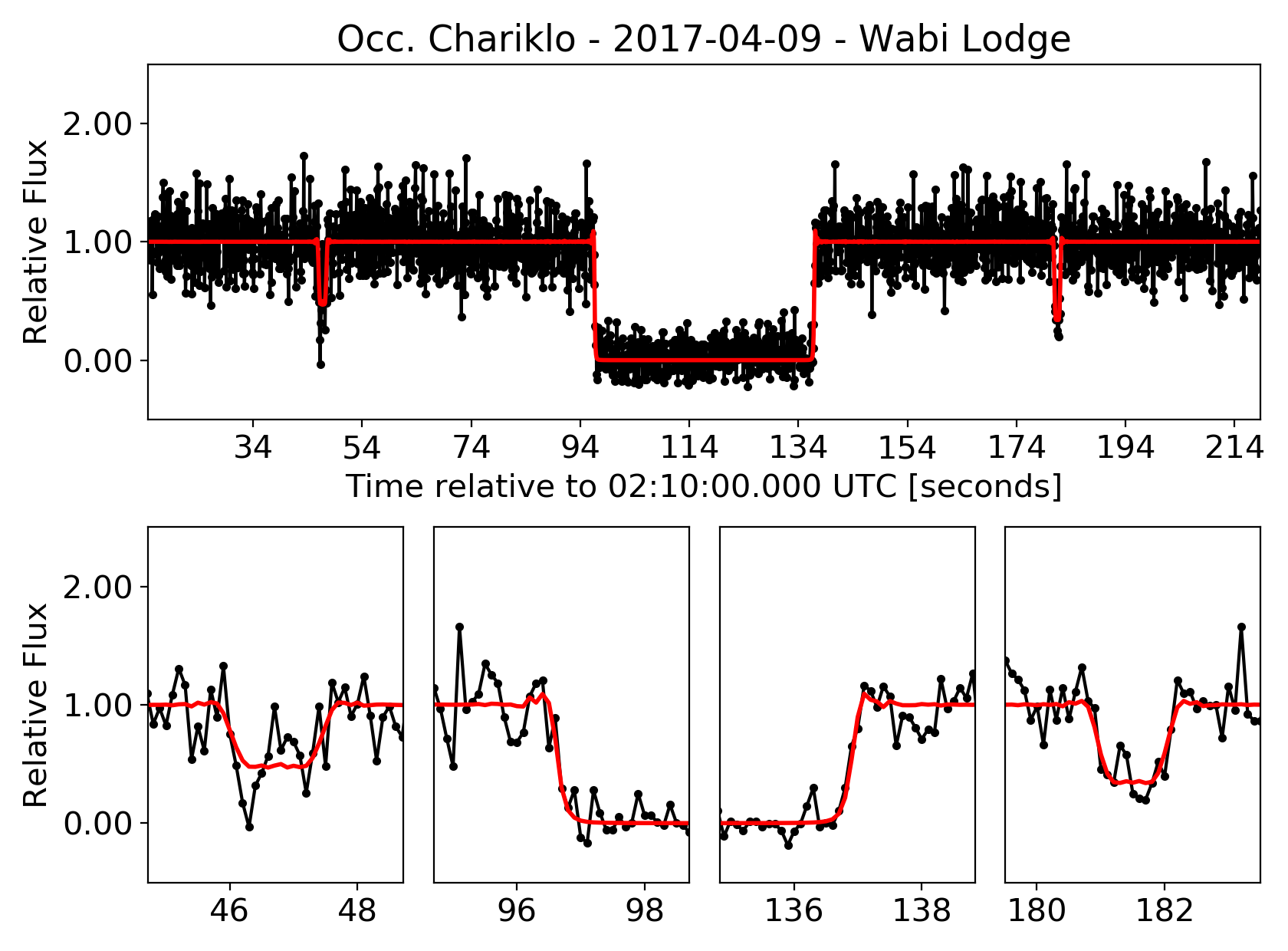}
\caption{Light curve obtained at Wabi Lodge on 2017-04-09.}
\label{Fig:lc_all_1}
\end{figure}               

\begin{figure}[h]
\centering
\includegraphics[width=0.50\textwidth]{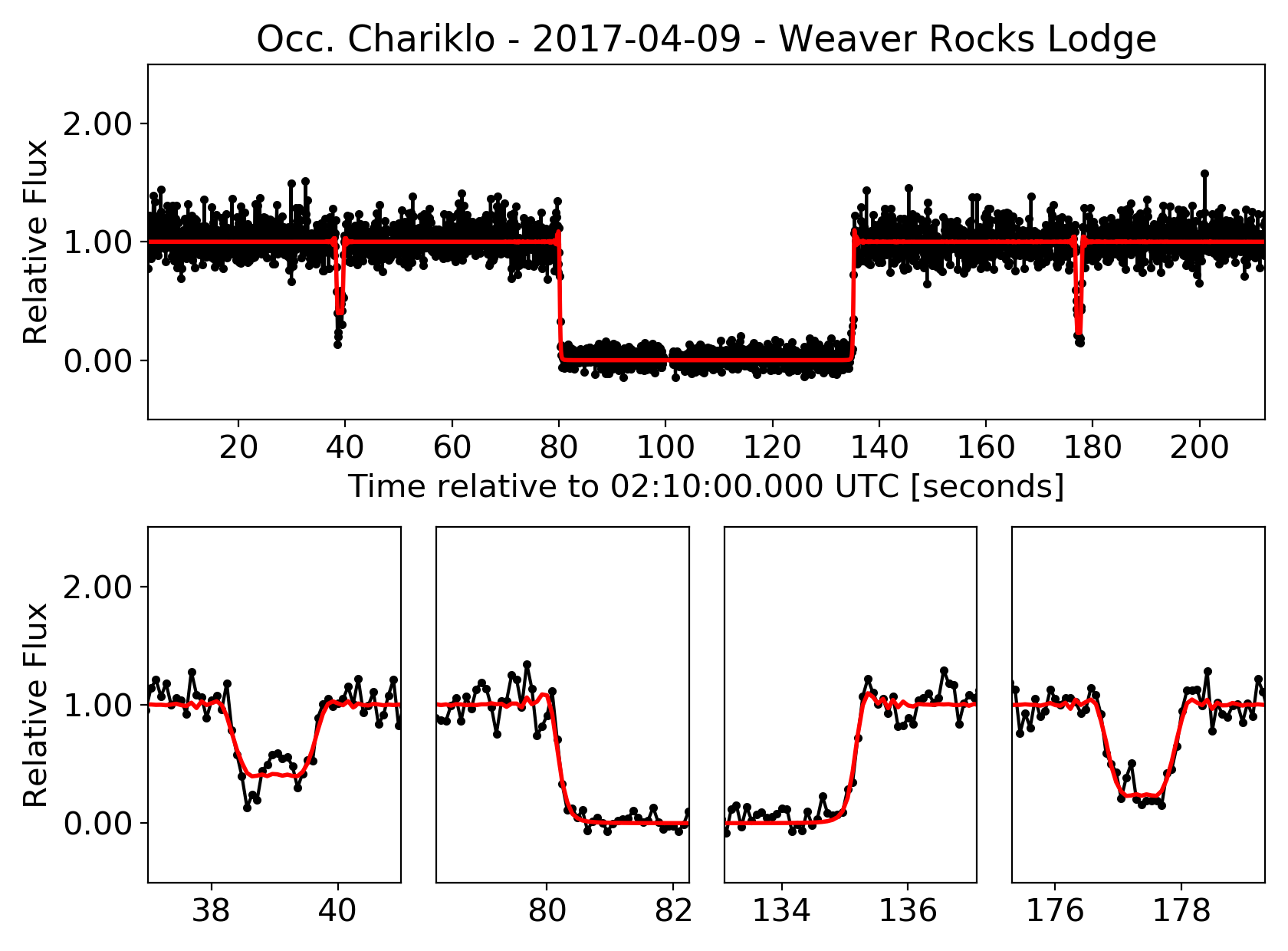}
\caption{Light curve obtained at Weaver Rocks Lodge on 2017-04-09.}
\label{Fig:lc_all_1}
\end{figure}               

\begin{figure}[h]
\centering
\includegraphics[width=0.50\textwidth]{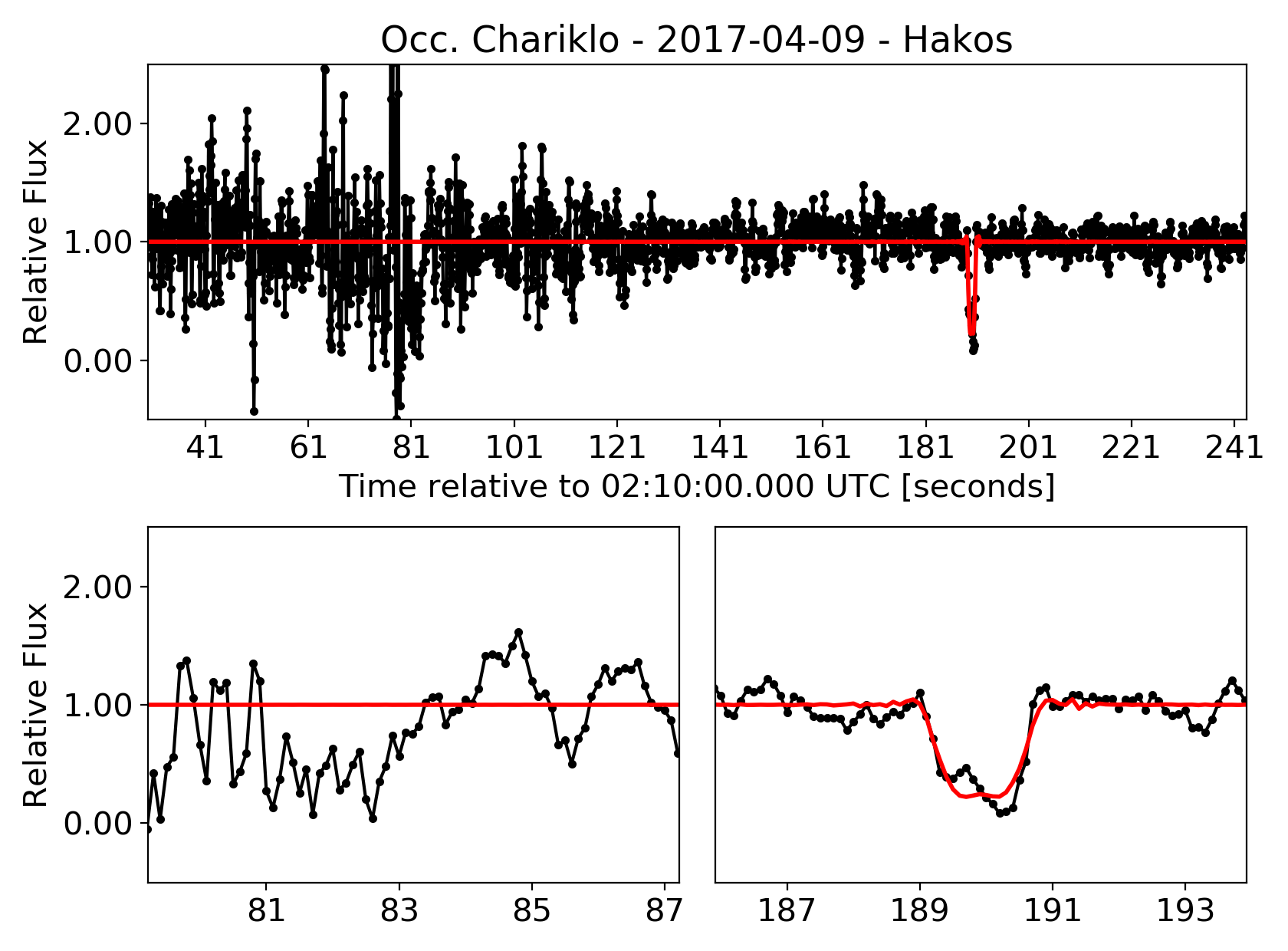}
\caption{Light curve obtained at Hakos on 2017-04-09.}
\label{Fig:lc_all_1}
\end{figure}

\end{appendix}

\end{document}